\documentclass{aa}  

\usepackage{graphicx}
\usepackage{txfonts}
\usepackage{natbib}
\bibpunct{(}{)}{;}{a}{}{,} 

\begin{document}

\title{A network of filaments detected by  
\textit{Herschel} \thanks{\textit{Herschel}
is an ESA space observatory with science instruments provided by European-led Principal Investigator
consortia and with important participation from NASA.} in the Serpens Core:
}
\subtitle{A laboratory to test simulations of low-mass star formation.}
   \author{V.~Roccatagliata\inst{1}, J. E.~Dale\inst{1,2}, T.~Ratzka\inst{3}, L.~Testi\inst{4,5,2}, A.~Burkert\inst{1,2},C.~Koepferl\inst{6,7},  A. Sicilia-Aguilar\inst{8,9}, 
   C.~Eiroa\inst{9}, B. Gaczkowski\inst{1} 
          }

\institute{Universit\"ats-Sternwarte M\"unchen, Ludwig-Maximilians-Universit\"at,
Scheinerstr.~1, 81679 M\"unchen, Germany\\
 \email{vrocca@usm.uni-muenchen.de} 
\and
Excellence Cluster `Universe', Boltzmannstr. 2, 85748 Garching bei M\"unchen, Germany
\and
Institute for Physics / IGAM, NAWI Graz, Karl-Franzens-Universit\"at, Universit\"atsplatz 5/II, 8010 Graz, Austria
\and
ESO, Karl-Schwarzschild-Strasse 2 D-85748 Garching bei M\"unchen, Germany
\and
INAF-Osservatorio Astrofisico di Arcetri, Largo E. Fermi 5, I-50125 Firenze, Italy
\and
Max Planck Institute for Astronomy, K\"onigstuhl 17, 69117 Heidelberg, Germany
\and
Max Planck International Research School for Astronomy and Cosmology, Heidelberg, Germany
\and
SUPA, School of Physics and Astronomy, University of St Andrews, North Haugh, St Andrews KY16 9SS, UK
\and
Departamento de F\'isica Te\'orica, Facultad de Ciencias, Universidad Aut\'onoma de Madrid, 28049 Cantoblanco, Madrid, Spain
}
   \date{Received 31 October, 2014; Accepted 10 June 2015}

  \abstract
   {Filaments represent a key structure during the early stages of the star formation process. Simulations show filamentary structure commonly formed before and during the formation of cores. }
   { The Serpens Core represents an ideal laboratory to test 
   the state-of-the-art of  simulations of 
   turbulent Giant Molecular Clouds. }
   {We use {\it Herschel} observations of the Serpens Core to compute temperature and column density maps of the region. 
   Among the simulations of \citet{Daleetal2012a}, we select the early stages of their Run I, before stellar feedback is initiated, 
    with similar total mass and physical size as the Serpens Core.  
   We derive temperature and column density maps also from the simulations. 
   The observed distribution of column densities of the filaments has been analysed first including and then masking the 
   cores. 
   The same analysis has been performed on the simulations as well. }
   {A radial network of filaments has been detected in the Serpens Core. The analysed simulation shows a striking morphological 
   resemblance to the observed structures. 
   The column density distribution of simulated filaments without cores shows only a log-normal distribution, while the observed filaments show a 
power-law tail. The power-law tail becomes evident in the simulation 
if one focuses just on the column density distribution
of the cores. In contrast, the observed cores show a flat distribution.
}
   {
    Even though the simulated and observed filaments are subjectively similar--looking, we find that they behave in very different ways. The simulated filaments are turbulence-dominated regions, the observed filaments are instead
self-gravitating structures that will probably fragment into cores. }


   \titlerunning{{\it Herschel} observations of the Serpens Core. } 
   \authorrunning{V. Roccatagliata et al.}

  \maketitle
   %

\section{Introduction}
Filamentary molecular structures are commonplace in giant molecular clouds (GMCs). Large-scale filaments have been detected in extinction maps of low-mass star forming regions \citep[e.g.][]{Cambresy1999, Kainulainenetal2011} or in the sub-/millimeter wavelength range in high-mass star forming regions \citep[e.g.][]{JohnstoneBally1999, gutermuthetal2008}. In the last years observations have been carried out using the  {\it Herschel} telescope in nearby 
star-forming regions, as well as in regions where star formation is not currently active \citep[e.g. the Polaris Flare region, ][]{Andreetal2010}.  A characteristic filamentary structure consists of a main filament with some sub-filaments converging on it. This is the case in low-mass star-forming regions  \citep[e.g.~the B213 filament in Taurus, or the Corona Australis region; ][]{Palmeirimetal2013, Sicilia-Aguilaretal2013}, as well as in high-mass star-forming regions \citep[e.g.~the DR21 ridge and filaments in Cygnus; ][]{Hennemannetal2012}.

\noindent 
Filaments are thought to be the consequence of thermal,  dynamical, gravitational or magnetic instabilities. These can be caused by collisions 
of large-scale warm flows that are themselves generated by the turbulent velocity fields that are also ubiquitous in GMCs. 
The  filaments originated by gravitational instabilities can be either produced in dense two-dimensional structures \citep[][]{BurkertHartmann2004} or 
resulting from the expansion of a feedback--driven bubble  \citep[][]{Krauseetal2013c}, or from the collision of two GMCs. 
An extensive review on the possible mechanisms of filament formation is presented in  \citet{Andreetal2014} and \citet{Molinarietal2014}. 

\noindent Numerical simulations of star formation on GMC-scales have in recent years become sophisticated enough that detailed comparisons with observations, beyond simple statistics such as star formation efficiencies and stellar mass functions, are required to move forward. 
Attempting to model a given object or system in detail is fraught with peril, since the initial conditions from which to begin the model are of course not known, and even the present-day state of the 
system is unlikely to be understood in enough detail. However, many simulations appear to be able to reproduce general features of star-forming clouds, such as filamentary structures in the 
gas \citep[e.g.][]{MoeckelBurkert2014}. To determine if this resemblance is more than merely subjective requires state-of-the-art observations of candidate systems and detailed comparisons with suitable 
simulations.

\noindent 
Recently, a comparison between GMC simulations and observations has been presented by \citet{Smithetal2014} in order to investigate the general observational result 
that filaments appear to have a surprising constant width of $\approx0.1$pc regardless of the structure or properties of their host cloud \citep[][]{Arzoumanianetal2011, Andreetal2014}. 
The authors performed hydrodynamical simulations of molecular clouds with turbulent velocity that is responsible for complex filamentary 
fields.  
They found that the widths of the filaments depend on the data range fitted by a Gaussian function, with an average full width half maximum (FWHM) of 0.3 pc. 
This value is in agreement with theoretical predictions for accreting filaments  \citep[][]{Smithetal2014}. 

\noindent 
A different approach, presented here in this paper, consists in the comparison between a specific GMC numerical simulation from \citet{Daleetal2012a} and a particular real object, namely a region of the Serpens molecular cloud.

\noindent The Serpens molecular cloud represents an ideal nearby laboratory to study on-going low-mass star formation. The distance 
to the Serpens Core of 415$\pm5$\,pc has been recently revised with astrometric observations obtained with the VLBA \citep[][]{Dzibetal2010}. The authors also estimated 
a distance for the entire Serpens molecular cloud of 415$\pm25$\,pc. 

\noindent
The region is located north of the Galactic plane and part of the  Aquila Rift. \citet{Konyvesetal2010} and \citet{Bontempsetal2010} presented the analysis of the {\it Herschel} observations of the Aquila Rift molecular complex and Serpens South, which are both in the southern part of the Serpens molecular complex. Most  of the protostars resolved in the {\it Herschel}  observations were newly discovered. In the Aquila rift the spatial distribution of protostars indicates three star-forming regions, the richest of which is the W40/Sh2-64 HII region.

\noindent
The new {\it Herschel} maps of Serpens presented in this paper are dominated by the so-called `Serpens Core Cluster' and the complex around a group of four T Tauri stars, called G3 to G6 (see Figure~\ref{rgb_pacs_spire}). In their Spitzer study  \citet{Harveyetal2006} called the Core Cluster `Cluster A' and the southern cluster around Ser G3-G6 `Cluster B'. Located further south, the prominent Herbig Be star VV Ser \citep[][]{HarveyDunham2009} is visible in the 70\,$\mu$m {\it Herschel}  map as a point source. 
As highlighted in the review of \citet[][]{Eiroaetal2008}, a peculiarity of the Serpens Core is that all the phenomena related to 
active star formation have been discovered, from still in-falling gas to pre-protostellar gaseous and dusty condensations, as 
well as already-formed Class 0 to Class II young stars. A similar scenario is found also in other young star-forming regions as, 
for example, the Coronet cluster \citep[][]{Sicilia-Aguilaretal2011}, or IC1396A \citep{Sicilia-Aguilaretal2014a, Sicilia-Aguilaretal2015}. 

\noindent 
In the millimeter regime \citet{Enochetal2007} detected sources in Cluster A, but also in Cluster B and in 
the bright filament  north-east of it. 
Several sub-millimeter sources have been detected in the  Core Cluster. They are associated with a south-eastern and a north-western 
clump \citep[][]{Davisetal1999, Casalietal1993}.  Continuum sub-millimeter observations carried out with the SCUBA bolometer array camera on the James Clerk Maxwell Telescope (JCMT) at 450 and 850\,$\mu$m revealed a prominent diffuse filamentary structure in the south and east direction from the south-eastern cluster of sub-millimeter sources. 
The outflows driven by these objects have been locally constrained close to the sources by CO observations \citep[][]{Whiteetal1995, Duarte-Cabraletal2010}. 
Observations of molecular gas  obtained with the IRAM 30m telescope 
revealed two velocity components between the two sub-clusters in the Serpens Core \citep[][]{Duarte-Cabraletal2010}. 
In their paper the authors proposed that after an initial triggered collapse, a cloud-cloud collision between the northern and southern clouds is responsible for 
the observed characteristics of the two sub-clusters. 
Observations of $N_2$H$^+$, HCO$^+$, and HCN toward the Serpens Main molecular cloud have been recently obtained using CARMA 
\citep[][]{Leeetal2014}. The authors find that  while the north-west (NW) sub-cluster in the Serpens Core shows a relatively uniform velocity, the south-east (SE) sub-cluster shows a more complicated velocity pattern. In the SE 
sub-cluster they also identified six filaments: those located in the northeast of the SE sub-cluster, have larger velocity gradients, smaller 
masses, and nearly critical mass-per-unit-length ratios, while the opposite properties have been found in the filaments in the  southwest of the 
SE sub-cluster.  
 
\noindent In this paper we present the continuum {\it Herschel} observations of the Serpens Core and compare them to a numerical simulation 
performed by \citet{Daleetal2012a}. 
We organize the paper as follows: in Section~\ref{obs} we summarize the  
{\it Herschel} observations and the data reduction. 
In Section~\ref{an} we present the analysis carried out including the  derivation of the temperature and the column density maps, while 
in Section~\ref{sim} the simulations used in our study are described. In Section~\ref{discussion} we compare the properties of the observed filaments with 
the filaments formed in the simulations.    
A summary and conclusions are given in Section~\ref{conclusions}. 

\section{Observations \& data reduction}
\label{obs}
The {\it Herschel}  data of Serpens were taken from the {\it Herschel}  Science Archive (HSA) and are part of the {\it Herschel}  Gould Belt survey \citep[][]{Andreetal2010}. 
The observations (obsIDs: 1342206676 and 1342206695) were made on October 16th, 2010 and cover an area of $\approx 2.5\deg \times 2.5\deg$. The Serpens cloud was simultaneously imaged with {\it Herschel} \citep[][]{Pilbrattetal2010} in five wavelengths, using the two cameras PACS \citep[][]{Poglitschetal2010} at 70 and $160\,\mu$m and SPIRE \citep[][]{Griffinetal2010} at 250, 350, and $500\,\mu$m. Two orthogonal scan maps were obtained by mapping in the parallel fast scan mode with a  speed of 60\arcsec/s. We  reduced the data using HIPE v10.3.0 for the calibration of the Level~0 PACS and SPIRE data. Then the Level~1 data of both instruments were taken to produce the final maps with the Scanamorphos package v23 \citep[][]{Roussel2013}, using the {\it /parallel} option for the reduction of observations with PACS and SPIRE in parallel mode and the {\it /galactic} option to preserve
brightness gradients over the field. To properly take care of glitches and artefacts that occurred in the PACS maps, we canceled the HIPE glitch-mask at the conversion of the Level~1 data to the Scanamorphos format. Additionally, we switched the {\it /jumps\_pacs} option on for the production of the final maps in order to first detect and then remove the jumps in the PACS mosaics. Those discontinuities can be caused 
either by glitches or by electronic instabilities. The pixel-sizes for the five maps at 70, 160, 250, 350, and $500\,\mu$m were chosen as 3.2\arcsec, 4.5\arcsec, 6\arcsec, 8\arcsec, and 11.5\arcsec, respectively. 
%
The recent version of Scanamorphos, in particular, takes into account the PACS distortion flat-fields. 

\noindent
%
 The final mosaics have been calibrated applying the {\it Planck offsets}. 
We adopt the rigorous approach presented by e.g. \citet{Bernardetal2010} which consists in combining the {\it Planck} data
 with the IRAS data  of the {\it Herschel} fields and extrapolating the expected total fluxes to the {\it Herschel} filter bands.
The difference between the expected and the observed fluxes provides the offsets to correct  
the {\it Herschel} mosaics. 
 The Planck offsets obtained at each filter are presented in Table~\ref{planck_corr}. 
%
These values are added to the Level 2 mosaics obtained with Scanamorphos. 
 \begin{table}
\caption{ Planck offsets of the {\it Herschel}  mosaics of the Serpens Core obtained using the method presented in \citet{Bernardetal2010}.}
\label{planck_corr}      
\centering               
\begin{tabular}{l  rrr}  
\hline\hline                
\noalign{\smallskip}
Instrument   &Filter & Offset\\
\noalign{\smallskip}
& [$\mu$m] & [MJy/sr]\\
\noalign{\smallskip}
\hline                
\noalign{\smallskip}
PACS	&70 & 12.2\\
		&160 & 111.6\\
SPIRE	&250 & 74.9\\
		&350 & 37.6\\
  		&500 & 15.8\\
 \noalign{\smallskip}
\hline                        
\noalign{\smallskip}
\end{tabular}
\end{table}

\section{Analysis}
\label{an}

In this section we present the analysis carried out on the {\it Herschel} observations of the Serpens molecular cloud. In particular, Section\,\ref{cloud} 
presents the morphological description of the  ``Serpens Core''  (Cluster A) and  the cluster ``Ser G3-G6'' (Cluster B).  
Section\,\ref{temp} presents temperature and column density  maps 
of the  region.  
 
 \begin{figure*}
\centering
\includegraphics[trim=0.5cm 0.1cm 1cm 0.1cm,width=18cm]{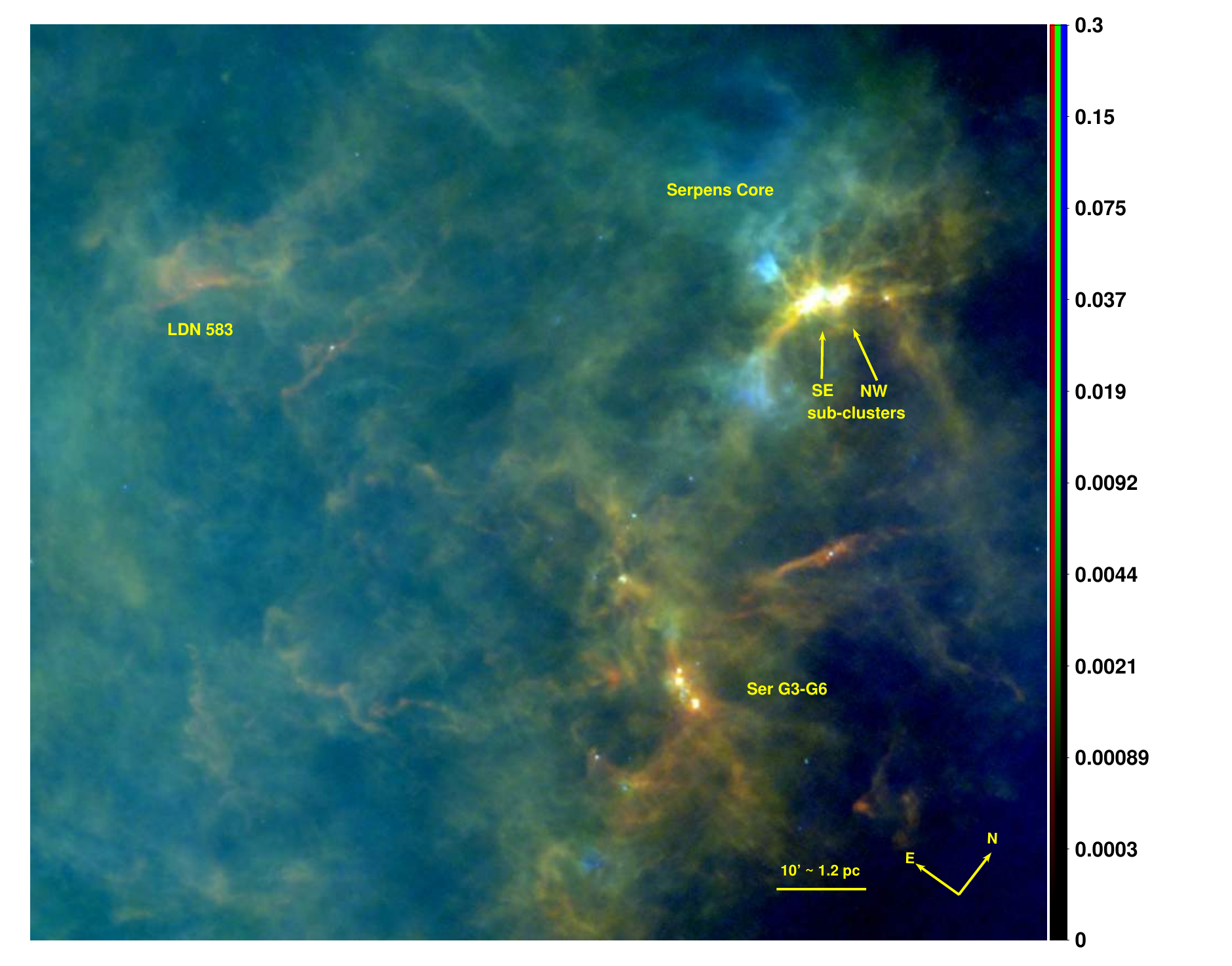}
  \caption{ Composite {\it Herschel}/PACS 70\,$\mu$m (blue), 160 \,$\mu$m (green) and {\it Herschel}/SPIRE 350\,$\mu$m (red) 
  mosaics in logarithmic scale. 
  The ranges are between 0 and 0.3 Jy/arcsec$^2$, 0 and 0.3 Jy/arcsec$^2$, and 0 and 0.05 Jy/arcsec$^2$  for the blue, 
  the green, and the red image, respectively. 
    The center of the figure corresponds to the position 
    $(\alpha_{\rm J2000},\, \delta_{\rm J2000})\, =\,(18^{\rm h}\,30^{\rm m}\,49^{\rm s},\, 
    + 00\degr\,39'\,24'')$ and the size of the region is $1.7\degr\times\,1.9\degr$. }
     \label{rgb_pacs_spire}
\end{figure*}

\subsection{Molecular cloud morphology}
\label{cloud} 
Figure\,\ref{rgb_pacs_spire} shows the composite of  {Herschel}/PACS 70\,$\mu$m, 160 \,$\mu$m, and  SPIRE 350\,$\mu$m. 
All the cloud structures become brighter at longer wavelengths. 
The entire {\it Herschel} maps in the eastern part extend up to the far-infrared bright region LDN 583.  This region appears to 
be almost empty, with only a cloud structure, prominent at the SPIRE wavelengths, with a point source, in the South-West direction 
from LDN 583.

\noindent
The Serpens star-forming region shows a highly filamentary morphology. The Serpens  Core Cluster seems to be connected  by several filaments to the underlying molecular cloud. These filaments have a radial structure similar to spokes of a wheel with the cluster forming the hub. This morphology is a common feature of star-forming regions \citep[e.g.][]{Myers2009, Hennemannetal2012}. \cite{Schneideretal2012rosette} note that  the junction of filaments seems to be intimately connected with the formation of star clusters, as was also found by \cite{DaleBonnell2011} in numerical simulations, suggesting that \emph{gas flows along the filaments and accumulates at the hubs.} The filaments are very structured showing kinks and condensations. In particular, the westernmost filament harbors a bright compact source. With respect to the colors, the eastern filaments show a remarkably bright emission at 70\,$\mu$m at  their tips. 

\noindent
 The Ser G3-G6 cluster 
seems to be aligned with a filament and associated with several bright sources in its vicinity. The filaments or cloud edges seen in this region 
are highly structured and reminiscent of turbulence. 

\subsection{Temperature and  column density of the clouds}
\label{temp}
The cloud temperature has been obtained using two different approaches: the first approach uses the ratio of the fluxes  at  70\,$\mu$m and 160\,$\mu$m, while the second is based on the fit of the SED 
fluxes between 70 and 500\,$\mu$m \footnote{ We highlight that this method can have some caveats in using also the 70\,$\mu$m map, since the emission might trace a hotter dust population compared to the colder dust mapped at longer wavelengths. }. 
 In both cases we used the {\it Herschel} maps after applying the {\it Planck} offsets. 
While the first approach conserves the angular resolution of the 160\,$\mu$m image \citep[as done in ][]{Preibischetal2012, 
Sicilia-Aguilaretal2014a}, the second uses the PACS and SPIRE images 
all convolved to the 500\,$\mu$m resolution \citep[as done in ][]{Roccatagliataetal2013}. 
The fit of the fluxes is obtained by using a black-body leaving as free parameters the temperature $T$ and the surface 
density $\Sigma\, [{\rm g/cm^2}]$, with the same procedure presented in \citet{Roccatagliataetal2013}. 
The dust properties are fixed by interpolating at the {\it Herschel}  bands the values of the dust mass absorption coefficient $\kappa_\nu$ from the model of \citet{OssenkopfHenning1994} with a $\beta$ value of 1.9, 
which adequately describes a dense molecular cloud. 
As already found and commented by \citet{Roccatagliataetal2013}, the SED fitting leads to 
slightly colder temperatures. 
 Since the temperature $T$ and the surface 
density $\Sigma\, [{\rm g/cm^2}]$ are two independent parameters, the best fit parameters have been obtained minimizing $\chi^2$.  
For the further analysis presented in the paper we use the temperature and surface 
density maps obtained  with this SED-fitting method.  

\noindent 

The column density is computed using the relation
\begin{equation} 
\label{col_density}
    N_{\rm H} = 2  N_{\rm H_2} = \frac{2\cdot \Sigma \cdot R}{m_{\rm H}\cdot\mu_{\rm H_2}}  
\end{equation} 
where $\Sigma$ is the surface density, m$_{\rm H}$ is the hydrogen mass (i.e. 1.6736e-24 $g$) and $\mu_{\rm H_2}$ is the 
mean molecular weight (i.e. 2.8). 
Multiplying by the gas-to-dust mass ratio $R$ (assumed to be 100), we obtain the total column density.

\noindent
 In the entire region covered by {\it Herschel} observations we find that the average temperature of most of the nebula is 
about 24\,K. It is interesting to notice that the coldest part  of the cloud, of about 19\,K, corresponds to the filamentary material converging into the 
central core.

\noindent
The total mass for each region considered in our study is computed summing up the column densities of all the pixels. 
Using a distance of the Serpens molecular cloud of 415\,$\pm$25\,pc, the total mass (of dust and gas) over a region of $2.0^\circ\times2.2^\circ$ is 
 3213\,$_{-	375	}^{+	399	}$\,M$_\sun$. 
The errors have been computed considering the  error of 25\,pc on the distance \citep[][]{Dzibetal2010}, which represents the dominant error in 
this estimate.

\noindent
The total mass of the Ser G3-G6 cluster \footnote{computed in a box of $7.3'\times\,13.4'$ in size, center 
$(\alpha_{\rm J2000},\, \delta_{\rm J2000})\, =\,(18^{\rm h}\,28^{\rm m}\,58^{\rm s},\, 
    + 00\degr\,37'\,55'')$, and orientation 37$\degr$ in North-East direction.}
 and the LDN 583 region
 \footnote{computed in a box of $21.7'\times\,9.3'$ in size, center 
$(\alpha_{\rm J2000},\, \delta_{\rm J2000})\, =\,(18^{\rm h}\,33^{\rm m}\,37^{\rm s},\, 
    + 00\degr\,30'\,07'')$, and orientation 40$\degr$ in North-East direction.}
 is $98	^{+	12	}_{-	12	}$\,M$_\sun$ and $67	^{+	8	}_{-	8	}$\,M$_\sun$, respectively. The temperatures range between 11\,K and 16\,K in the case of the Ser G3-G6 cluster, while in the case of  the LDN 583 region they range between 
14\,K and 18\,K. 

\noindent
A zoom of the temperature and column density map of the Serpens Core is shown in Figure~\ref{pos_fil_obs}. 
The total mass present in the Serpens Core is  494	$_{-	58	}^{+	61	}$\,M$_\odot$,  
where the errors are computed from the
error in the distance. The region considered for the total mass of the Serpens Core is highlighted with a large box in Figure~\ref{pos_fil_obs} and is 42$'\times$42$'$, which corresponds to about 5\,pc\,$\times$\,5\,pc. 

\noindent
Using the extinction thresholds in the study of \citet[][]{Ladaetal2012}, we computed the following total masses  in the region of 5\,pc\,$\times$\,5\,pc of the Serpens Core:
\noindent
for $A_K>0.1$~mag  ($A_V\simeq0.9$~mag), the total cloud mass is  470$^{+60}_{-55}$\,M$_\sun$,  
and for $A_K>0.8$~mag  ($A_V\simeq7.3$~mag) 
 100$^{+13}_{-10}$\,M$_\sun$. 
The fraction of dense material present in the Serpens Core (using as threshold $A_K>0.8$~mag) is $\sim\,20$\%.
The star formation rate of the Serpens Core has been computed by \citet{Harveyetal2007}. Rescaling  their value with the updated distance of 415\,pc of the 
Serpens Core, we obtain 28\,M$_\odot$/Myr. 
%
%
  These values for the Serpens Core agree well with the relation derived for other low-mass molecular clouds by \citet[][]{Ladaetal2012, Ladaetal2010} and \citet[][]{Heidermanetal2010}. 
%
This relation has been further debated by \citet{BurkertHartmann2013}, who do not find  
a particular threshold for star formation but instead a strong continuous increase of the star formation efficiency with density.

\subsubsection{Comparison with previous mass determinations: dust vs. gas masses}
We compare the total mass computed for the Core region with previous dust and gas mass estimates. 
Assuming a distance of the Serpens molecular cloud of  311~pc \citep{DeLaraetal1991}, \citet{Whiteetal1995} estimated a lower mass limit  of 1450\,M$_\sun$ from the 
C$^{18}$O (J=2-1) data and 400\,M$_\sun$ from the C$^{17}$O data. 
  We scaled their masses to the new distance of 415 pc and computed the masses of the same  regions from our data. 
We found a total mass about 7 times lower  than the gas mass estimate from the C$^{18}$O. A possible reason for this discrepancy might be   a different value of the gas-to-dust ratio which is always assumed to be 100 in our calculations.   
 
\noindent
\citet{TestiSargent1998} computed the total mass from the 3 mm continuum emission in the inner part of the 
Serpens Core within 5\arcmin$\times$5\arcmin. Out of the 4 cores detected by  \citet{TestiSargent1998}, only the mass of 3 of them is reported in \citet{Testietal2000}.
 We scaled their masses to the 415\,pc  distance  and we computed the masses using our column density map on the same positions of their cores. 
Only Core D, which corresponds to the position of S68N, has the same mass using both approaches. Cores B and C are both three times less massive than the masses derived at 3\,mm. 
\citet{Leeetal2014} presented the properties of 18 cores detected in the Serpens Core in the 3\,mm continuum data obtained with CARMA. 
The  Core B and Core C in  \citet{TestiSargent1998} correspond to the cores S11 and S7, respectively,  in 
\citet{Leeetal2014}. They computed a mass of about 6\,M$_\sun$ for S11 and 11\,M$_\sun$ for S7. We highlight that the mass of S11 from \citet{Leeetal2014} corresponds to the mass we obtained. The mass of S7 differs only marginally,  
while other cores are about two times smaller than the values  computed by \citet{TestiSargent1998} and \citet{Testietal2000}.
%

\noindent
\citet{Duarte-Cabraletal2010} measured gas temperatures and masses of the clumps identified in the Serpens Core. 
The clumps on the C$^{17}$O $J$ = 1 - 0 channel maps  have been derived by \citet{Duarte-Cabraletal2010}  using the 2D{\sc CLUMPFIND} algorithm,  
assuming a constant temperature of 10\,K, a mean molecular weight of 2.33, and a C$^{17}$O fractional abundance of 
4.7$\times$10$^{-8}$ with respect to H$_{\rm 2}$.  They computed also the values of the virial mass of the clumps  
from the observed velocity dispersion. The ratio between the virial mass ($M_{\rm virial}$) 
and the gas clumps mass ($M_{\rm gas}$)  gives the information whether the structure is gravitationally 
bound {\bf($M_{\rm virial}$/$M_{\rm gas}$$ \le$1)} or unbound ($M_{\rm virial}$/$M_{\rm gas}>$1). According to this definition, cores D, E, F, and G 
are unbound. All the dust masses computed from 
our {\it Herschel} maps  in correspondence to the clump positions  agree  well with the 
estimates based on the sub-millimeter and millimeter fluxes. We highlight that the total mass that we compute for the dust mass assumes that the gas-to-dust ratio is 100. However the adopted canonical value has an intrinsic uncertainty of at least a factor of 2, especially locally. This can explain the small differences found in 
the presented comparison.
\section{Star formation simulations. }
\label{sim}

\indent \citet{Daleetal2012a} simulated the evolution of a group of model GMCs distributed in (mass, radius, velocity dispersion) parameter space so as to mimic the properties of Milky Way clouds inferred by \citet{Heyeretal2009}. The clouds range in mass from $10^{4}$--$10^{6}$\,M$_{\odot}$ and in initial radius from $5$--$90$\,pc. The initial condition of the simulations is an idealised spherical cloud with mild Gaussian density profiles giving a degree of intrinsic central concentration, seeded with a supersonic turbulent velocity field. The velocity field initially has the power spectrum of Burgers turbulence with $P(k)\propto k^{-4}$ and wave numbers between 4 and 128 are  initially populated. The normalisation of the velocity field is adjusted to give all the clouds an initial virial ratio of 0.7, so that they are all formally bound. This results in a range of velocities of $\approx2$--$10$\,km s$^{-1}$.

\noindent The density field responds rapidly to the velocity field and the clouds develop complex structures of sheets and filaments from the interactions of turbulent shocks. Despite being formed by turbulence, many of these structures are persistent, although they are often advected through the clouds by the flows from which they formed. On timescales comparable to the cloud freefall times ($\sim1$\,Myr), sufficient mass accumulates in some dense structures for subregions of the clouds to become self--gravitating and to begin forming stars.

\noindent The purpose of the simulations was to examine how the effects of massive stellar feedback in the form of photoionisation and winds vary with the properties of the host clouds. However, these calculations have other uses. In particular, one can search through the parameter space (including the time dimension) to find a simulation, or a part of a simulation, which is similar to a given real system. Serendipitous similarities are likely to be more trustworthy than targeted simulations because no assumptions have been made to achieve a desired result and one can be confident that the result is not there merely because it was inserted from the beginning.

\noindent  Amongst the calculations of \citet{Daleetal2012a}, we find that the innermost 5pc of their Run I, at a time that we refer to as $t_{\rm SC}$ (0.25 Myr after the initiation of star formation but before stellar feedback is initiated), bear a striking (subjective) morphological resemblance to the Serpens Core. The gross physical characteristics such as mean column and volume densities, and temperatures are also numerically sufficiently similar to make a more detailed comparison worthwhile. We therefore treat this subregion of the Run I calculation as a model of the Serpens Core, in the same way that the real Serpens Core is a subregion of its parent molecular cloud.

\noindent   A $5\times5\times5$pc box (shown in projection in Figure 3) near the simulation center of mass at this epoch exhibits a pc-scale core containing a small cluster of three stars with a total mass of 14 M$_{\odot}$, with several dense filaments radiating from it, one of which joins with a smaller core approximately one pc distant in projection (and in three dimensions, in fact).  
The masses of the larger and smaller cores are $\sim\,100\,M_\odot$  (Box 1 in Figure~\ref{reg_sim}) and $\sim\,30\,M_{\odot}$  (Box 2  in Figure~\ref{reg_sim}), and  total gas mass column and volume densities  are $\approx1$g cm$^{-2}$ and $\sim10^{4}$-$10^{5}$cm$^{-3}$, respectively. There are also several other filamentary structures connected with the two cores.

\noindent 
The filaments in the simulation were originally created by colliding flows in the supersonic turbulent velocity field imposed as part of the simulation initial conditions. The clouds in the simulations of \citet{Daleetal2012a} are all initially mildly centrally condensed and gravitationally bound, so that the densest structures tend to form at the bottom of the clouds' gravitational wells near their barycenters. However, owing to the different mean densities and turbulent velocities of the clouds, the properties of the filaments vary between simulations. Run I is the simulation with the smallest mean turbulent velocity and a fairly low mean density, so the densities of the filaments are also comparatively low. They form a hub structure and material flows \emph{along} them towards the common hub, where it accumulates and forms a dense cluster of stars. The timestep,  t$_{\rm SC}$, chosen for this comparison occurs at the very beginning of the formation of the cluster, when there are only a few stars present. 
We also use a second snapshot 0.4\,Myr further forward in time to examine how the characteristics of the central region of the simulation vary on timescales of the order of the freefall time, in  particular the column density distribution.

\noindent 
In contrast to what occurs in other simulations from \citet{Daleetal2012a}, the filaments in Run I do not become gravitationally unstable and do not fragment to form stars themselves because the freefall timescale for fragmentation to occur within them is longer than the timescale on which flows along the filaments transport material into the hub where they join. They are long-lived structures whose function over $\sim$Myr timescales is largely to deliver fresh gas to the forming cluster located near the cloud's center of mass.

\subsection{Analysis of the Simulations}

\noindent  We identify in Figure~\ref{reg_sim} three filamentary structures of the Run I simulation at the selected time t$_{\rm SC}$, one of which physically connects the two cores. The masses of the filaments are in the range 20--40$\,M_{\odot}$, comparable to the mass of the small core and a few times less than that of the larger core  (see Table~\ref{table_filam_1}). The flow velocities of the gas are of the order of 1 km s$^{-1}$. The filaments do not fragment and most of the gas they contain is accreted by the cores (and the clusters they eventually form). It is simplistic, however,  to describe the gas as only flowing along the filaments. At early times when the cloud velocity field is still dominated by the original imposed turbulence, the filaments are being advected by larger--scale flows in directions which are not necessarily parallel or perpendicular to their long axes. The velocity field in the neighbourhood of the two cores is extremely complex and, at this early stage in the simulation, has not yet settled down into the simpler configuration of several filaments feeding a common hub.

\noindent The larger and smaller cores begin to fragment into stars and merge into a single object after $\approx1$Myr and, after this occurs, the 
cluster continues to grow by accretion in a morphologically simpler filament--hub configuration. From a control simulation without feedback, it is clear that this 
morphology would be stable for a protracted period of time, even while the central protocluster continues to grow and form stars. 
It rapidly destroys both the core and the filament network, leaving an eroded pillar and a fully--exposed cluster.
We analyse the raw temperature and column density maps extracted directly from the hydrodynamical simulations without performing any radiative transfer calculations. 
 We rescale the pixel scale and convolve the surface density maps of the simulations with the Point Spread Function (PSF) of SPIRE at $500\,\mu$m \citep{Gordonetal2008}. We apply the {\sc FluxCompensator} \citep[as done in ][]{Koepferletal2015} to perform this task \footnote{The  {\sc FluxCompensator} was designed to produce realistic synthetic observations from radiative transfer modelling \citep[e.g. Hyperion by ][]{Robitaille2011}. By modelling the effects of convolution with arbitrary PSFs, spectral transmission curves, finite pixel resolution, noise and reddening, hydrodynamical simulations - through radiative transfer modelling - are directly comparable with real observations (\url{www.mpia-hd.mpg.de/~koepferl/}). The tool will be publicly available in the future. }. 
We use the same conversion factor as in the observations to convert projected gas mass surface densities to column densities  as in Equation~\ref{col_density}. 

\begin{figure*}
\centering
\includegraphics[trim=1.0cm 0.05cm 1.0cm 0.8cm,width=9cm]{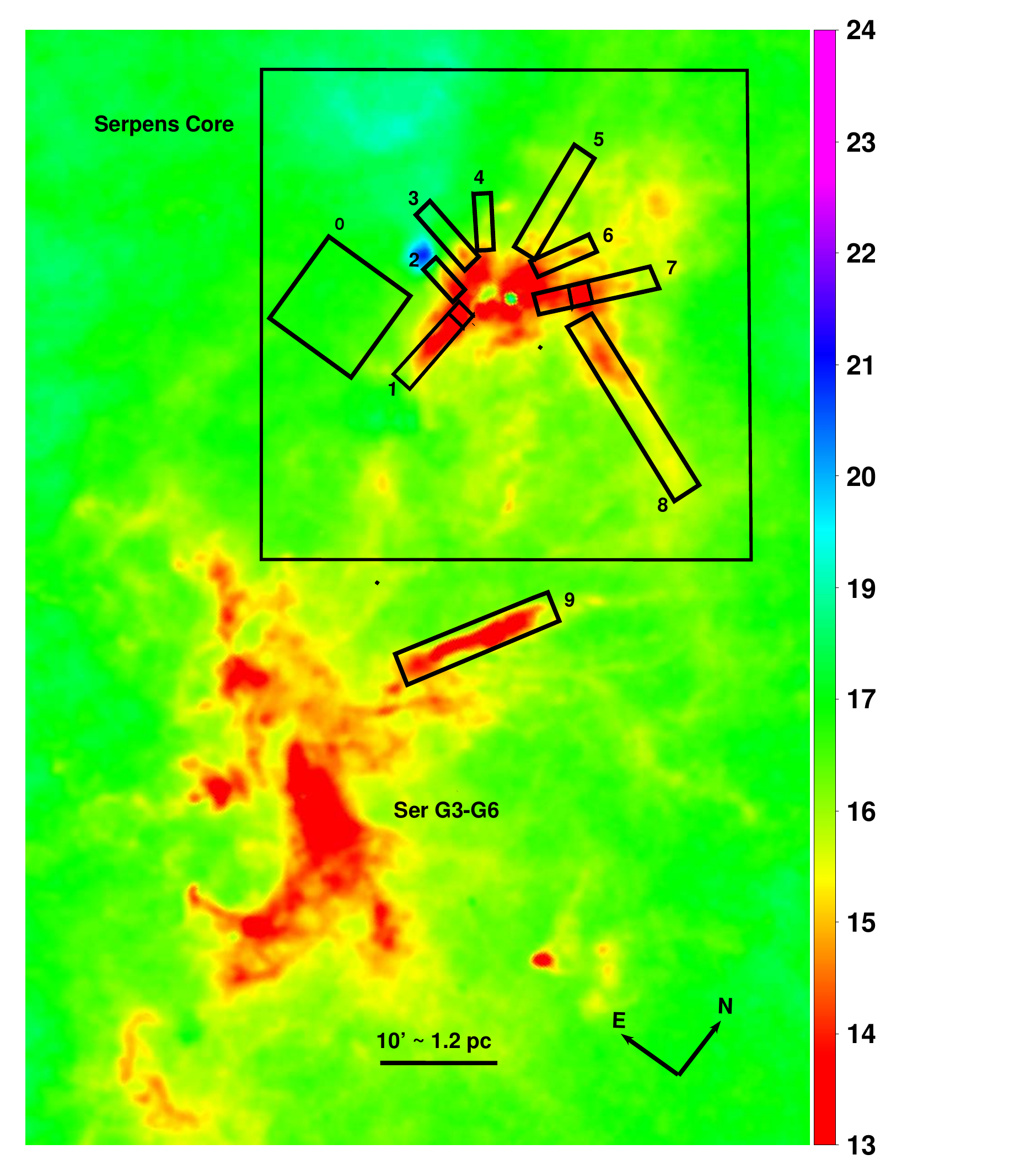}
\includegraphics[trim=1.0cm 0.05cm 1.0cm 0.8cm,width=9cm]{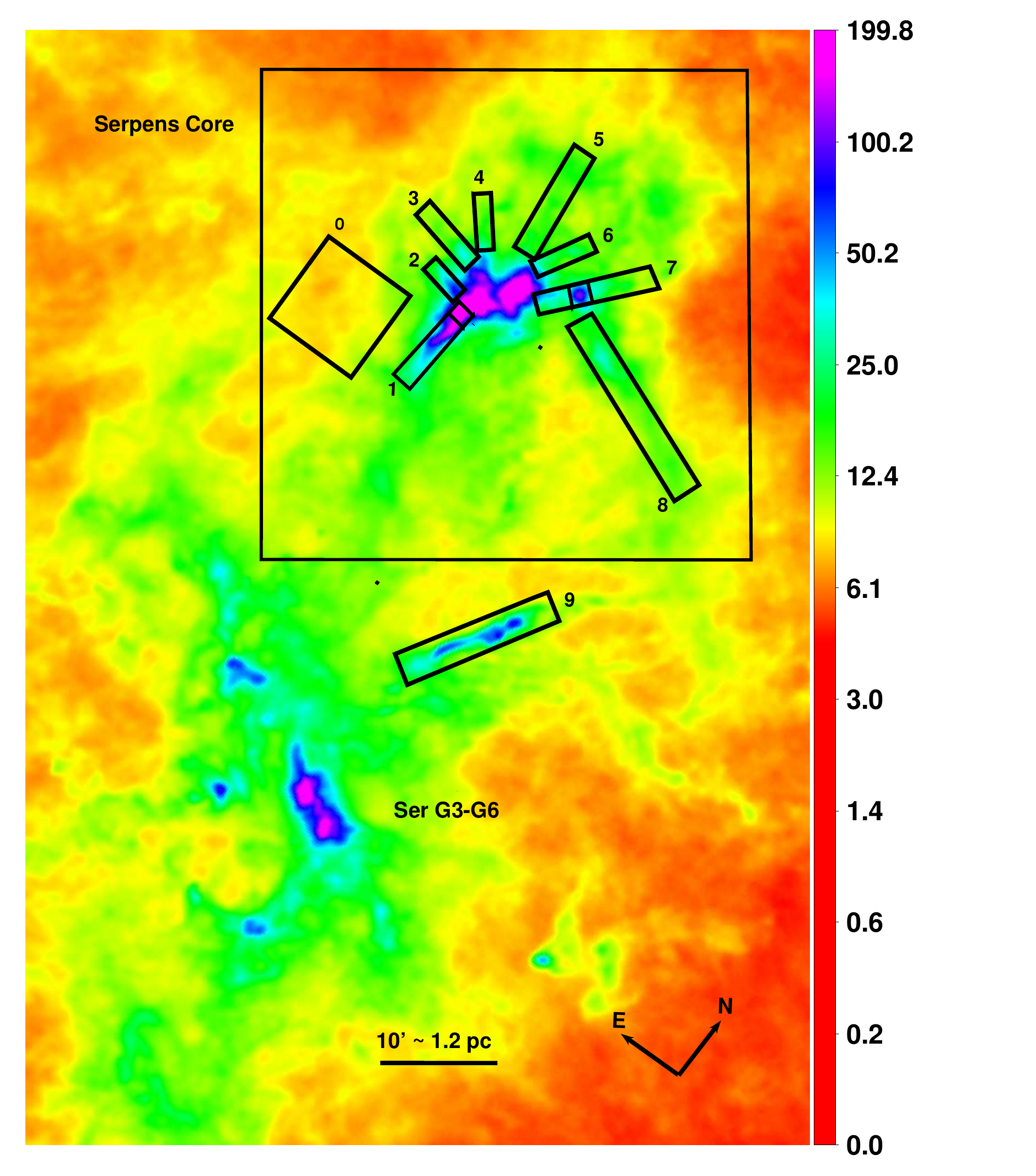}
   \caption{  {\it Left:} Temperature map of the Serpens Core in ${\rm K}$.  {\it Right:} Column density map ($N_{\rm H}$) of the Serpens  
   Core in [$\times$10$^{+20}cm^{-2}$]. The center of the image 
   corresponds to the position $(\alpha_{\rm J2000},\, \delta_{\rm J2000})\, =\, (18^{\rm h}\,29^{\rm m}\,18^{\rm s},\, 
   + 00\degr\,51'\,08'')$ and the size is 94$'\,\times\,$67$'$. 
   In both panels the large boxes define the position of the whole Serpens Core. The nine filaments considered in our analysis are highlighted by 
   numbered boxes. In filaments 1 and 7 the small squared boxes highlight the positions of the cores. Box 0 represents the control field.}
     \label{pos_fil_obs}
\end{figure*}

\section{Discussion: observed filaments versus simulated filaments.}
\label{discussion}

In order to compare the properties of the filaments in the observations of the Serpens Core to those formed in the simulation, we carry out in parallel the same 
analysis on the observations and simulations. The pixel scale of both, observations and simulations, is about 0.02 pc/px. Both  are convolved to the spatial resolution of  SPIRE at 500\,$\mu$m. 

\noindent
 In our analysis, we refer to compact enhancements in column density as `filaments' if they are long and linear, and as `cores' otherwise. Using this definition in the observations, the cores always contain point source counterparts at 70\,$\mu$m. The analysis of the observed filaments is carried out by defining by eye a box around each filament. The positions of the 9 filaments analysed are 
highlighted in Figure~\ref{pos_fil_obs}. The filaments numbered 1 to 8 are converging into the Serpens Core, while the ninth is placed between the Serpens Core and the 
Ser G3-G6 region.  In the appendix we zoom in on each filament, showing the {\it Herschel}, the column density, and the temperature maps. In order to define the boxes of the cores present 
in filaments 1 and 7, we checked  the increase in column density and the emission at 70 and 160~$\mu$m, where the point 
sources are resolved. The boxes which include the cores are much larger than the point sources resolved  at 70~$\mu$m: this 
conservative approach allows to exclude any contamination by the cores in the analysis of the filamentary structures 
alone.

\begin{figure*}
\centering
\includegraphics[width=8cm]{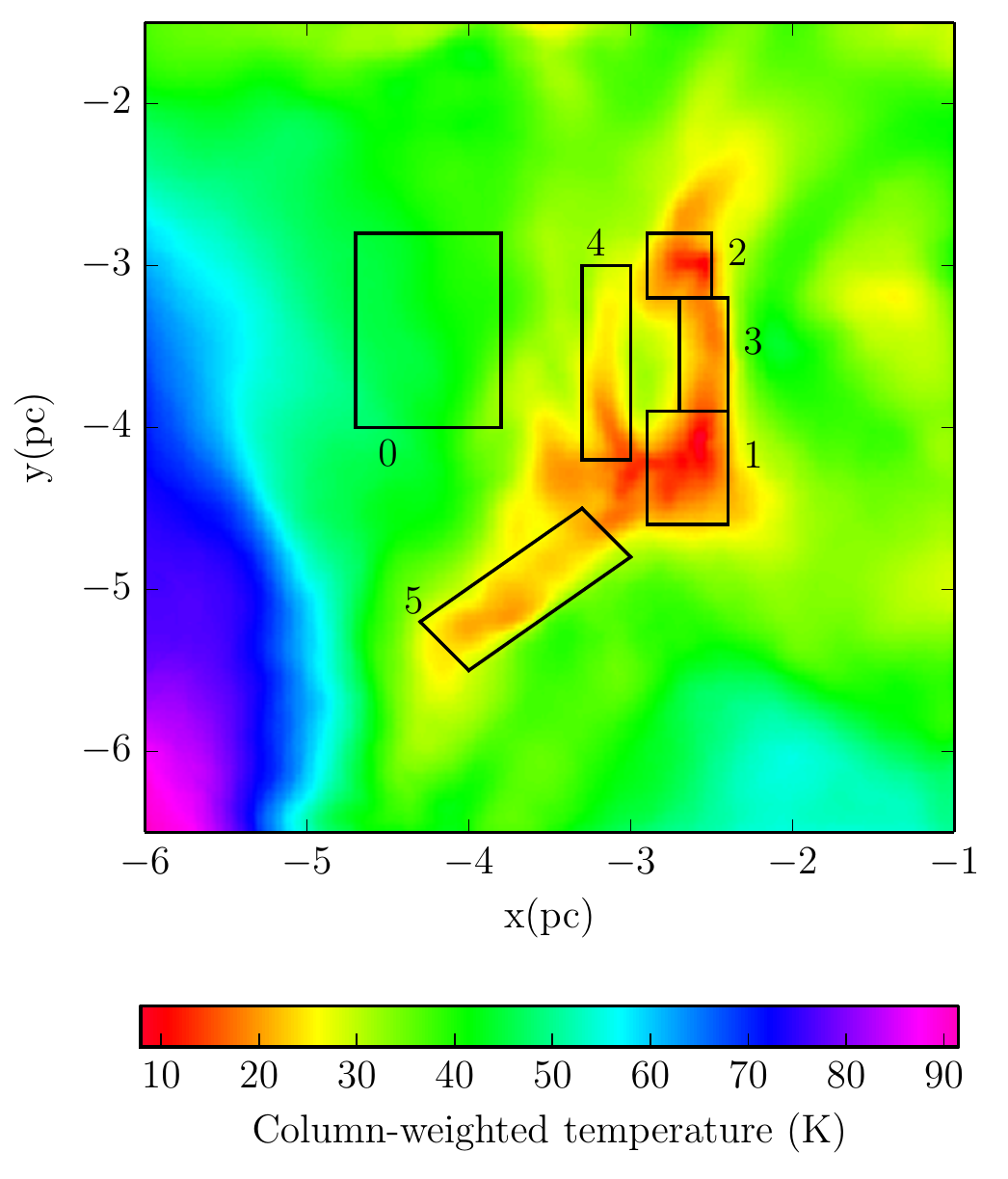}
\includegraphics[width=8cm]{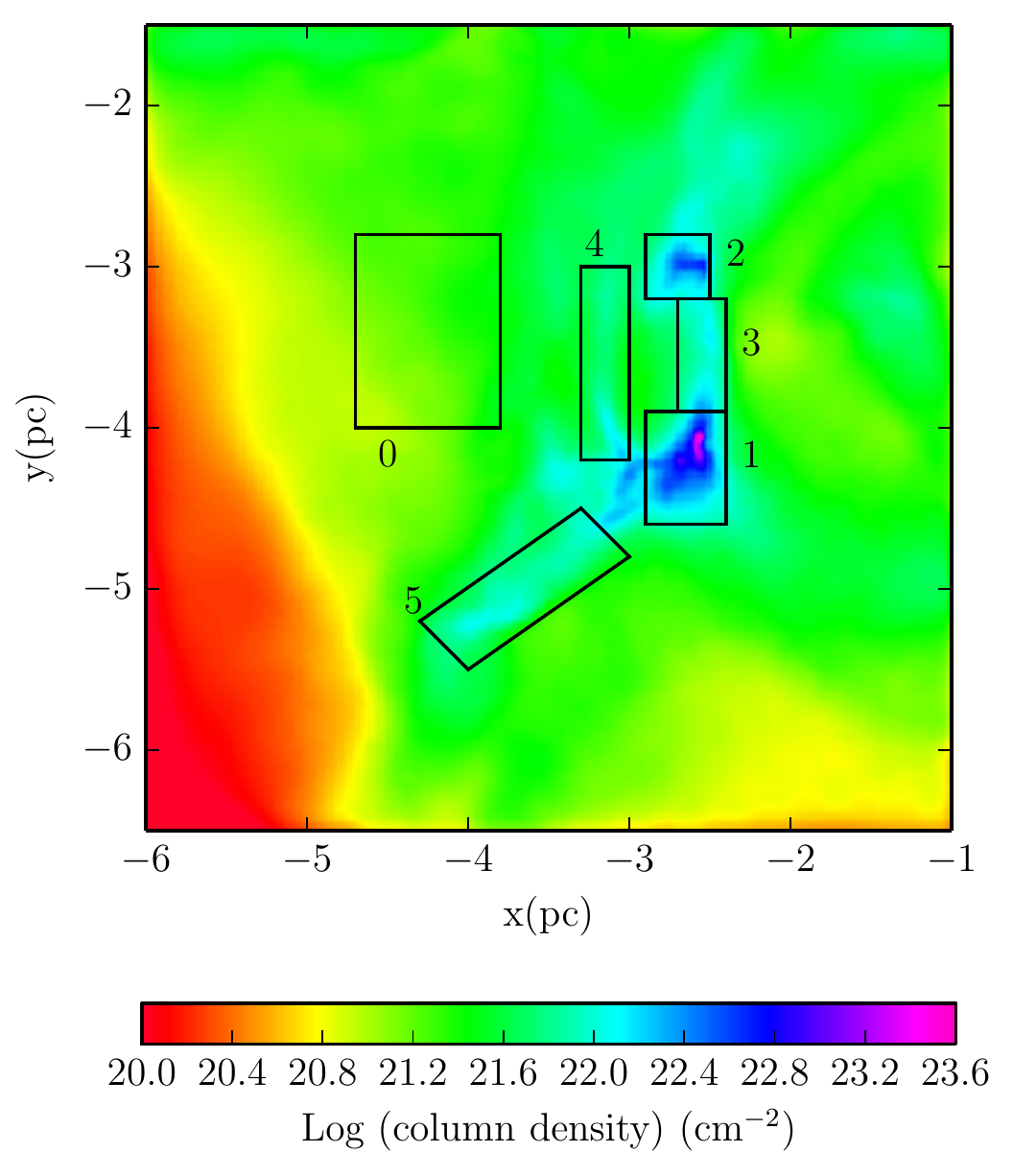}
   \caption{Temperature map and column density map  ($N_{\rm H}$) of the high-resolution simulation of Run\,I of \citet{Daleetal2012a}. The boxes represent the regions which contain filaments and cores. Box 0: control field; Boxes 1\,\&\,2: cores; Boxes 3,\,4\,\&\,5: filaments. The small protocluster is contained in Box 1.} 
     \label{reg_sim}
\end{figure*}

\noindent
Five boxes (highlighted in Figure~\ref{reg_sim}) have been drawn on the most prominent structures in the simulations. 
These include the main central core where a small cluster of four stars has already formed, 
the secondary core, a filament which bridges the gap between the main and secondary core, and two other prominent filaments.

\noindent
In order to understand the nature of the network of filaments present in the Serpens Core, we construct the histograms of the column densities of all the filaments combined 
together in order to minimize the error in the histogram (which corresponds to $\sqrt{N}$). 
%
In each histogram we report on the upper x-axis the visual extinction corresponding to the column density, computed adopting the canonical relation of, e.g.,~\citet[][]{Bohlinetal1978}: 
$A_{\rm V}=1$\,mag corresponds to  $N_{\rm H}\sim\,$2$\times$10$^{21}$ cm$^{-2}$. 

\noindent
We also compute the column density probability  density function (PDF) following the definition given in \citet{Schneideretal2015}\footnote{$\eta= \ln{\left(\frac{N_{\rm H}}{<N_{\rm H}>}\right)}$}.

\subsection{Mass and Temperature of the filaments}
We first compute the total mass and the temperature range in each observed region. The results are reported in Table \ref{table_filam}. 
For comparison, we also show the characteristics of filament number 9, which is not related to the network of filaments of the Serpens Core. 

\noindent
The masses range between  2 and 33\,M$_\odot$. 
Summing up the masses of the network of filaments converging into the Serpens Core a total mass of gas and dust of 95\,$_{-	10	}^{+	10	}$\,M$_\odot$ is computed  without considering 
 the massive filament (Number 9 in Fig.~\ref{pos_fil_obs}). This represents about 16\% of the total mass derived from the column density map in the Serpens Core region from the {\it Herschel} 
 wavelengths. 
 The most massive filaments have temperatures a few K colder than the less massive ones. 
 Dividing these masses by the lengths of each filament, we also computed the ``mass per length'' ($M_{\rm L}$). The comparison of $M_{\rm L}$ with the critical values of mass length ($M_{L, crit}$) computed for an isothermal, self-gravitating cylinder without magnetic support, allows to understand if the structure is thermally supercritical and possibly in gravitational contraction. 
Following \citet{Ostriker1964} we compute $M_{L, crit}$ as:
\begin{equation}
M_{L, crit} = 2c_s^2/G = 16.7\left(\frac{T}{10 K}\right)\,M_\odot pc^{-1}
\end{equation}
where the temperature $T$ is the average temperature of each filament. Only the filament number 1 has a $M_{\rm L}$ larger than $M_{L, crit}$ and might gravitationally contract. All the other filaments have  $M_{\rm L}$ lower than $M_{L, crit}$. 


\noindent
 Some of the prominent filaments analysed in our work have been recently resolved by \citet{Leeetal2014} with $N_2$H$^+$(1-0) observations carried out with CARMA. All their 
filaments are found to be associated with the SE sub-cluster in the Serpens Core, while our {\it Herschel}  coverage allowed us to 
detect 4 filaments (numbers 5, 6, 7, and 8) 
also close to the NW sub-cluster.  Their filaments FS1, FS2 and FS3 correspond to our filament number 1, while their structures FC1 and FN1 correspond to numbers 2 and 4, respectively. Our 
filaments number 1 and 2 are longer than the length considered by \citet{Leeetal2014}, while filament number 4 has the same length.   They identify the filaments based on their $N_2$H$^+$ emission, while our method is based on an enhancement of the column density  due to the dust emission. 
For the three filaments in common, the masses given in \citet{Leeetal2014}  agree within the errors  with the values we computed. 
Also the temperatures are consistent with the temperature range we measured (see Table~\ref{table_filam}). 
 We therefore agree with \citet{Leeetal2014} that filament number 1 is thermally supercritical.
 
 \begin{table*}
\caption{ Position, size, total mass, mass per length, critical mass per length, and temperature range derived for the filaments highlighted in Figure~\ref{pos_fil_obs}. For comparison we report the total mass and temperature range of the Serpens Core.}
\label{table_filam}      
\centering               
\begin{tabular}{r |c cc cc  r rrr r r }
\hline\hline                
\noalign{\smallskip}
\multicolumn{10}{c}{Observations of the Serpens Core}\\
\hline                
\noalign{\smallskip}
Filament   &\multicolumn{2}{c}{Coordinates$^{(1)}$}&\multicolumn{2}{c}{Size$^{(2)}$}& Angle$^{(3)}$& M$_{\rm dust+gas}$ &M$_{\rm L}$ &$M_{ L, crit}$ &T$^{(4)}$\\
\noalign{\smallskip}
 number &$\alpha_{\rm J2000}$&$\delta_{\rm J2000}$ &[$'$]&[pc]&$[^\circ$]&  [M$_\odot$]  & [M$_\odot$ pc$^{-1}$] & [M$_\odot$ pc$^{-1}$] & [K] \\
\noalign{\smallskip}
\hline                
\noalign{\smallskip}
1	&$18^{\rm h}\,30^{\rm m}\,02^{\rm s}$ & $+ 01\degr\,08'\,19''$ & 1.9$\times$8.3   & 0.2$\times$1.0& 353   & 32      $_{-    4       }^{+    4       }$ & 32& 24& 12 - 17 \\
2	&$18^{\rm h}\,30^{\rm m}\,12^{\rm s}$ & $+ 01\degr\,13'\,21''$ & 1.3$\times$3.7   & 0.2$\times$0.4& 79    &   6       $_{-    1       }^{+    1       }$  & 13& 27&  13 - 19\\
3	&$18^{\rm h}\,30^{\rm m}\,20^{\rm s}$ & $+ 01 \degr\,16'\,31''$ & 1.7$\times$6.3  & 0.2$\times$0.8&77      &  6       $_{-    1       }^{+    1       }$  & 7& 27&  14 - 18\\
4	&$18^{\rm h}\,30^{\rm m}\,13^{\rm s}$ & $+ 01\degr\,19'\,32''$ & 1.5$\times$4.8   & 0.2$\times$0.6&40       & 3      $_{-    0       }^{+    0       }$  & 5& 27&  15 - 17\\
5	&$18^{\rm h}\,29^{\rm m}\,57^{\rm s}$ & $+ 01\degr\,24'\,17''$ & 2.2$\times$10.0 & 0.3$\times$1.2&6 	&  8       $_{-    1       }^{+    1       }$  & 7& 26&  15 - 17\\
6	&$18^{\rm h}\,29^{\rm m}\,43^{\rm s}$ & $+ 01\degr\,20'\,58''$ & 1.5$\times$5.5   & 0.2$\times$0.7&330 	&  4       $_{-    1       }^{+    1       }$  & 6& 25&  14 - 16\\
7	&$18^{\rm h}\,29^{\rm m}\,28^{\rm s}$ & $+ 01\degr\,20'\,07''$ & 1.7$\times$10.4 &0.2$\times$1.3& 319 	&  13      $_{-    2       }^{+    2       }$  & 10& 24&  12 - 16\\
8	&$18^{\rm h}\,28^{\rm m}\,53^{\rm s}$ & $+ 01\degr\,14'\,06''$ & 2.2$\times$17.4 & 0.3$\times$2.1&68 	&  23      $_{-     3      }^{+    3       }$  &  11& 26& 14 - 17\\
9	&$18^{\rm h}\,28^{\rm m}\,51^{\rm s}$ & $+ 00\degr\,50'\,09''$ & 2.9$\times$14.3 & 0.3$\times$1.7&327	&  23      $_{-    3       }^{+    3       }$  &13 &24 &  12 - 17\\
\noalign{\smallskip}
\hline                
\noalign{\smallskip}
Serpens core& $18^{\rm h}\,29^{\rm m}\,50^{\rm s}$ & $+ 01\degr\,14'\,14''$ & 42$\times$42&5$\times$5&36& 494	$_{-	58	}^{+	61	}$ & & & 18 - 30\\
field&&&&&&&&&\\
\noalign{\smallskip}
\hline
\end{tabular}
\tablefoottext{$1$}{ The coordinates represent the center of the boxes.}
\tablefoottext{$2$,$3$}{ Size and orientation of the boxes. The angles are measured from North over East. }
\tablefoottext{$4$}{ Lowest and highest temperatures in the corresponding boxes.}
\end{table*}
 \begin{table}
\caption{ Mass and temperature range derived for the simulated filaments highlighted in Figure~\ref{reg_sim}. 
For comparison we report the total mass and temperature range of the entire region that appears in Figure~\ref{reg_sim}.}
\label{table_filam_1}      
\centering               
\begin{tabular}{cc cc}
\hline\hline                
\noalign{\smallskip}
\multicolumn{4}{c}{Simulation}\\
\noalign{\smallskip}
\hline                
\noalign{\smallskip}
Region&Size&M &T\\
\#&pc$\times$pc&[M$_\odot$] &[K]\\
\noalign{\smallskip}
\hline                
\noalign{\smallskip}
3	&0.3$\times$0.7&20 & 19 - 23   \\
4	&0.3$\times$1.2&22 & 21 - 28    \\
5	&0.4$\times$1.3&36 & 22 - 27   \\
\noalign{\smallskip}
\hline                
\noalign{\smallskip}
\noalign{\smallskip}
Entire Field&5$\times$5&801&23 - 50 \\
 \noalign{\smallskip}
\hline                        
\noalign{\smallskip}

\end{tabular}
\end{table}

\begin{figure*}
\centering
\vspace{0.8cm}
\includegraphics[trim=1.5cm 0.05cm 1.7cm 0.8cm,width=8cm]{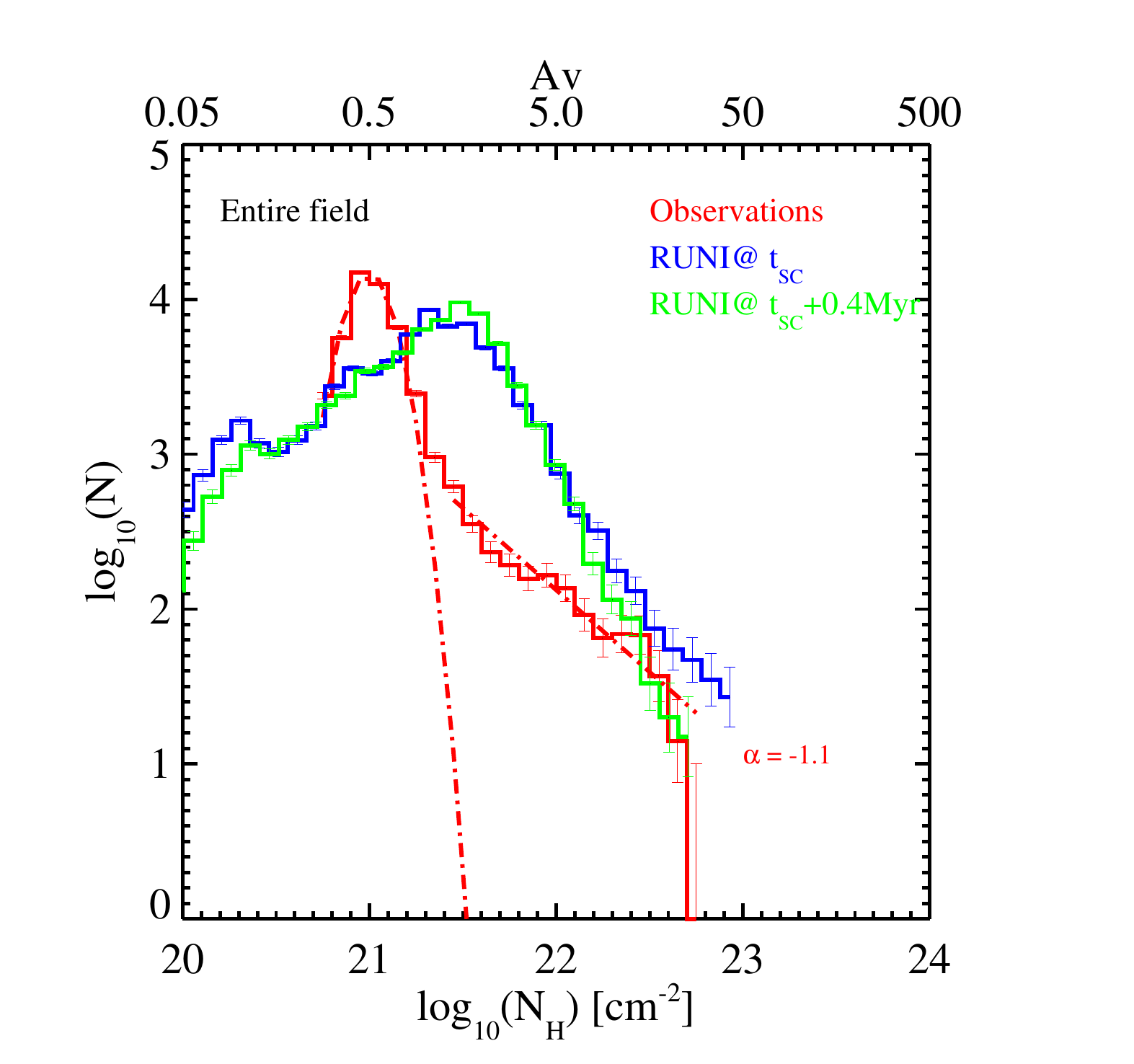}
\includegraphics[trim=1.5cm 0.05cm 1.7cm 0.8cm, width=8cm]{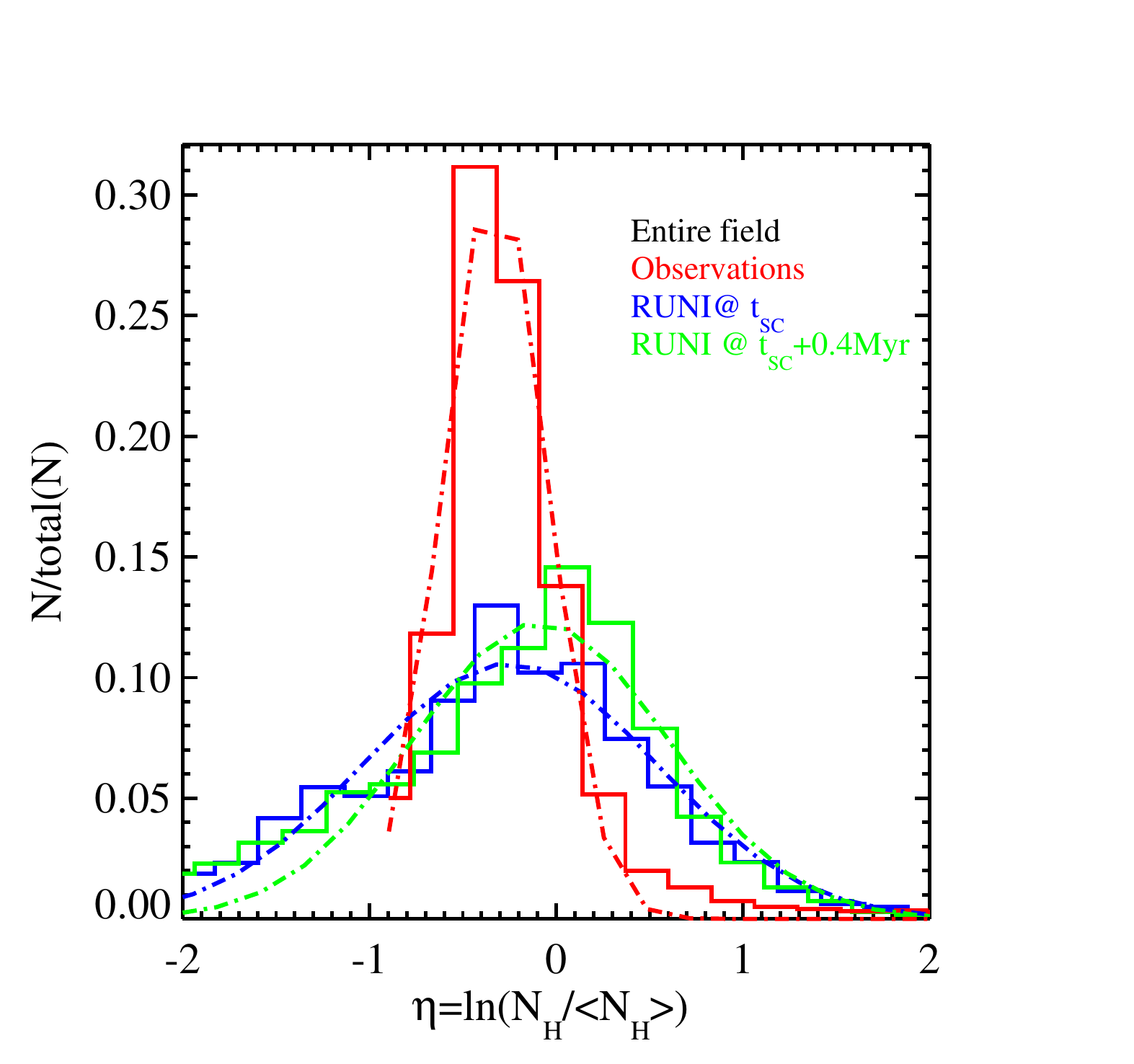}
 \caption{The left panel presents the column density distributions and the right panel the normalised PDFs of the Serpens Core (in red)  and two snapshots of Run I of the simulations of \citet{Daleetal2012a} that differ by 0.4\,Myr in time (in blue and green). The snapshot at $t_{SC}$ is shown in Figure\,\ref{reg_sim}.  The dashed-dotted red lines in the left panel represent the fit of a Gaussian distribution and a power-law.  In the right panel the dashed-dotted lines represent the best Gaussian fit of the observations (in red), Run I at $t_{SC}$ (in blue) and Run I at $t_{SC}$+0.4\,Myr (in green).}
     \label{entire_field}
\end{figure*}


\begin{figure*}
\centering
\includegraphics[trim=1.5cm 0.05cm 1.7cm 0.8cm, width=8cm]{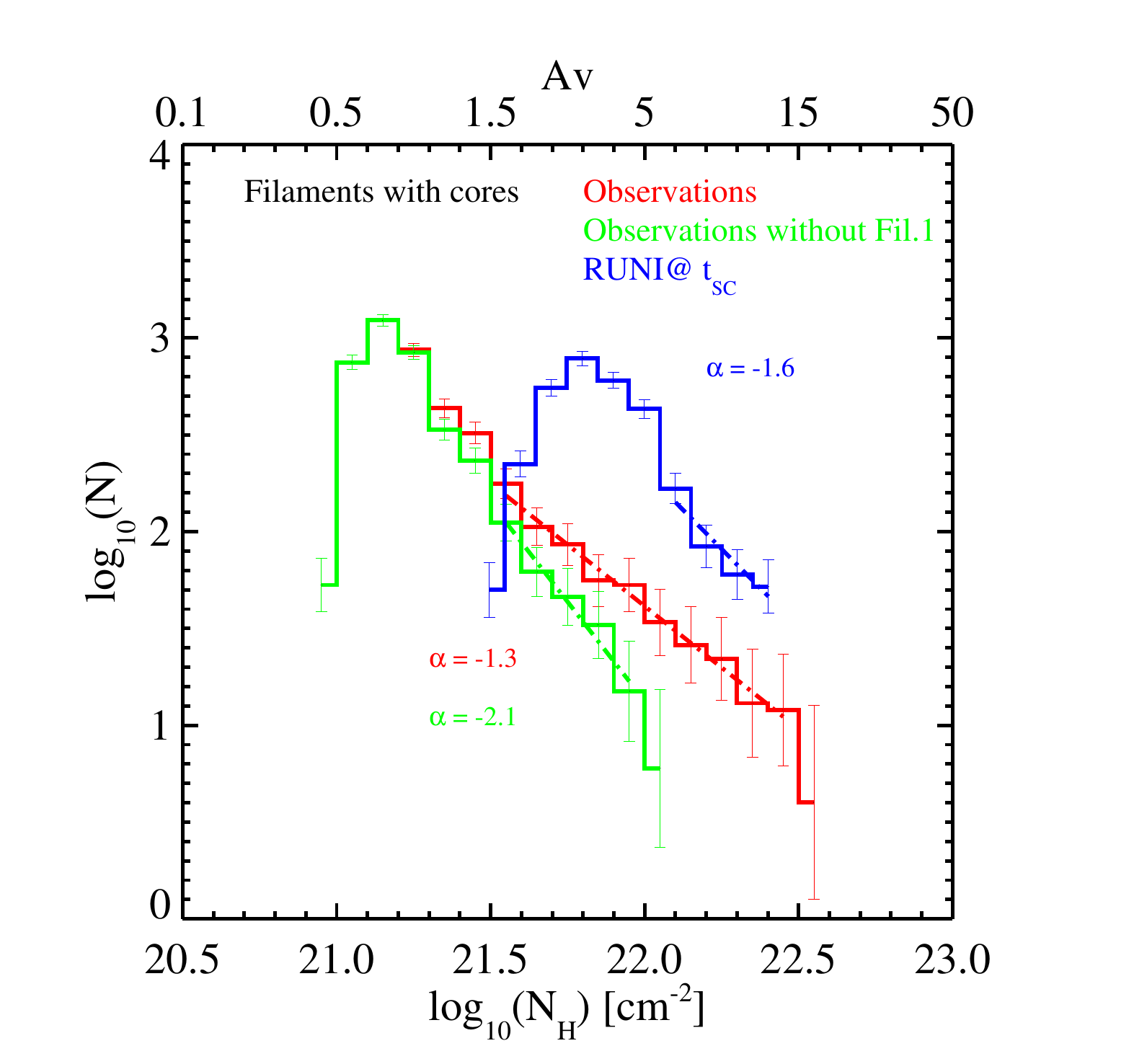}
\includegraphics[trim=1.5cm 0.05cm 1.9cm 0.8cm, width=8cm]{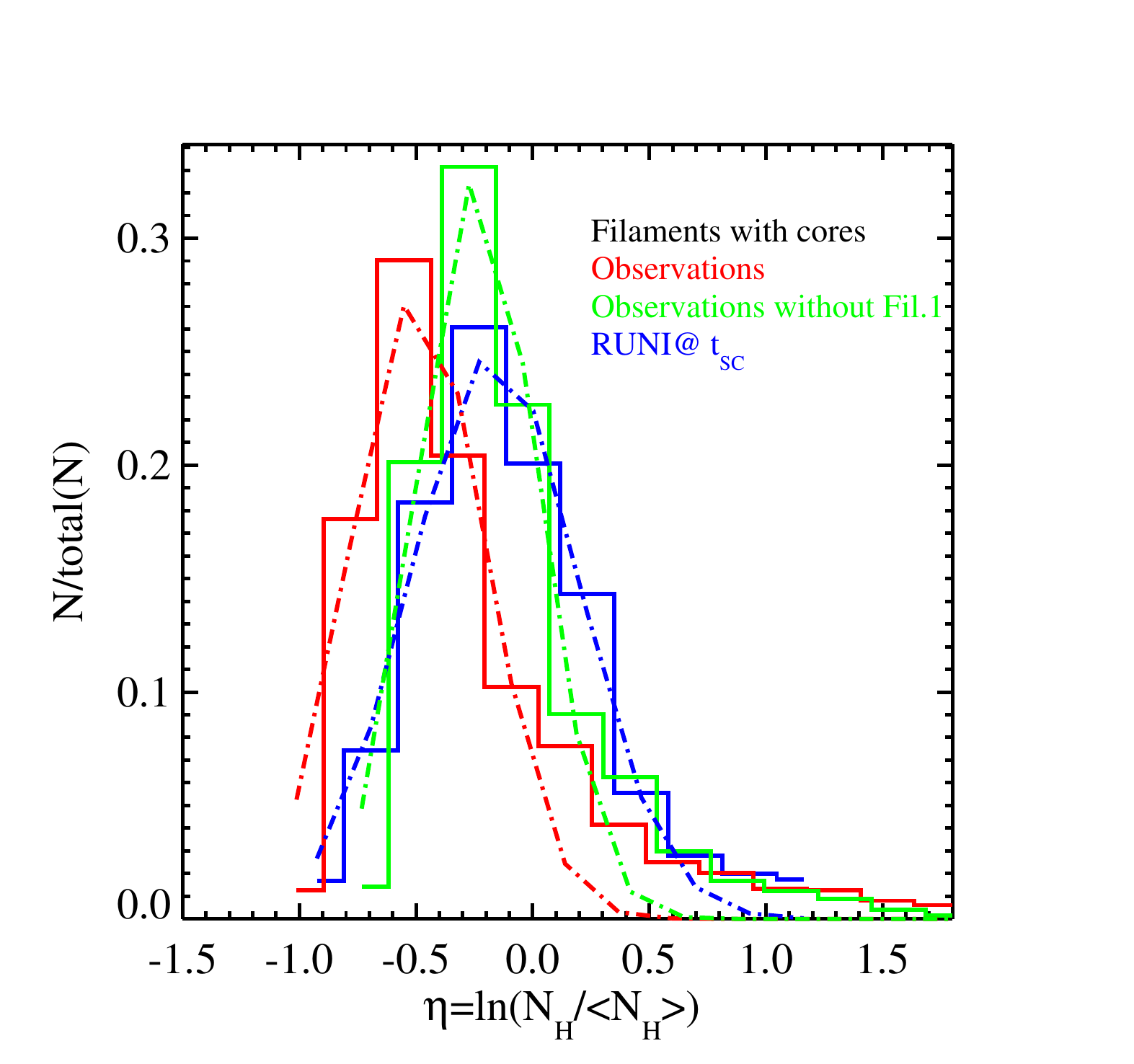}
 \caption{The left panel represents the  histograms of the column density distributions while the right panel shows the 
 normalised PDFs of the Filaments with cores. The dashed-dotted lines overplotted to the 
 PDFs represent the best Gaussian fit of the data. Red lines represent the observations, the blue ones represent the 
 simulations, and the green lines represent the observations without Filament number 1. The positions of the 
 observed filaments and cores are highlighted in Figure\,\ref{pos_fil_obs}, while the  simulated filaments and cores are outlined in Figure\,\ref{reg_sim}.}
     \label{network_core1}
\end{figure*}
\begin{figure*}
\centering
\includegraphics[trim=1.5cm 0.05cm 1.7cm 0.8cm, width=8cm]{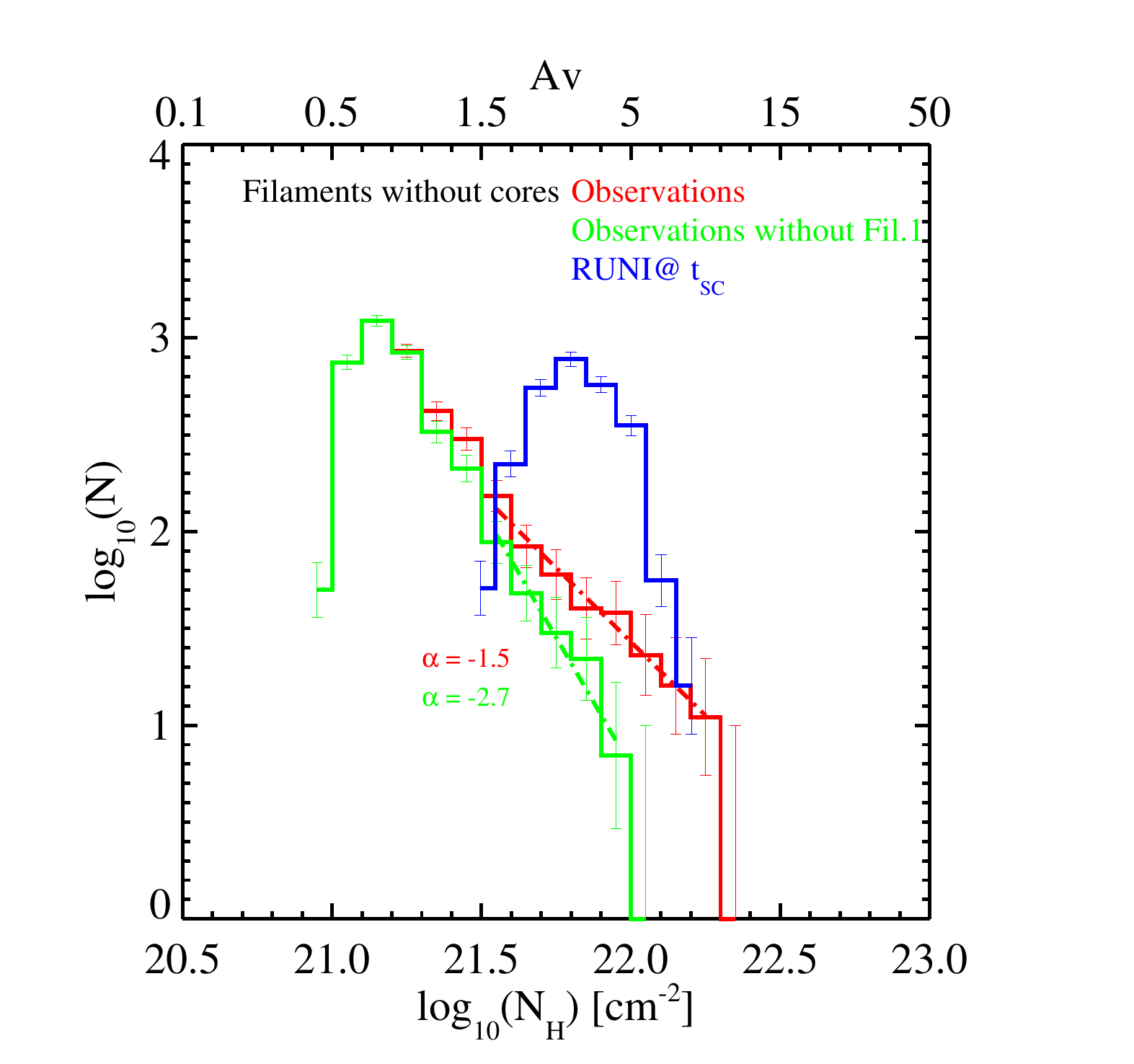}
\includegraphics[trim=1.5cm 0.05cm 1.7cm 0.8cm, width=8cm]{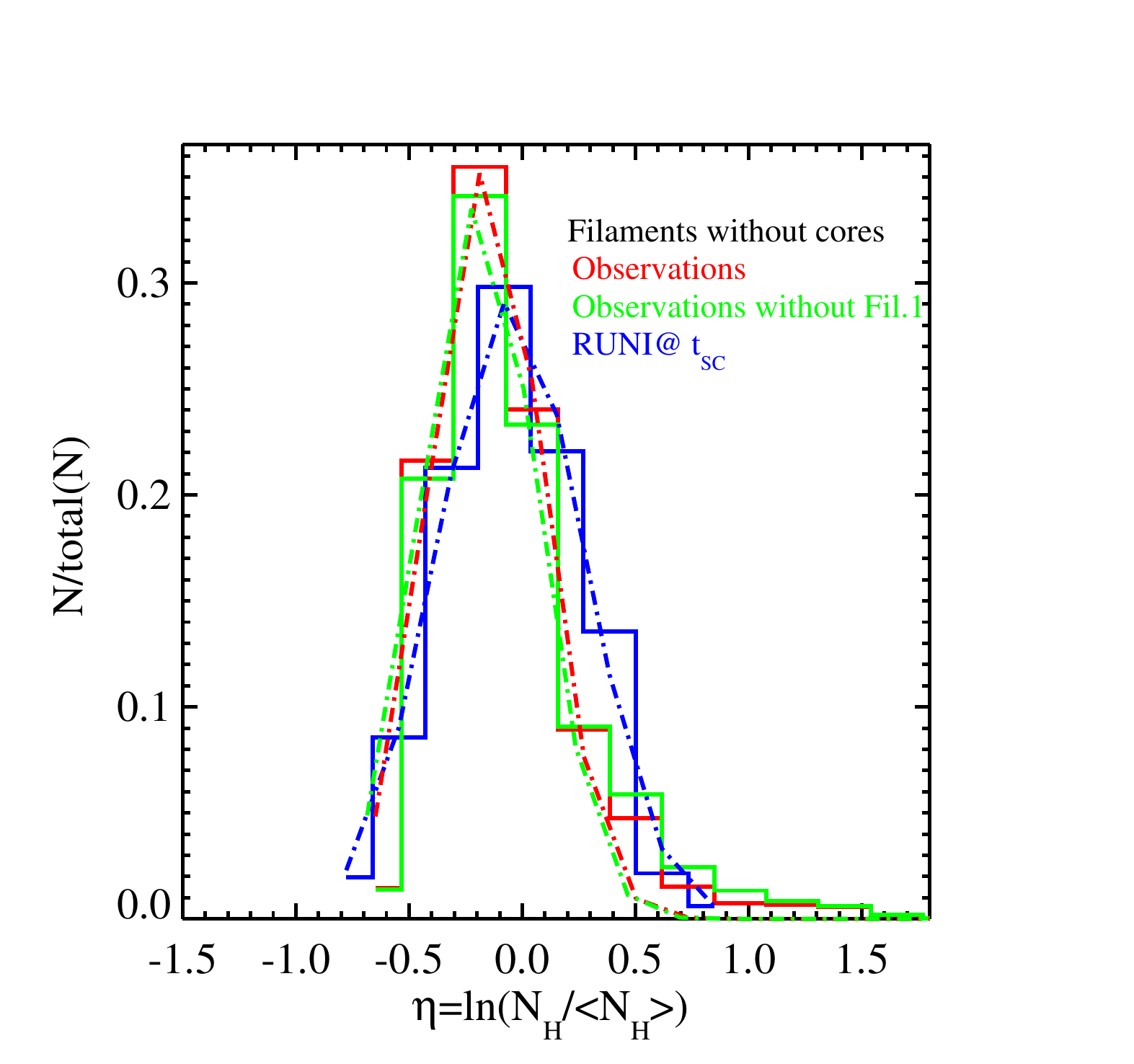}
 \caption{The left panel represents the  histograms of the column density distributions while the right panel shows the 
 normalised PDFs of the Filaments without cores.  The dashed-dotted lines overplotted to the 
 PDFs represent the best Gaussian fit of the data. 
 The positions of the observed filaments are highlighted in Figure\,\ref{pos_fil_obs}, while the 
 simulated filaments are outlined in Figure\,\ref{reg_sim}. The different colors of the lines are as in Figure\,\ref{network_core1}.}
     \label{only_fil}
\end{figure*}

\begin{figure*}
\centering
\includegraphics[trim=1.5cm 0.05cm 1.7cm 0.8cm,  width=8cm]{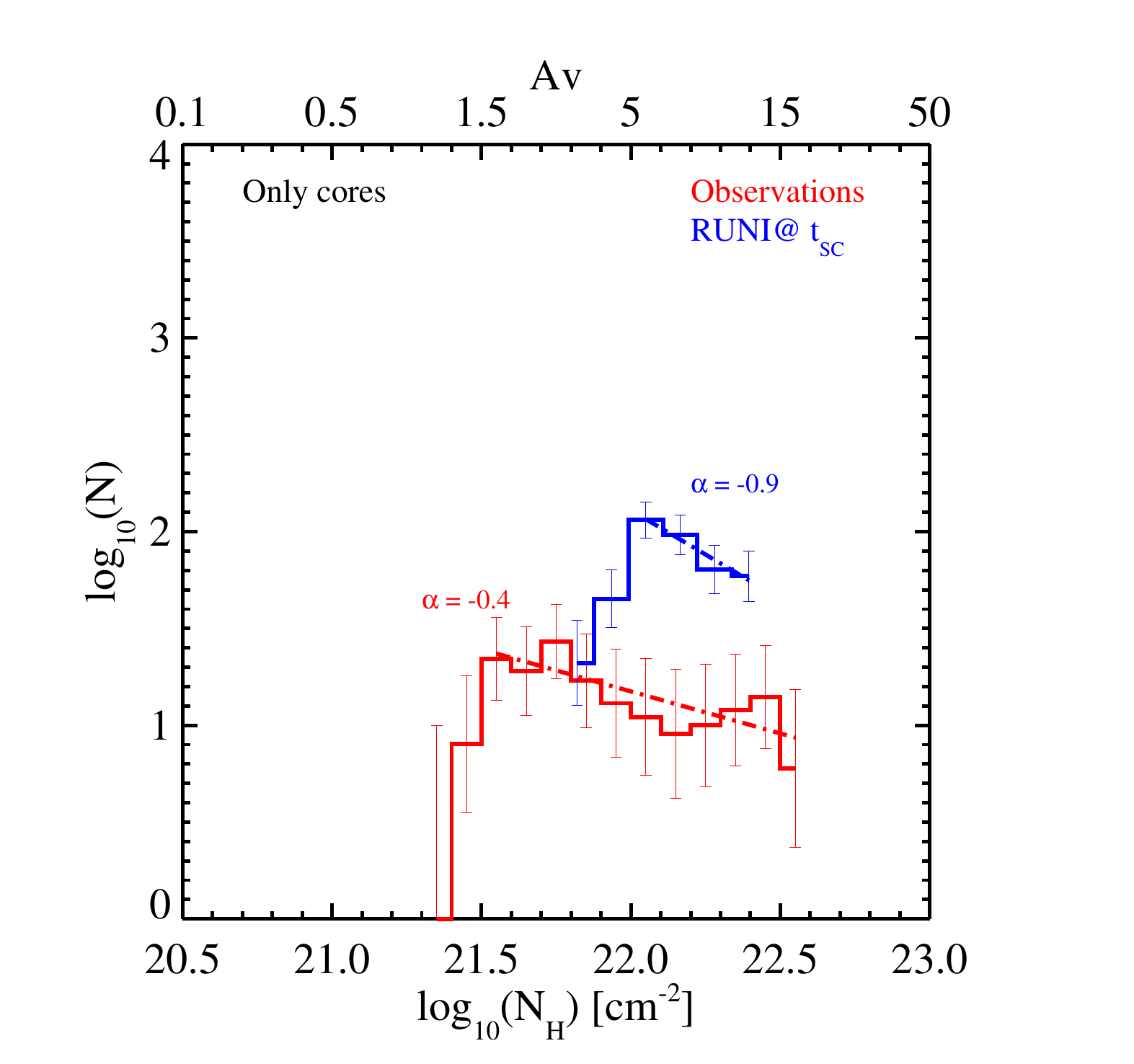}
\includegraphics[trim=1.5cm 0.05cm 1.7cm 0.8cm, width=8cm]{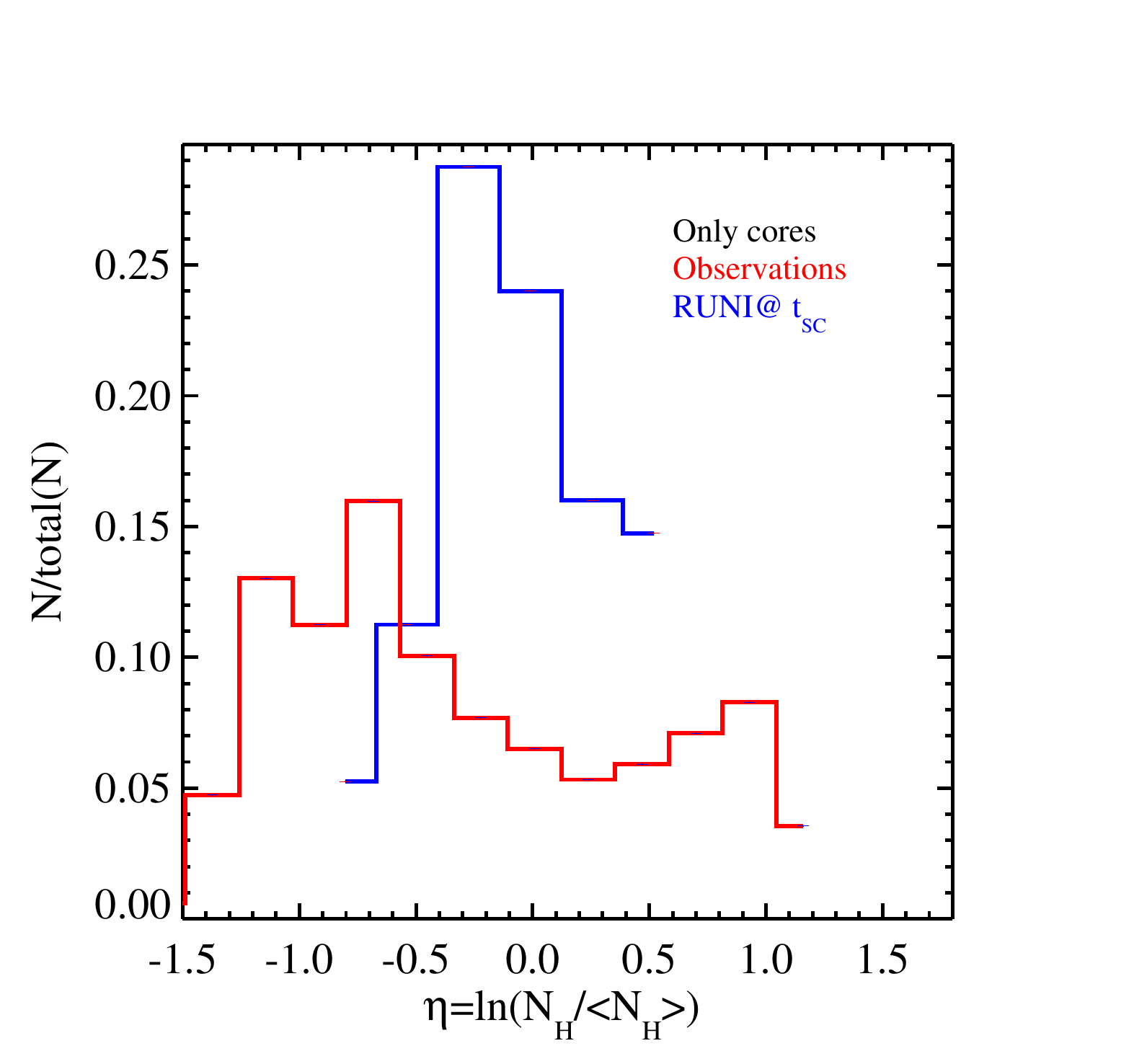}
 \caption{The left panel represents the histograms of the column density distributions while the right panel shows the 
 normalised PDFs of the cores alone. 
  The positions of the observed cores are highlighted in Figure\,\ref{pos_fil_obs}, while the 
 simulated cores are outlined in Figure\,\ref{reg_sim}. The different colors of the lines are as in Figure\,\ref{network_core1}.}     \label{network_core}
\end{figure*}

\subsection{column density distributions}
\label{col_den}
The column density distribution of the entire field that covers the Serpens Core (as highlighted in Figure~\ref{pos_fil_obs}) is shown in Figure~\ref{entire_field} as a red histogram. 
The observations show a cut at densities lower than 10$^{20.5}$ cm$^{-2}$, due to the sensitivity limit of the observations. The sensitivity limit is sufficient 
to well sample the peak in the observed column density distribution, which is more narrow than the peak in the simulations. 
Overplotted are the column density distributions from two simulation snapshots separated in time by $\approx0.4$Myr.  The 
distribution evolves relatively little during this period of time. 

\noindent
The slopes of the histograms of column densities have been compared between observations and simulations, as it is shown 
in Figures\,\ref{entire_field} to \ref{network_core}. The histograms of column densities of the simulations  and the 
PDFs 
have been constructed as in the analysis of the observations. 
 The dashed-dotted lines in the left panel of Figure\,\ref{entire_field} represent the fit of a  Gaussian  distribution and a power-law. We find that the deviation from a log-normal distribution corresponds to  A$_{\rm v}\sim$1.2-2.0\,mag. 
\noindent
Both column density distributions of the two snapshots of the simulation separated by 0.4\,Myr in time,  
show a very broad  peak resembling a log-normal distribution, but with a pronounced tail at high column densities, rather different from that seen in 
the observed cloud. In particular, there is a considerable amount of low-density material  that does not appear in the observed data (it is below the observational sensitivity limit),  
and the peak of both simulated distributions occurs at a higher column density. The quantity of high-column density material, between 
$\approx10^{22}$ and $10^{23}$ cm$^{-2}$ drops between the two simulation snapshots. The quantity of material which 
appears to be `missing' between the two distributions is $\sim10$\,M$_{\odot}$, corresponding to an order of one percent of the total 
mass of gas in the simulation volume, and to about thirty percent of all gas denser than $10^{22}$ cm$^{-2}$. In the 0.4\,Myr time 
interval between the two snapshots some of the high--density gas is consumed by star formation. 
The figure highlights the fact that, while a simulation snapshot and an observed field may 
appear subjectively similar, more careful analysis can reveal important differences between them.

\noindent
The distribution of the observed column densities of {\it all filaments, including the cores}, shows a peak at $A_{\rm V}\sim0.6$\,mag, with 
a maximum value of $A_{\rm V}\sim18$\,mag  (see left panel in Figure~\ref{network_core1}). 
  For $A_{\rm V}\sim$\,1.5\,-\,14\,mag, the distribution decreases as a power-law, following the relation 
 $d log(N) / d log(N_H) \sim \alpha$, 
 where  $\alpha \sim -1.3$  (red histogram in  the left panel of Figure~\ref{network_core1}). 
In the simulations the peak is found at a higher value i.e. $A_{\rm V}\sim3$\,mag. 
The slope computed between $A_{\rm V}\sim$\,6 -  12\,mag is $\alpha \sim -1.6$, i.e. steeper than the observed value.

\noindent
The core emission has been identified as an enhancement in the column density map, as well a point source emission at least at one {\it Herschel}  wavelength. In  filament number 1 in the Serpens Core, we found two cores (both included in the same box) and  in filament number 7 one core. Since all the cores do show a point source counterpart at 70\,$\mu$m, this means that we are selecting the more evolved protostellar cores, and not the cores at an earlier evolutionary stage (when they do not have yet 70\,$\mu$m emission). In the simulations two cores are identified.  
The histogram of the  {\em filaments without cores}, obtained by drawing a box around the cores and excluding them, is 
computed for both observations and simulations (see left panel of Figure~\ref{only_fil}). In the observations the maximum 
column density reaches $A_{\rm V}\sim9$\,mag, and its peak is at $A_{\rm V}\sim0.7$\,mag. While the observations show  a peak with a power-law tail with a slope $\alpha \sim-1.5$, the simulations show only a peak  at $A_{\rm V}\sim3$\,mag. The slope in the observed filaments becomes steeper excluding the cores. 

\noindent
We also did the analysis of the filaments with and without cores excluding the filament number 1, that is the only 
supercritical filament in the network of filaments of the Serpens Core. In Figures ~\ref{network_core1} and ~\ref{only_fil} the
histograms without filament number 1 are drawn in green.  In both cases the histogram still shows a power law tail, but with a steeper slope. In the case of the {\em filaments with cores} a steeper slope was expected, since by excluding filament 1,  
we are not considering the two cores that are present in this filament. It is even more interesting to notice that even without 
the only supercritical filament, the observed filaments still show a power law tail.

\noindent
A histogram showing the {\it cores alone} has been constructed excluding the filaments in both observations and simulations (see left panel of Figure~\ref{network_core}).  
In the case of the simulation the maximum column density reaches a value of $A_{\rm V}\sim12$\,mag, and the slope decreases to an almost flat slope of 
$\alpha \sim - 0.9$.  In this case of  the {\it cores alone} there are not many pixels which sample the cores and the errors are much larger than in the histograms of the column densities of the filaments with and without cores.  



\subsection{Probability density functions (PDFs) of the filaments.}
%
%
%
%
%
%
The PDFs of the same regions described in Section~\ref{col_den} have been computed for the observations and simulations and they are 
shown in the right panels of Figures\,\ref{entire_field} to \,\ref{network_core}. 

\noindent
 A quantitative comparison between the PDFs of the observations and the simulations has been carried out using the 
 standard non-parametric Kolmogorov-Smirnov statistic test. The computed probability that the observed and simulated distributions are drawn from the same distribution
is give in Table~\ref{gaussfit}. We notice that the lowest probability is found in the case of filaments with cores. The value is, however,  
not sufficiently low to conclude 
that simulations and observations are from different parent distributions. 

\noindent
We perform a Gaussian fit 
of each PDF in order to characterise the log-normal function:
\begin{equation}\label{lognorm_eq}
p_\eta d\eta=\frac{1}{\sqrt{2\pi\sigma^2_\eta}}\exp{\Big[-\frac{(\eta-\mu)^2}{2\sigma^2_\eta}\Big]} d\eta
\end{equation}
 where $\sigma_\eta$ is the dispersion, and $\mu$ is the mean logarithmic column density. 
 The widths, $\sigma_\eta$, resulting from the fitting are reported in Table~\ref{gaussfit}. 
 We notice that the values of $\sigma_\eta$ in the observations of the Serpens Core are always lower than the values in the simulations. In the case of the cores alone, the  
 values of $\sigma_\eta$ are not meaningful since both, the observed and simulated PDFs do not show any peak.  

\noindent
As for  the column density distributions, also the  PDFs of the {\it filaments alone} in the simulations show only a log-normal distribution while the observed filaments do show a  power-law tail (see  right and left panel of Figure~\ref{only_fil}). This is in agreement with the finding that the filaments themselves in the simulations are not fragmenting, but are instead serving as conduits of gas into the main cluster. The simulated cores at higher column densities show instead the power-law tail, while the observed cores have an almost  flat distribution (see Figure~\ref{network_core}). 
 In the PDFs the log-normal component always dominates compared to the power-law tail and the simulations show always a higher FWHM than the observations,  apart from the case of the {\em cores alone}. Since the FWHM of the PDFs is a measure of the turbulent dispersion velocity  \citep[e.g.][]{FederrathKlessen2013}, this implies that the turbulent velocities in the observed filaments are always lower than those in the simulation.

\noindent
 A power-law tail in the column density distribution has 
been interpreted as a transition between a turbulence-dominated gas and a self-gravitating gas \citep[e.g.][]{Klessen2000,  
FederrathKlessen2013}. 
Studies of the probability distribution functions, computed from the column density maps, have been used to constrain star formation processes. 
Using the column density maps from {\it Herschel},  \citet[][]{Schneideretal2015} compared the PDFs of low-mass and high-mass star-forming regions, finding a common ``threshold'' value of A$_{\rm V}\sim$4-5 mag (corresponding to $N_{\rm H}=$3.8-4.7$\times$10$^{21}$ cm$^{-2}$) between turbulence-dominated gas and self-gravitating gas.  
This ``threshold'' is higher than the value of A$_{\rm V}\sim$1.2-2.0 mag that we find in the column density distribution of the Serpens Core 
(see Figure~\ref{entire_field}). 
\noindent
 The column density PDFs presented suggest that the simulated filaments are likely turbulence-dominated, while the observed filaments have the characteristic power-law tails indicating that they are most likely in the self-gravitating regime and that they will probably eventually fragment into cores. A more stringent criterion for gravitational instability which indicates that filaments are \emph{already in the process of sub-fragmentation}  \citep[e.g. ][]{Andreetal2010, Andreetal2014} is whether their masses per unit length are supercritical. Recent observations by \citet{Leeetal2014} found that filaments FS1, FS2 and FS3 (which correspond to our filament number 1) do indeed have supercritical values of M/L, while their filaments FC1 and FN1 (our filaments 2 and 4, respectively) have M/L values close to the critical threshold (Table~\ref{table_filam}), suggesting that they will begin sub-fragmentation very soon. 
\begin{small}
\begin{table}
\caption{Standard deviation $\sigma_\eta$ (defined in Equation~\ref{lognorm_eq}) of the log-normal distributions overplotted in  Figures\,\ref{entire_field},\,\ref{network_core1},\,\ref{only_fil}\,\&\,\ref{network_core}.  The last column shows the probability that the simulated and observational data (the observed filaments refer to numbers 1 to 8)  are drawn from the same distribution using the K-S test. 
}
\label{gaussfit}      
\centering               
\begin{tabular}{l rrrr}  
\hline\hline                
\noalign{\smallskip}
   &Serpens Core&Run I @\,t$_{\rm SC}$& P$_{\rm K-S}$\\
\noalign{\smallskip}
\hline                
\noalign{\smallskip}
Entire Field                    & 0.27&0.78\tablefootmark{a}&0.54\\
\noalign{\smallskip}
Filaments with cores      &  0.28\tablefootmark{b} (0.25\tablefootmark{c})&0.36&0.03\\
\noalign{\smallskip}
Filaments without cores  & 0.24\tablefootmark{b}  (0.25\tablefootmark{c})& 0.32&0.19\\
\noalign{\smallskip}
Control field                    &  0.13&0.38&0.36\\
\noalign{\smallskip}
\hline                        
\noalign{\smallskip}
\end{tabular}
\tablefoottext{a}{$\sigma_\eta$\,of Run I at\,t$_{\rm SC}$+0.4\,Myr is 0.68. }
\tablefoottext{b}{Filaments numbers 1-8.}
\tablefoottext{c}{Filament numbers 2-8 (excluding Filament number 1).}
%
\end{table}
\end{small}

\section{Summary and Conclusions }
\label{conclusions}

In this paper we analyze continuum {\it Herschel} observations of the Serpens Core and compare them to a numerical simulation from the suite of calculations performed by \citet{Daleetal2012a}. Since the Serpens Core does not harbor any high-mass star, we choose an early state of a particular simulation of \citet{Daleetal2012a} before any O--stars have formed and therefore without any feedback.  It is possible that this region will form high--mass stars in the future, since it has a substantial reservoir of gas available and its maximum column density is only a factor of a few lower than the canonical 1 g\,cm$^{-2}$ threshold for massive star formation proposed by \citet{KrumholzMcKee2008}. 
In the chosen simulation, the action of the filaments is mainly to deliver more gas into the cluster which forms at the hub where they join, without fragmenting. 
We selected a sub-region of Run I with the most similar total mass, spatial scale and morphology to the Serpens Core. We then compared the properties, such as temperature and surface density in regions identified by the presence of a filamentary structure or a core. Note that the simulation was \emph{not} designed to produce an object resembling Serpens (or any other star--forming region). The resemblance is serendipitous and owes to the aim of \citet{Daleetal2012a} to cover a realistic parameter--space with general simulations. 
The picture which emerges is rather complex. With the additional time dimension available in the simulations, we saw that, while the simulations and observations may appear similar in morphology at a given time, they may differ in other respects, such as in the form of the column density distribution. Advancing the simulation in time resulted in the morphological disparity increasing, but the two datasets becoming more similar from the point of view of their column density distributions. This indicates that quantitatively comparing simulations and observations is more difficult than it may firstly appear. We defer a more detailed analysis to a later paper in which we construct synthetic observations of the simulations in order to make such a comparison more robust.

\noindent
The main results and conclusions of our study can be summarized as follows:

\begin{enumerate}
\item {\it Herschel} continuum observations between 70 and 500\,$\mu$m reveal  a radial network of filaments converging into the central cluster of the Serpens Core: this low-mass star forming region has been used to test the state-of-the-art of simulations of GMCs.
\item In the filaments the dust temperatures are found to lie in the range between 12\,K and  19\,K in the observations and  few Kelvin  
lower than  the gas temperatures in the simulations: this might reflect a difference in temperature between gas and dust in the region. 
\item The peak in the observed column density distribution of the entire field of the Serpens Core is more narrow than the peak in the simulations. 
 Both distribution show a power-law tail at higher column densities but with different slopes. This can be explained by a different evolutionary stage of the simulation in comparison to the real observations.
\item The peaks in the column density distributions of the filaments are about $10^{0.5}$\,cm$^{-2}$ higher in the simulation than those in the observations. 
\item  From the measured widths, $\sigma_\eta$, of the PDFs, we found that the turbulent velocities in the observed filaments are likely lower than those in the simulations.

\item The column density distribution and PDFs of simulated filaments shows only a log-normal distribution, while the observed filaments show a 
power-law tail. The power-law tail becomes evident in the simulation in the column density distribution of the cores, while in the observations this 
shows a flat distribution.  This means that while the simulated filaments are likely turbulence-dominated regions, the observed filaments are probably turbulence-dominated
self-gravitating structures which will fragment into cores. 

In the simulation, the main function of the filaments is to feed gas into the central cluster which is where most of the star formation takes place. Most of the gas in the filaments falls into the cluster before fragmenting. It is therefore possible that, despite the physical and morphological similarities between the chosen simulation timesteps and the Serpens Core, the filaments in the real and simulated clouds are behaving in very different ways. Whether this is due to neglected physics in the simulations (e.g.~magnetic fields), to unknown differences between the early phases of the Serpens Core and the initial conditions of the calculations, or just to the extreme difficulty of comparing two very complex dynamical systems 
is not clear.  The simulated and observed filaments are subjectively similar--looking, but behave differently, the former being turbulence-dominated and the latter being self-gravitating. 
\end{enumerate}
\begin{acknowledgements}
V.R. was partly supported by the DLR grant number 50~OR~1109 and  
 by the {\it Bayerische Gleichstellungsf{\"o}rderung} (BGF). This research was partly supported by the Priority Programme 1573 "Physics of the Interstellar Medium" of the German Science Foundation (DFG), the DFG cluster of excellence `Origin and Structure of the Universe' and by the Italian Ministero dell'Istruzione, Universit\`a e Ricerca through the grant Progetti Premiali 2012 -iALMA (CUP C52I13000140001).  C.E. is partly supported by Spanish Grants AYA 2011-26202 and
AYA 2014-55840-P. 
 We acknowledge the HCSS / HSpot / HIPE, which are joint developments by the {\it Herschel} 
Science Ground Segment Consortium, consisting of ESA, the NASA {\it Herschel} Science Center, and the HIFI, 
PACS and SPIRE consortia.  We thank Jean-Philippe Bernard for computing the {\it Planck} offset for {\it Herschel} mosaics. V.R. thanks the useful discussions at MPIA with A.Stutz and J. Kainulainen. V.R. thanks also L.Mirzagholi and the support from the mentoring program at LMU.  
 \end{acknowledgements}

\bibliographystyle{aa}
\bibliography{references}

\begin{appendix}
\section{ The Control field}
\begin{figure*}
\centering
\includegraphics[trim=1.5cm 0.05cm 1.7cm 0.05cm,  width=8cm]{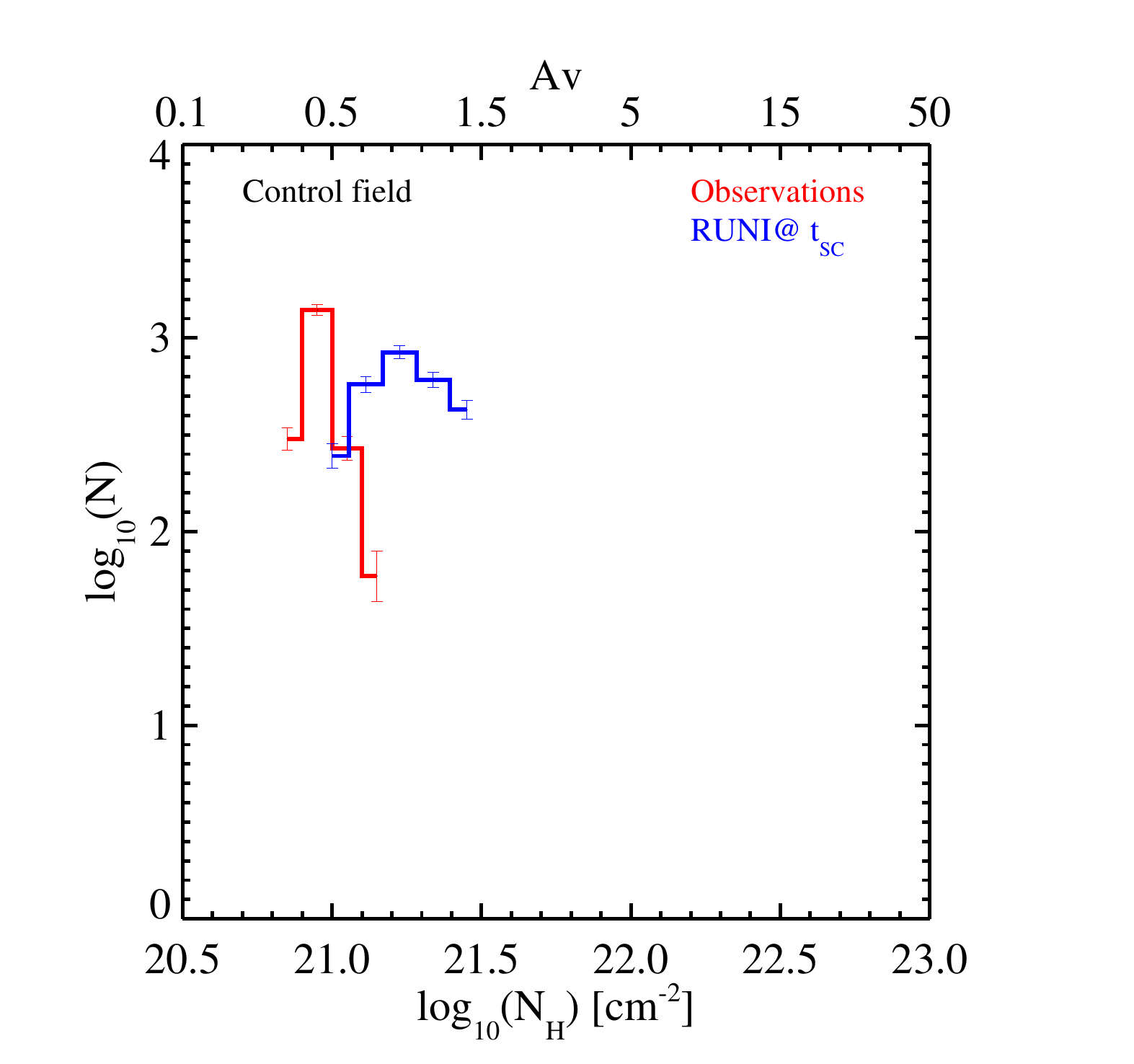}
\includegraphics[trim=1.5cm 0.05cm 1.7cm 0.8cm,, width=8cm]{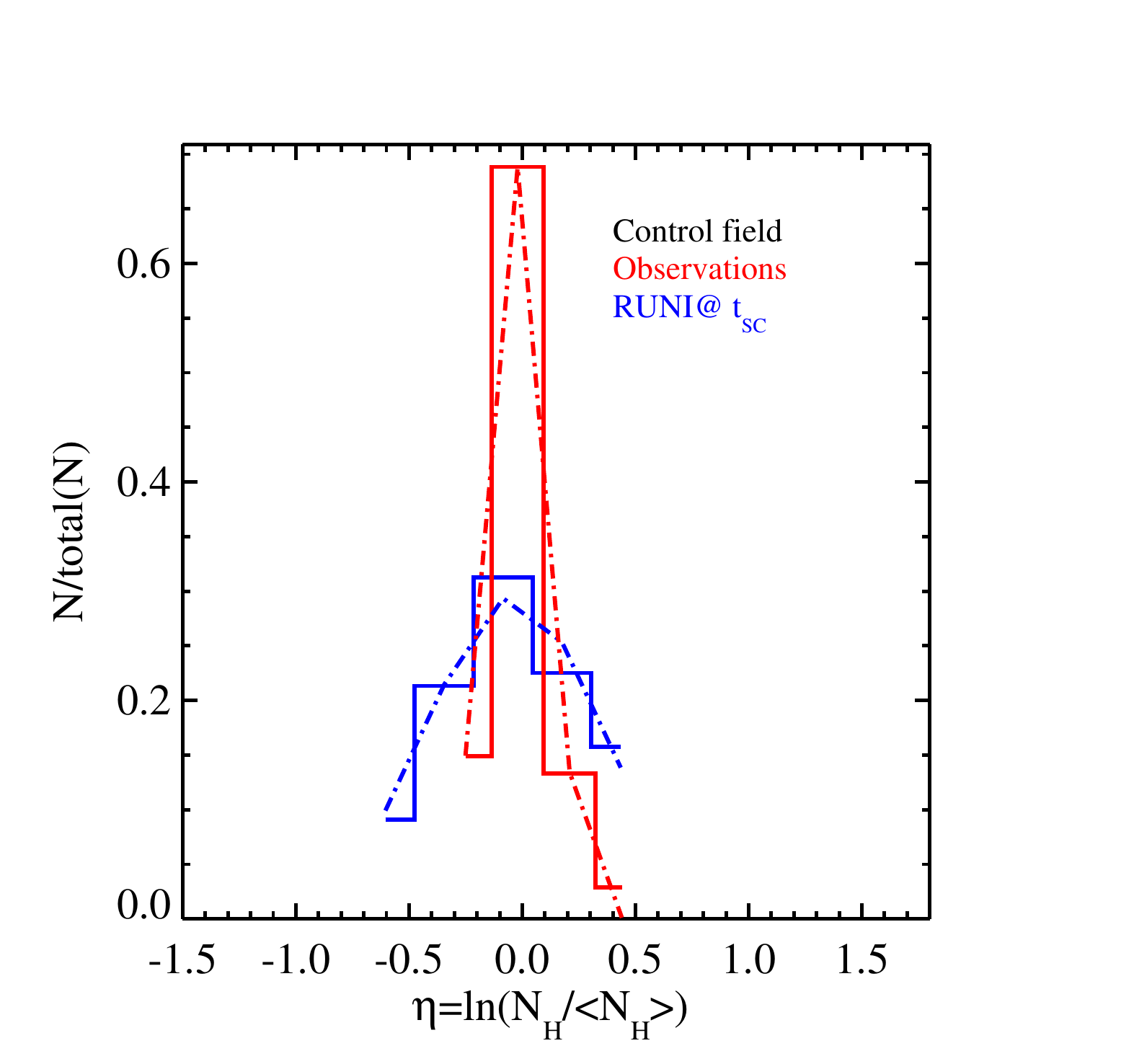}
 \caption{The left panel represents the  histograms of the column density distributions of the control field, while the right panel the PDFs.  The dashed lines overplotted to the 
 PDFs represent the best Gaussian fit of the data. Red lines represent the 
 observations, while the blue ones represent the simulations.  }
     \label{obs2t_empty}
\end{figure*}

\noindent
 We consider in both observations and simulations a {\em control} region where neither filamentary structures nor cores are present (box 0 in Figures\,\ref{pos_fil_obs} and \ref{reg_sim}). 
Figure~\ref{obs2t_empty} shows the histograms of the column density distributions on the left, and the PDFs on the right. 
 Since only few bins are covered in the column density distribution, these plots are shown in the appendix. However, the small 
error-bars in each bin highlight that the column density distributions show only a peak without any additional structure.  
This peak at $A_{\rm V}\sim0.5$\,mag is very narrow in the observations, while  
in the simulations the peak is 
much broader. 

\section{ Filamentary structures in the Serpens Core.}
\begin{figure*}
\centering
  \includegraphics[trim=0.8cm 0.05cm 0.8cm 0.2cm,width=2.55cm]{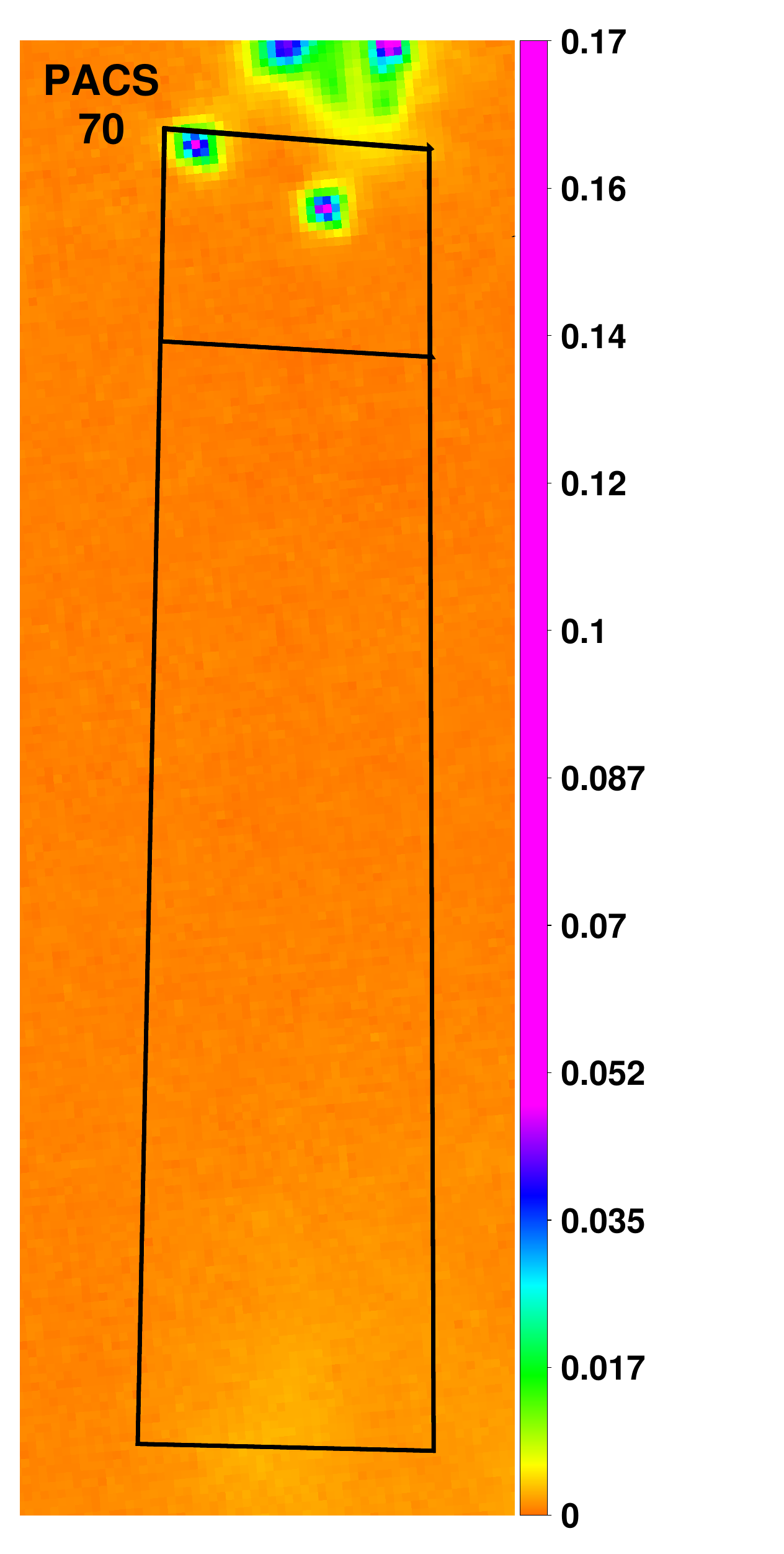}
  \includegraphics[trim=0.8cm 0.05cm 0.8cm 0.45cm,width=2.55cm]{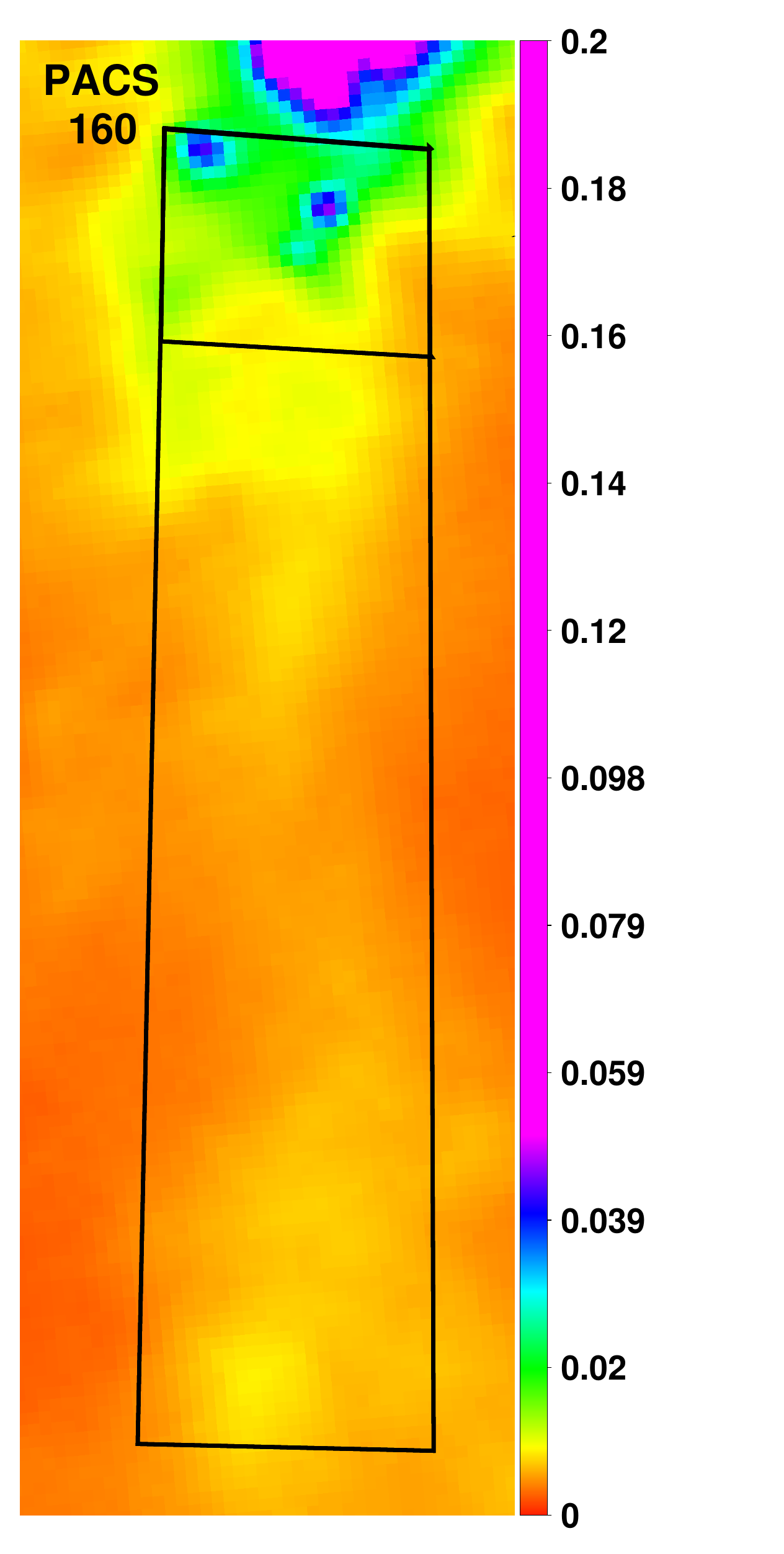}
  \includegraphics[trim=0.8cm 0.05cm 0.8cm 0.45cm,width=2.55cm]{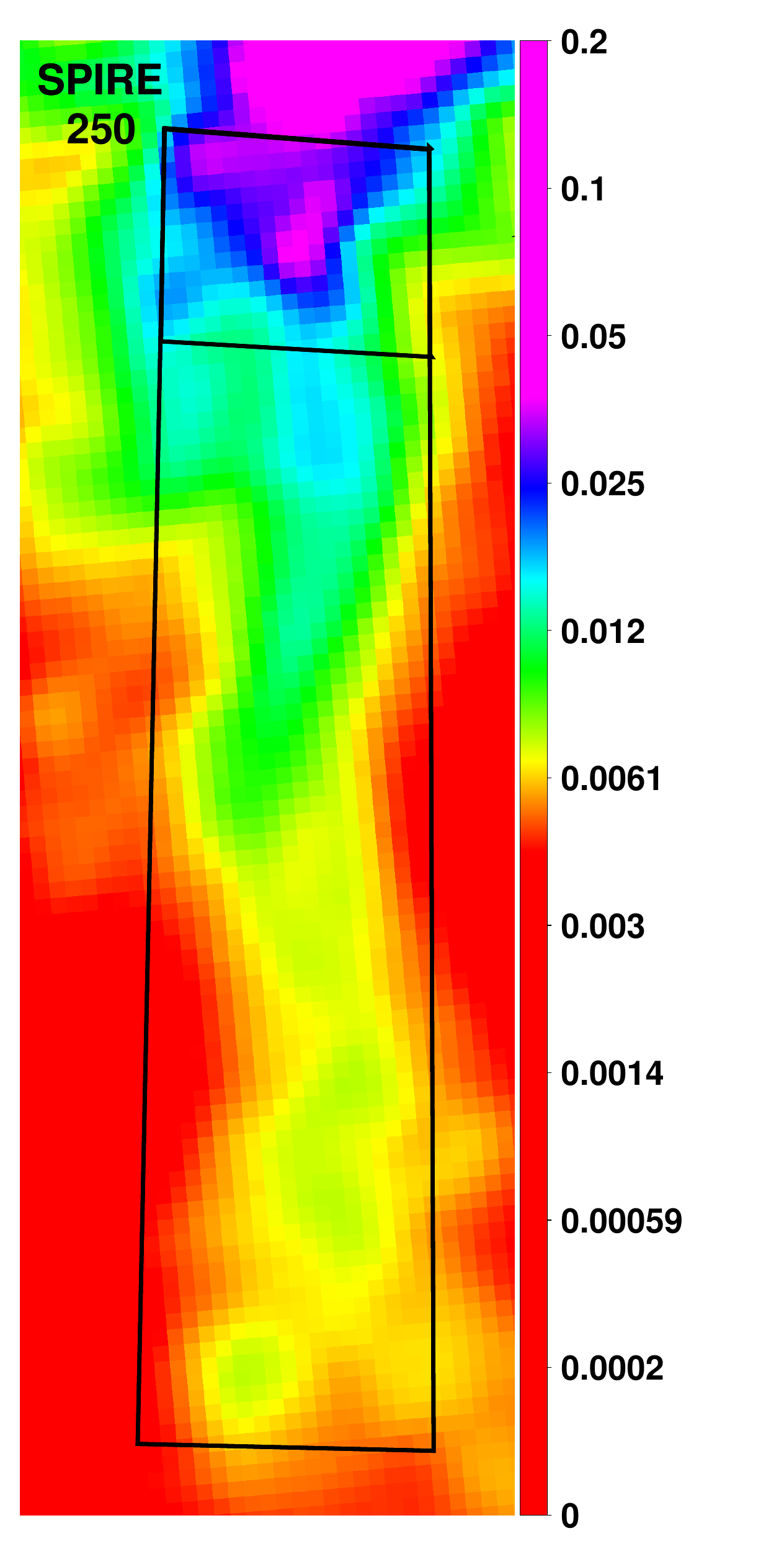}
  \includegraphics[trim=0.8cm 0.05cm 0.8cm 0.45cm,width=2.55cm]{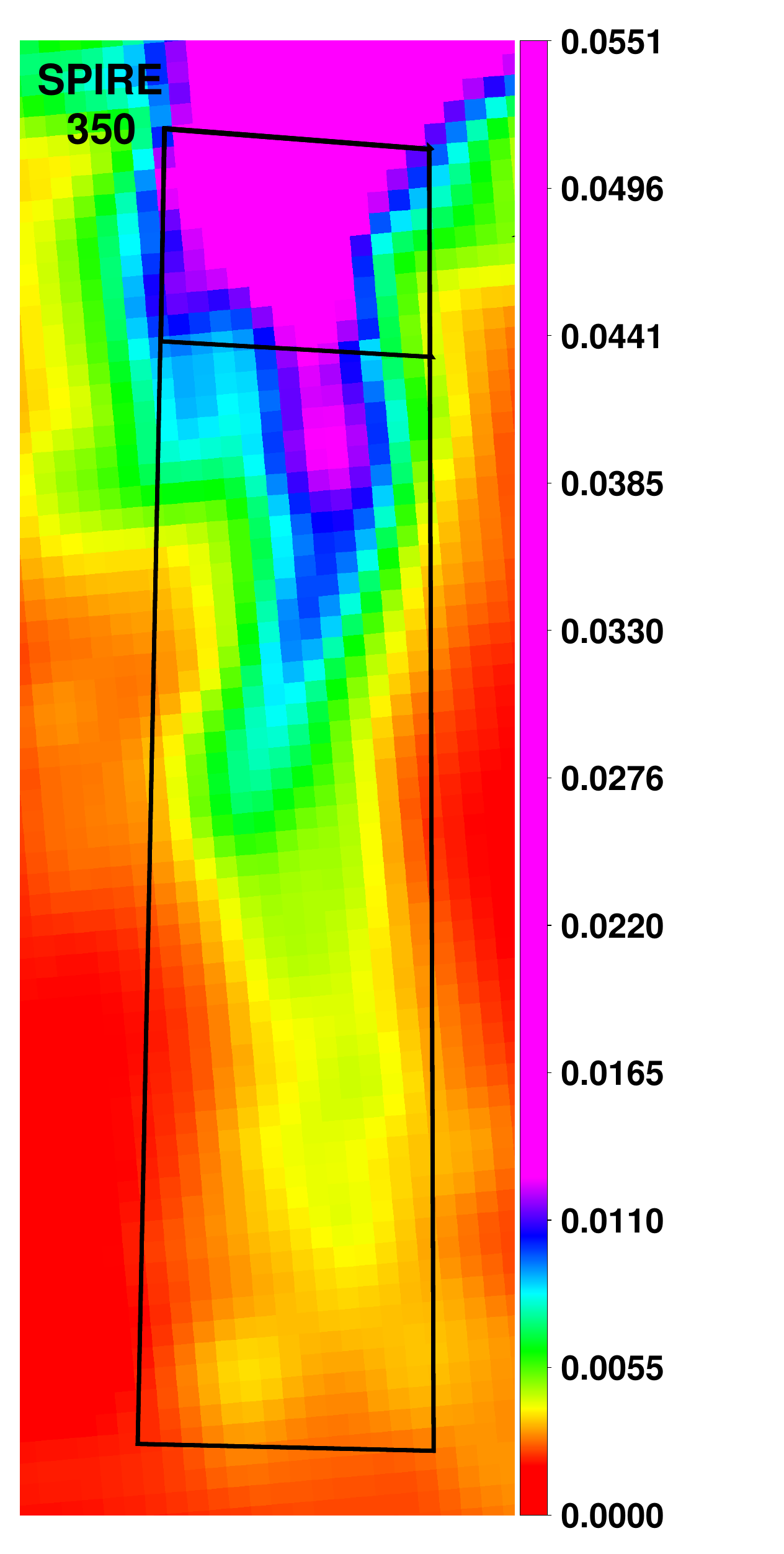}
  \includegraphics[trim=0.8cm 0.05cm 0.8cm 0.45cm,width=2.55cm]{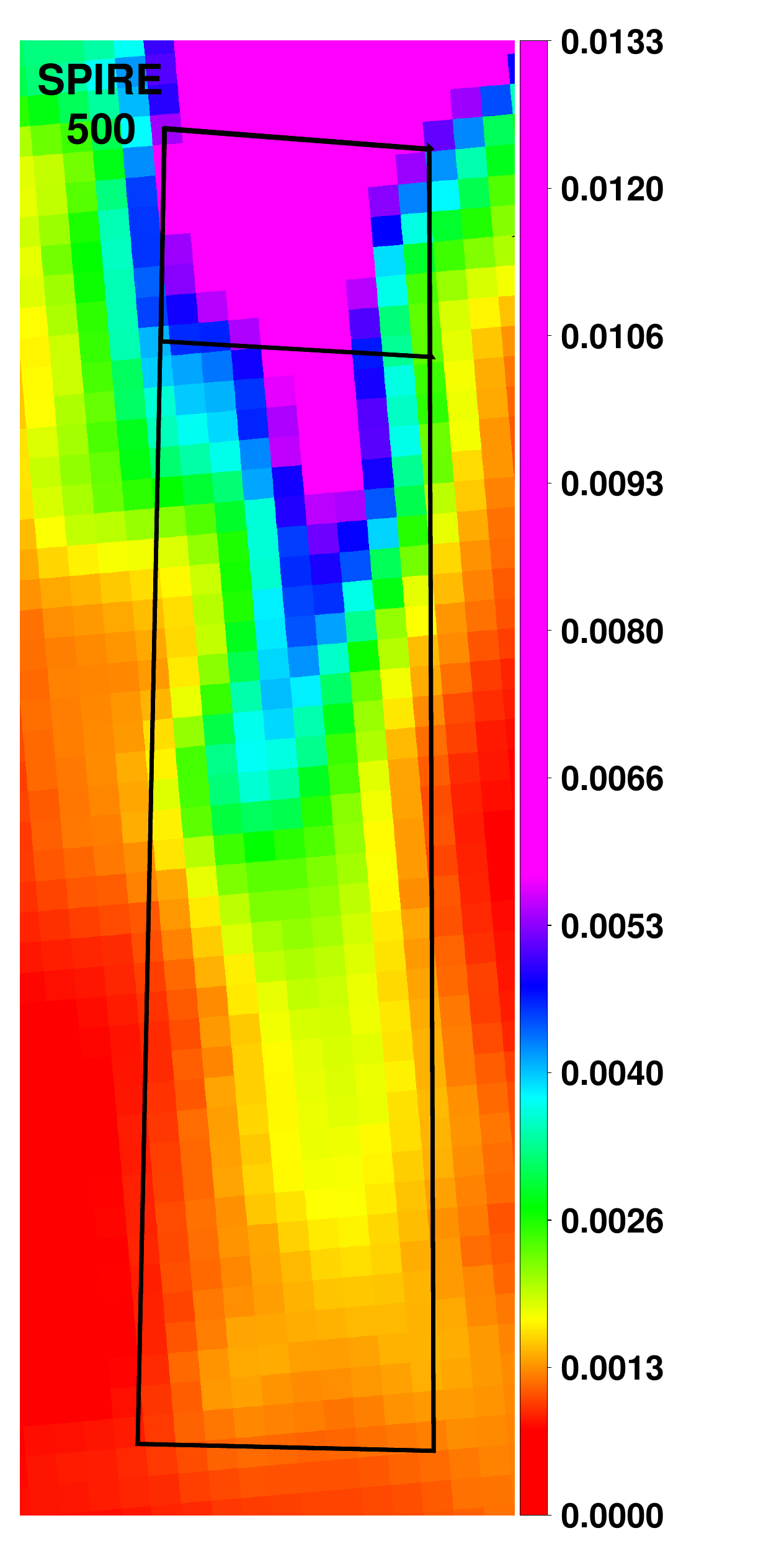}
   \includegraphics[trim=0.8cm 0.05cm 0.8cm 0.45cm,width=2.55cm]{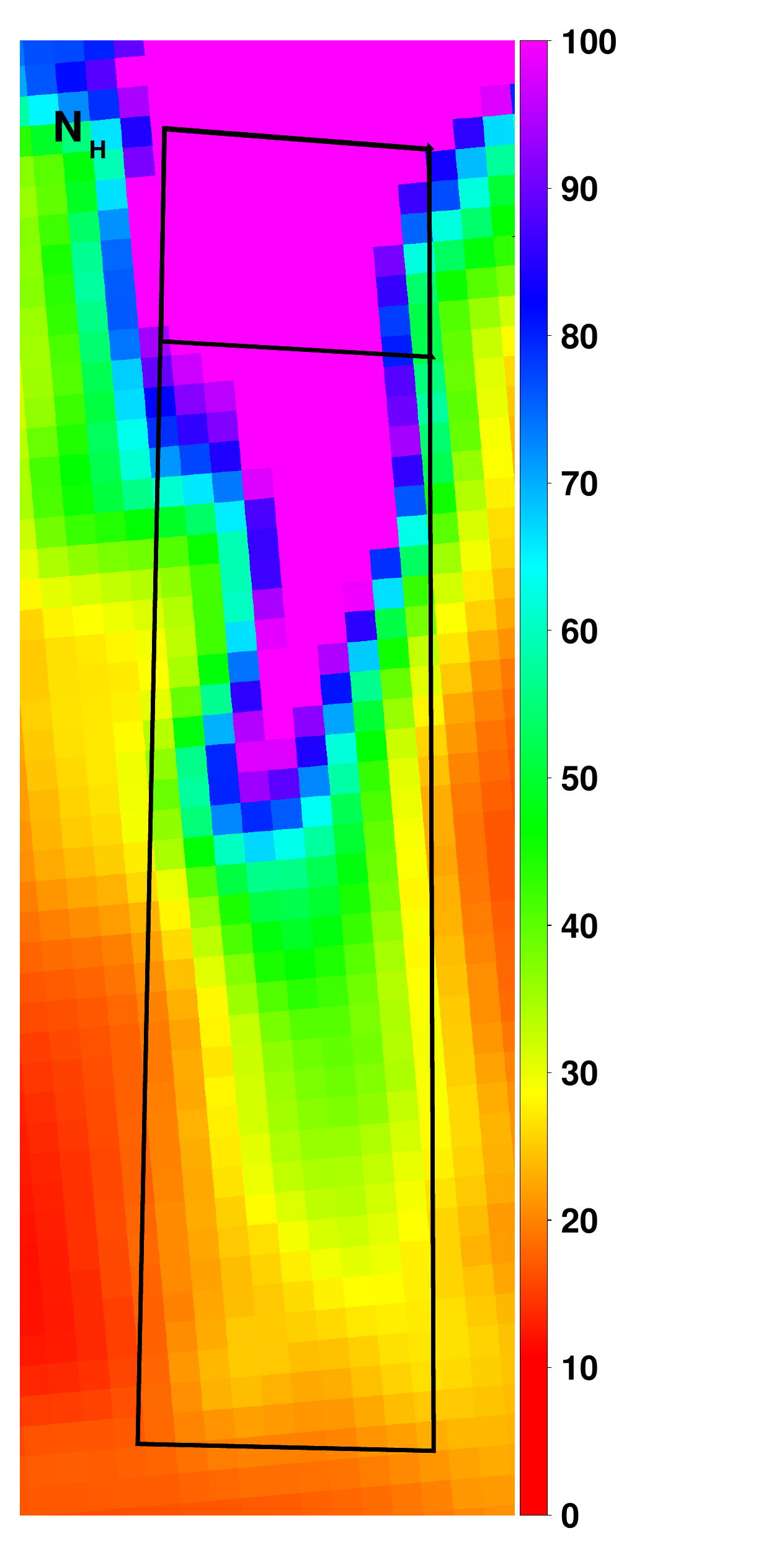}
  \includegraphics[trim=0.8cm 0.05cm 0.8cm 0.45cm,width=2.55cm]{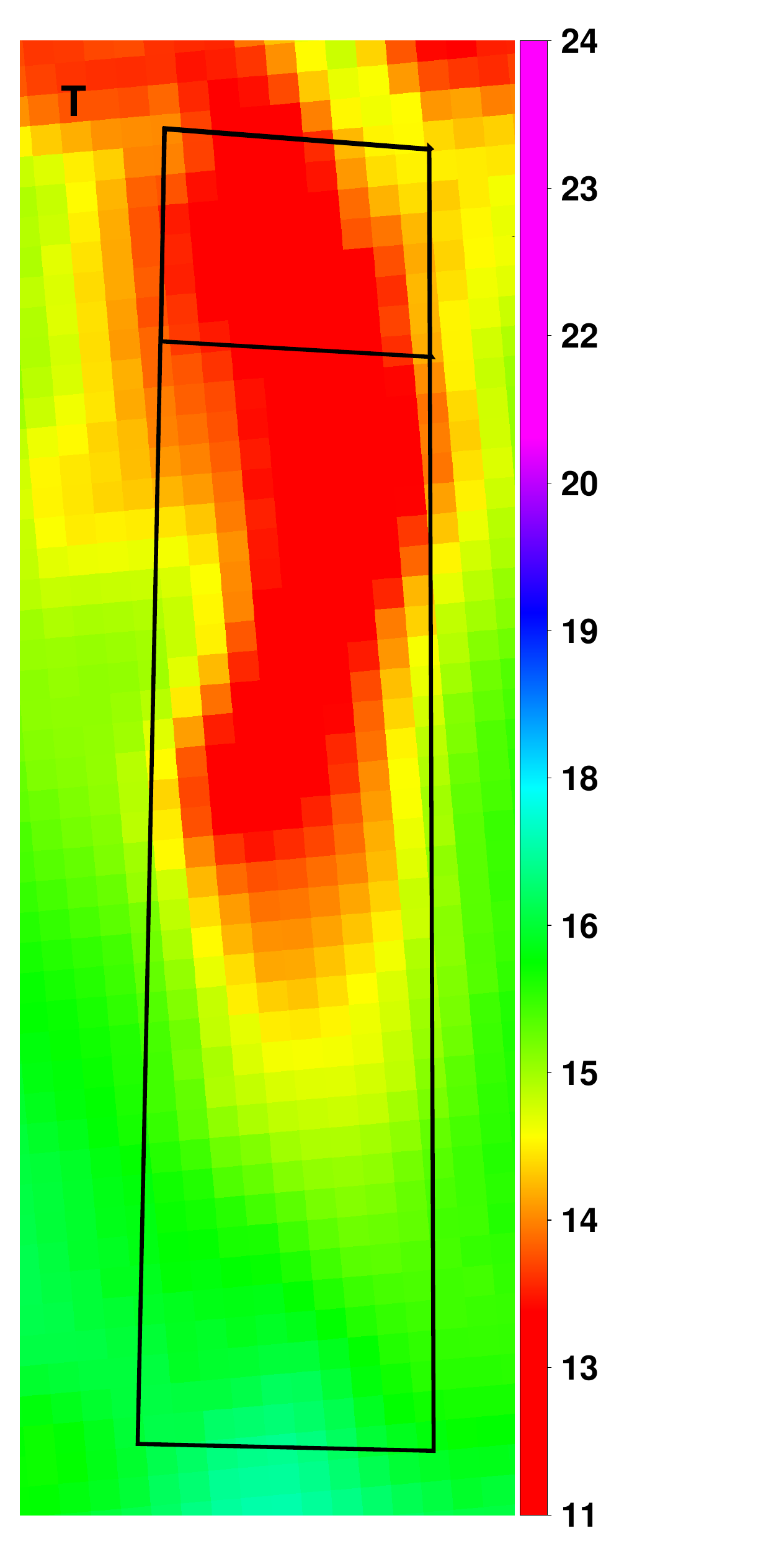}
    \caption{Filament number 1. Starting from the left side, the panels show  {\it Herschel}/PACS 70\,$\mu$m, 160 \,$\mu$m, and  
    SPIRE 250\,$\mu$m, 350\,$\mu$m, and 500\,$\mu$m, respectively. The unit of all the images is Jy/arcsec$^2$. The last two 
    panels show the column density map ($N_{\rm H}$) in [$\times$10$^{+20}cm^{-2}$] and the temperature map  
    in ${\rm K}$. The box identifies the extension of the filament and its size is listed in Table~\ref{table_filam}. The 
    smaller box identifies the position of the cores among the filament.}
     \label{fil1}
\end{figure*}

 Figure~\ref{fil1} shows the enlargements on Filament number 1: starting from the left side, the first five panels show the {\it Herschel} PACS and SPIRE mosaics, while the last two panels show the column density and the temperature map. 
An additional box identifies the position of a core structure among the filament. 

\noindent
In the online material,  Figures~\ref{fil1_on}-\,\ref{fil9} show the enlargements on each filament as in Figure~\ref{fil1}.
\onlfig{
\setcounter{figure}{7} 
\begin{figure*}
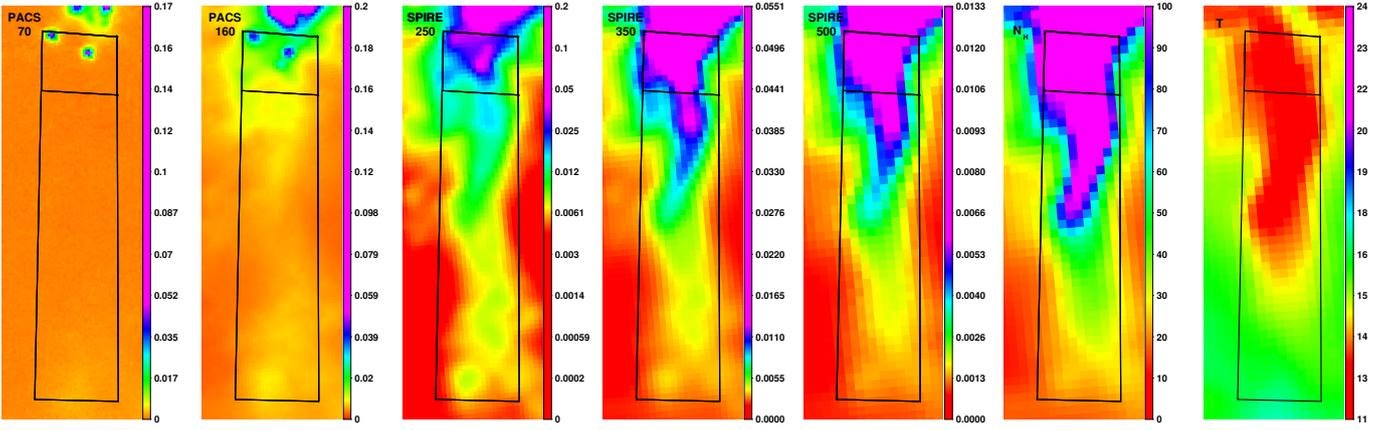

\centering
  \includegraphics[trim=0.8cm 0.05cm 0.8cm 0.2cm,width=2.55cm]{fil1_70-eps-converted-to.pdf}
  \includegraphics[trim=0.8cm 0.05cm 0.8cm 0.45cm,width=2.55cm]{fil1_160-eps-converted-to.pdf}
  \includegraphics[trim=0.8cm 0.05cm 0.8cm 0.45cm,width=2.55cm]{fil1_250-eps-converted-to.pdf}
  \includegraphics[trim=0.8cm 0.05cm 0.8cm 0.45cm,width=2.55cm]{fil1_350-eps-converted-to.pdf}
  \includegraphics[trim=0.8cm 0.05cm 0.8cm 0.45cm,width=2.55cm]{fil1_500-eps-converted-to.pdf}
   \includegraphics[trim=0.8cm 0.05cm 0.8cm 0.45cm,width=2.55cm]{fil1_Nh-eps-converted-to.pdf}
  \includegraphics[trim=0.8cm 0.05cm 0.8cm 0.45cm,width=2.55cm]{fil1_T-eps-converted-to.pdf}
    \caption{Filament number 1. Starting from the left side, the panels show  {\it Herschel}/PACS 70\,$\mu$m, 160 \,$\mu$m, and  
    SPIRE 250\,$\mu$m, 350\,$\mu$m, and 500\,$\mu$m, respectively. The unit of all the images is Jy/arcsec$^2$. The last two 
    panels show the column density map ($N_{\rm H}$) in [$\times$10$^{+20}cm^{-2}$] and the temperature map  
    in ${\rm K}$. The box identifies the extension of the filament and its size is listed in Table~\ref{table_filam}. 
    The smaller box identifies the position of the cores among the filament.}
     \label{fil1_on}
\end{figure*}

\begin{figure*}
\centering
  \includegraphics[trim=0.8cm 0.05cm 0.8cm 0.2cm,width=2.55cm]{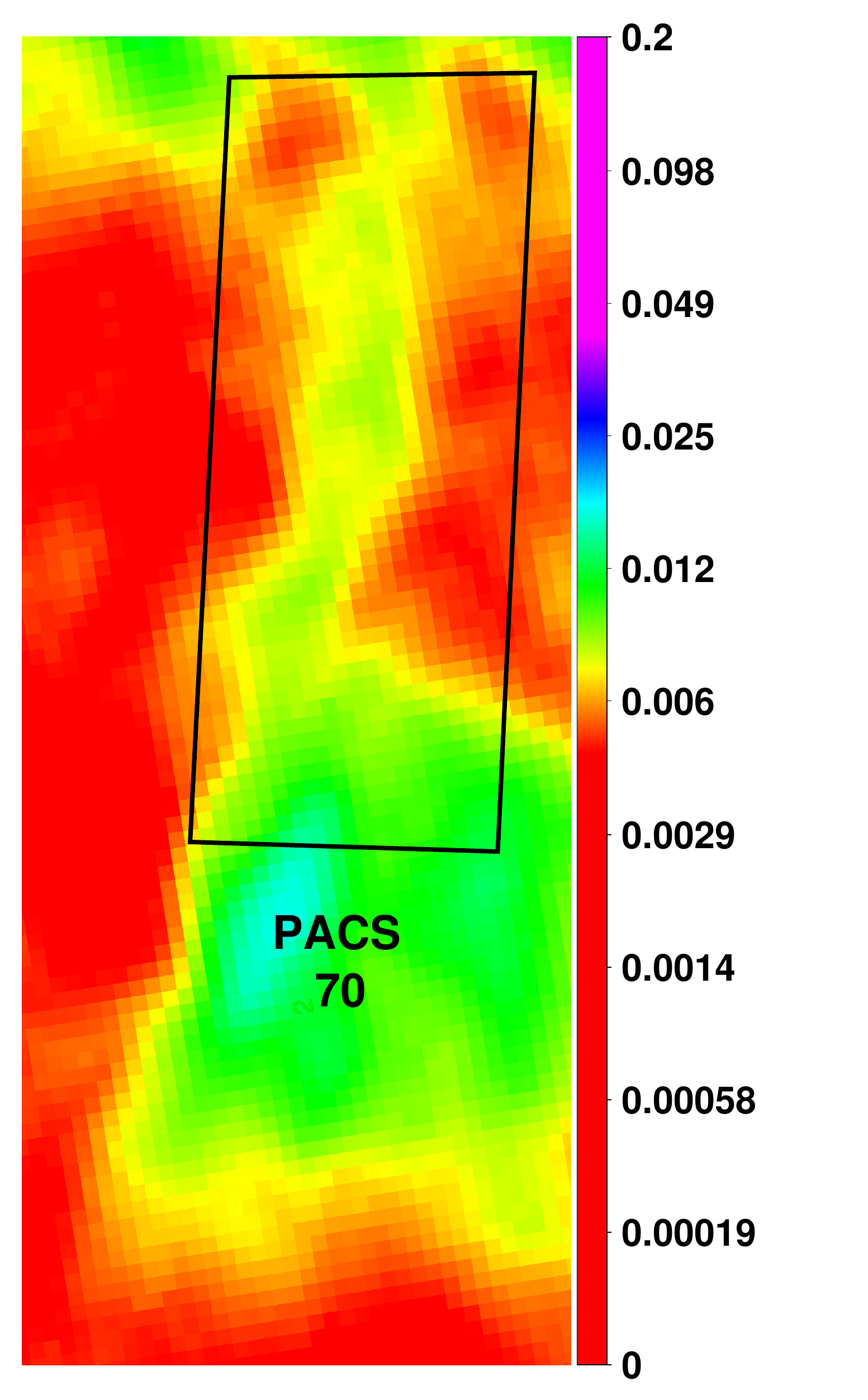}
  \includegraphics[trim=0.8cm 0.05cm 0.8cm 0.45cm,width=2.55cm]{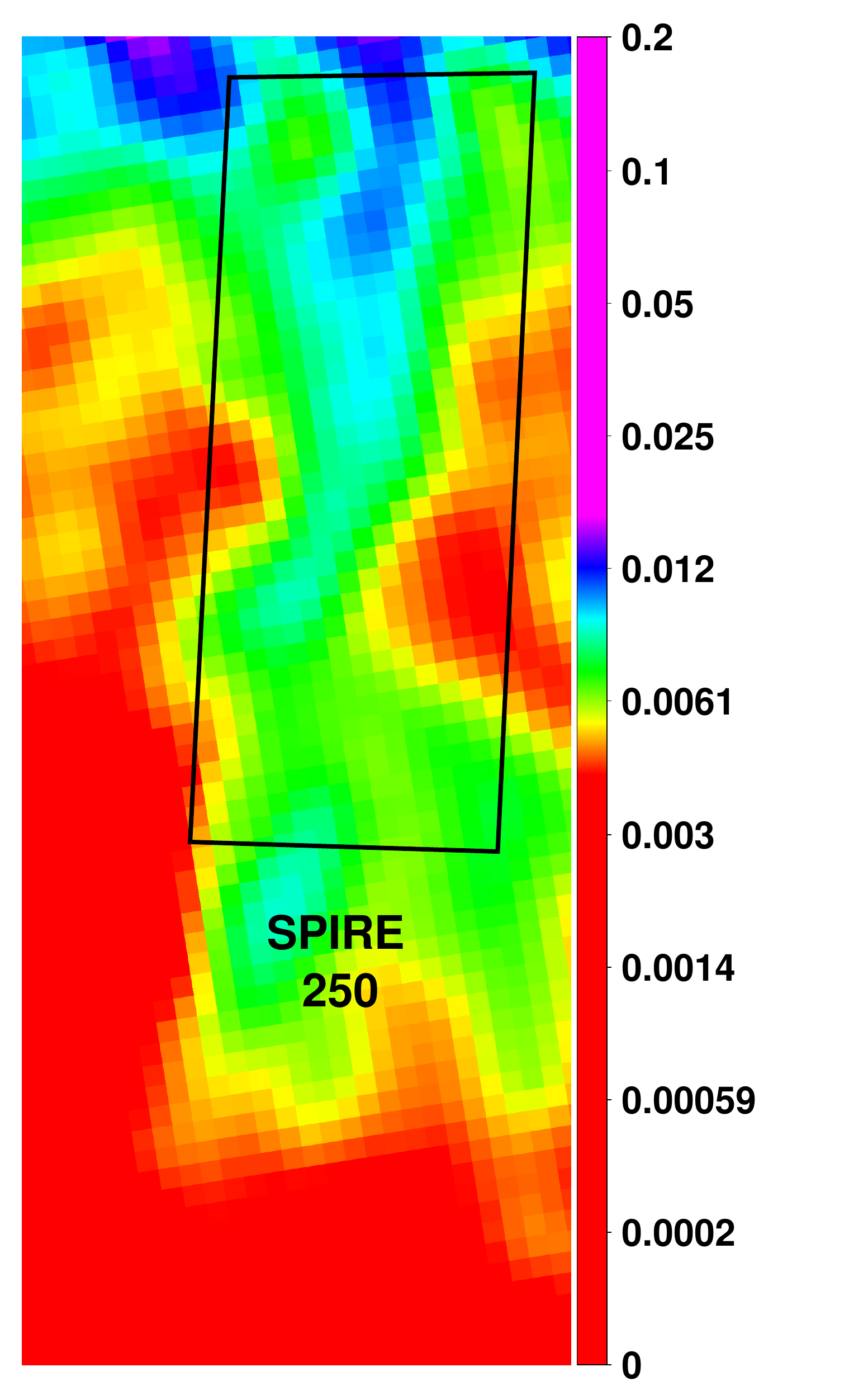}
  \includegraphics[trim=0.8cm 0.05cm 0.8cm 0.45cm,width=2.55cm]{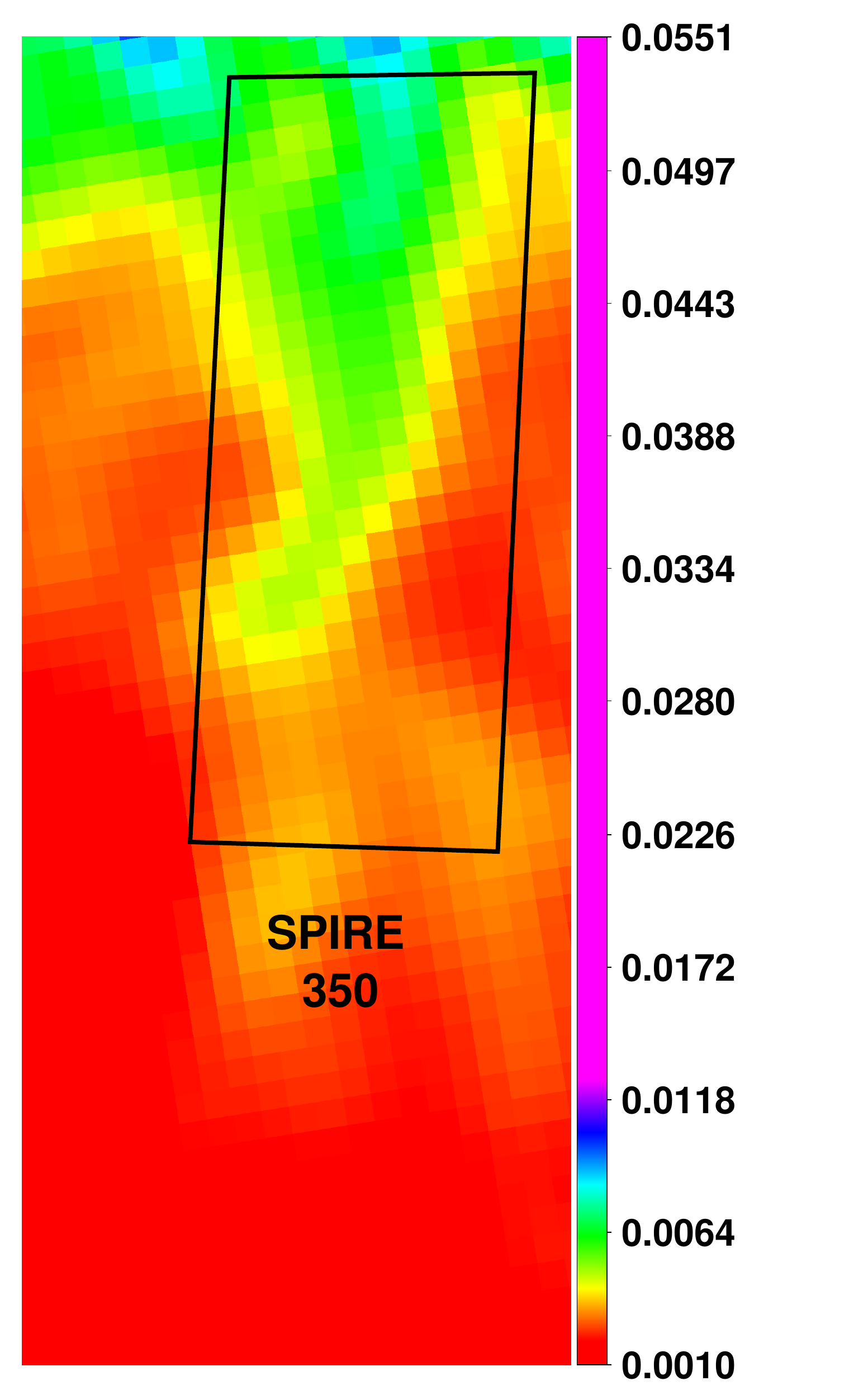}
  \includegraphics[trim=0.8cm 0.05cm 0.8cm 0.45cm,width=2.55cm]{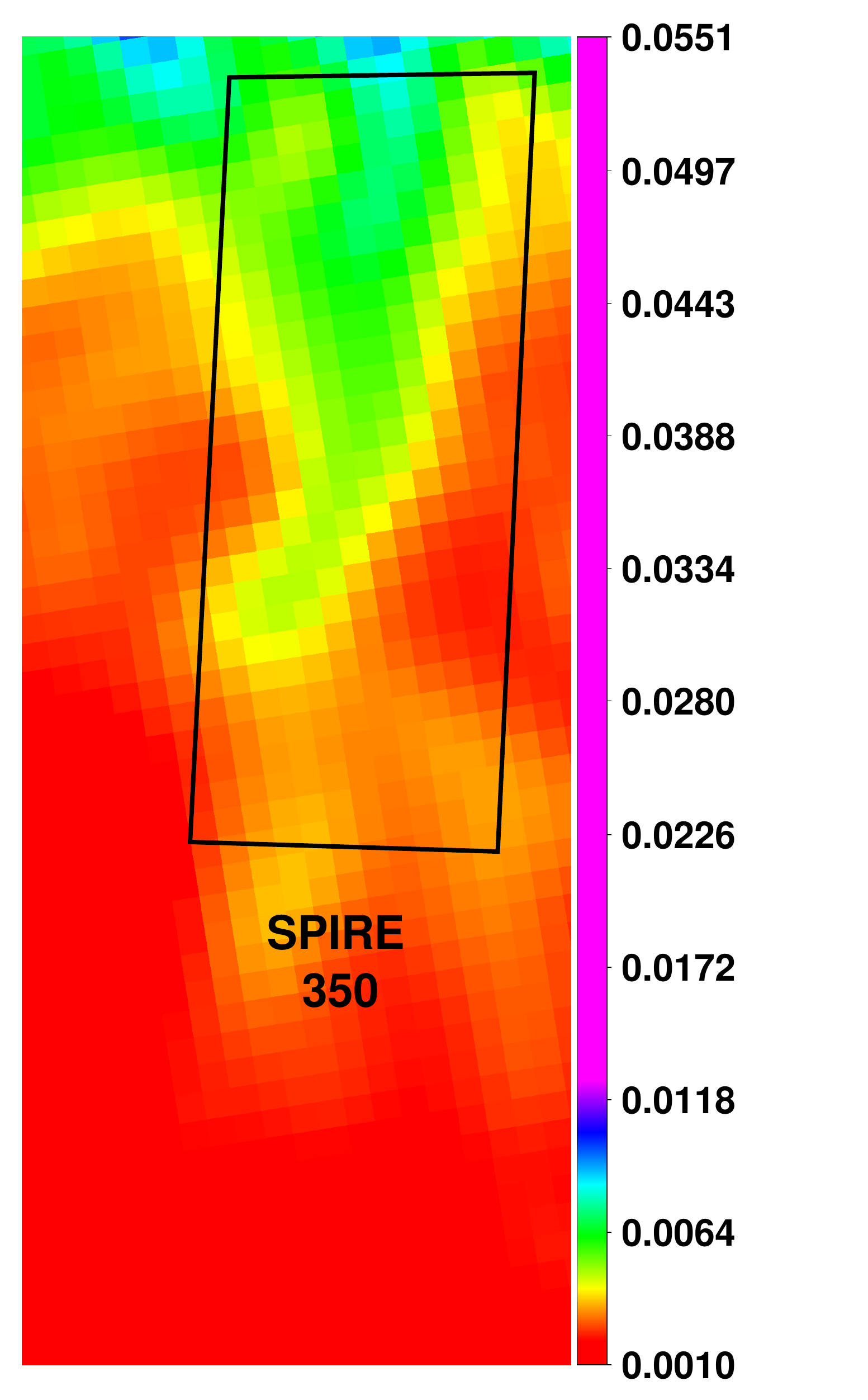}
  \includegraphics[trim=0.8cm 0.05cm 0.8cm 0.45cm,width=2.55cm]{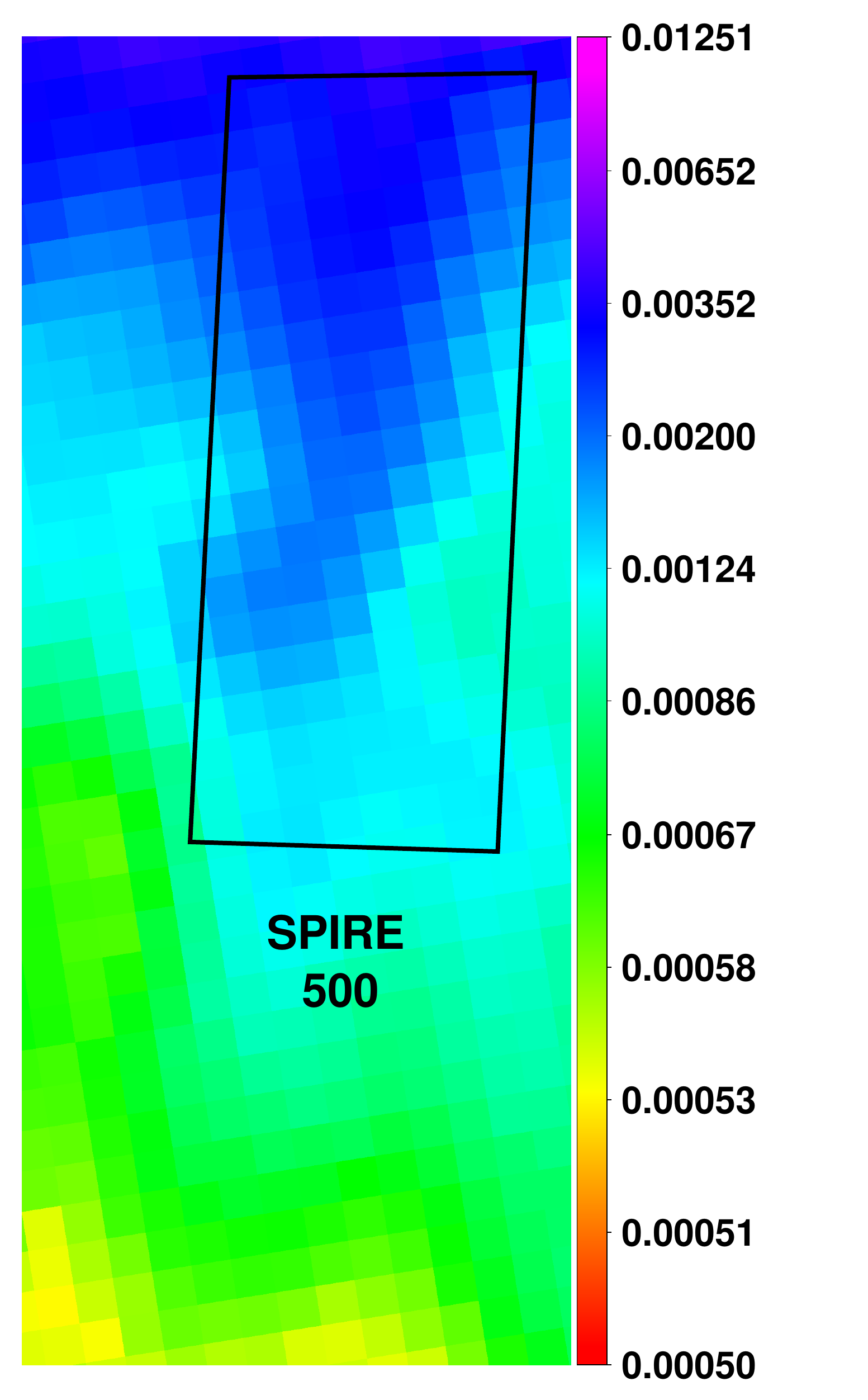}
   \includegraphics[trim=0.8cm 0.05cm 0.8cm 0.45cm,width=2.55cm]{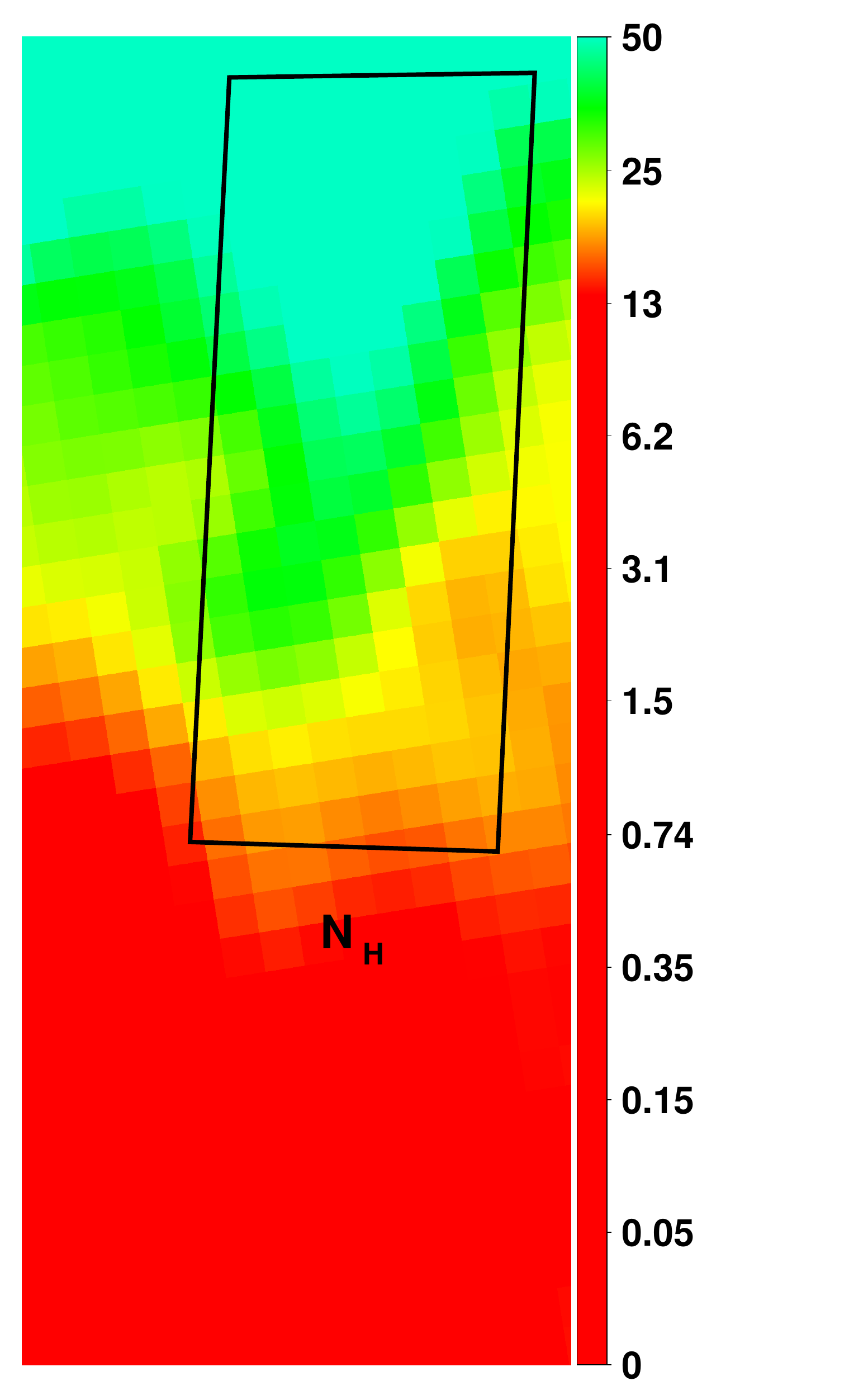}
  \includegraphics[trim=0.8cm 0.05cm 0.8cm 0.45cm,width=2.55cm]{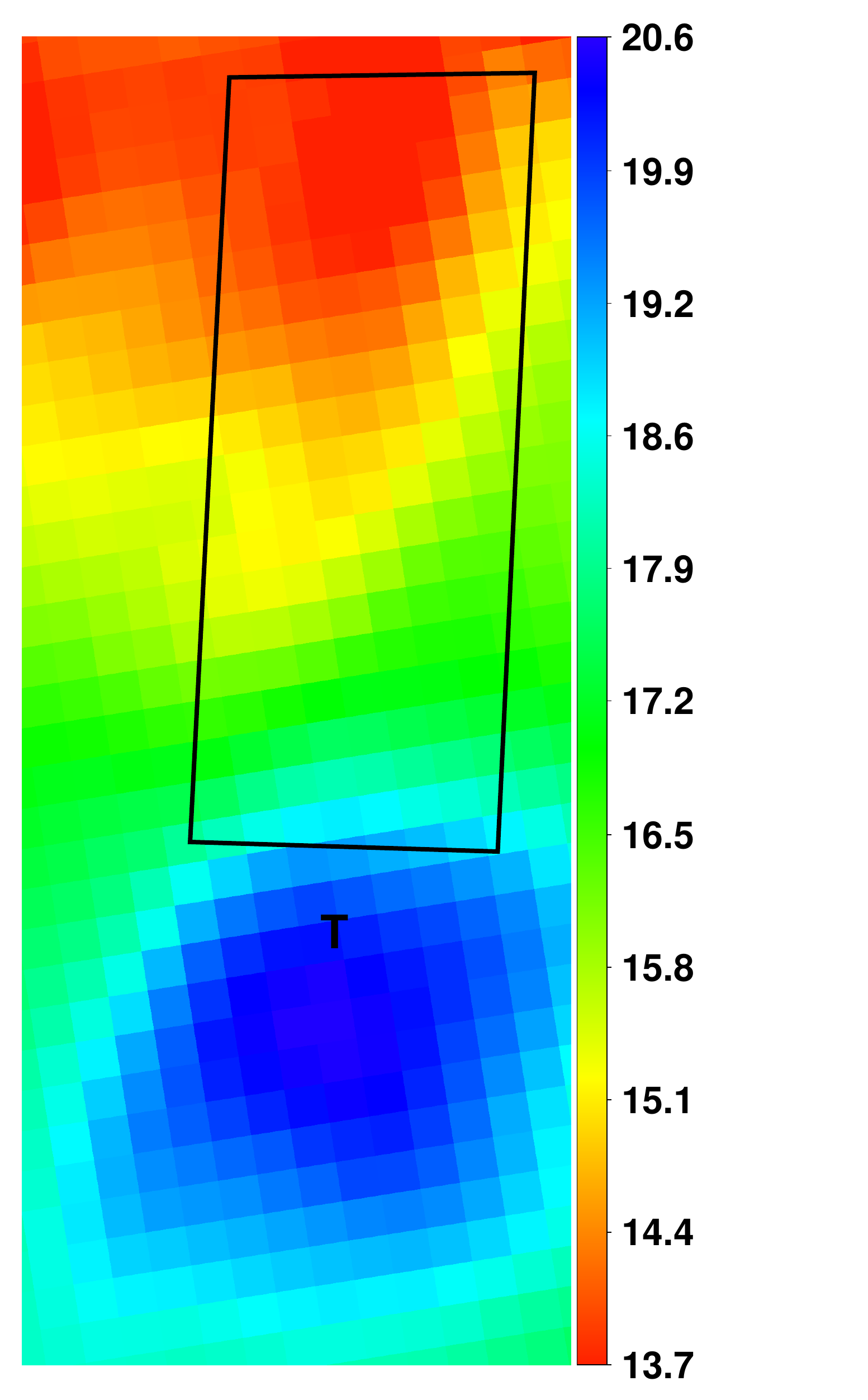}
    \caption{Filament number 2. The panels and the units are as in Figure~\ref{fil1_on}.}
     \label{fil2}
\end{figure*}

\begin{figure*}
\centering
  \includegraphics[trim=0.8cm 0.05cm 0.8cm 0.2cm,width=2.55cm]{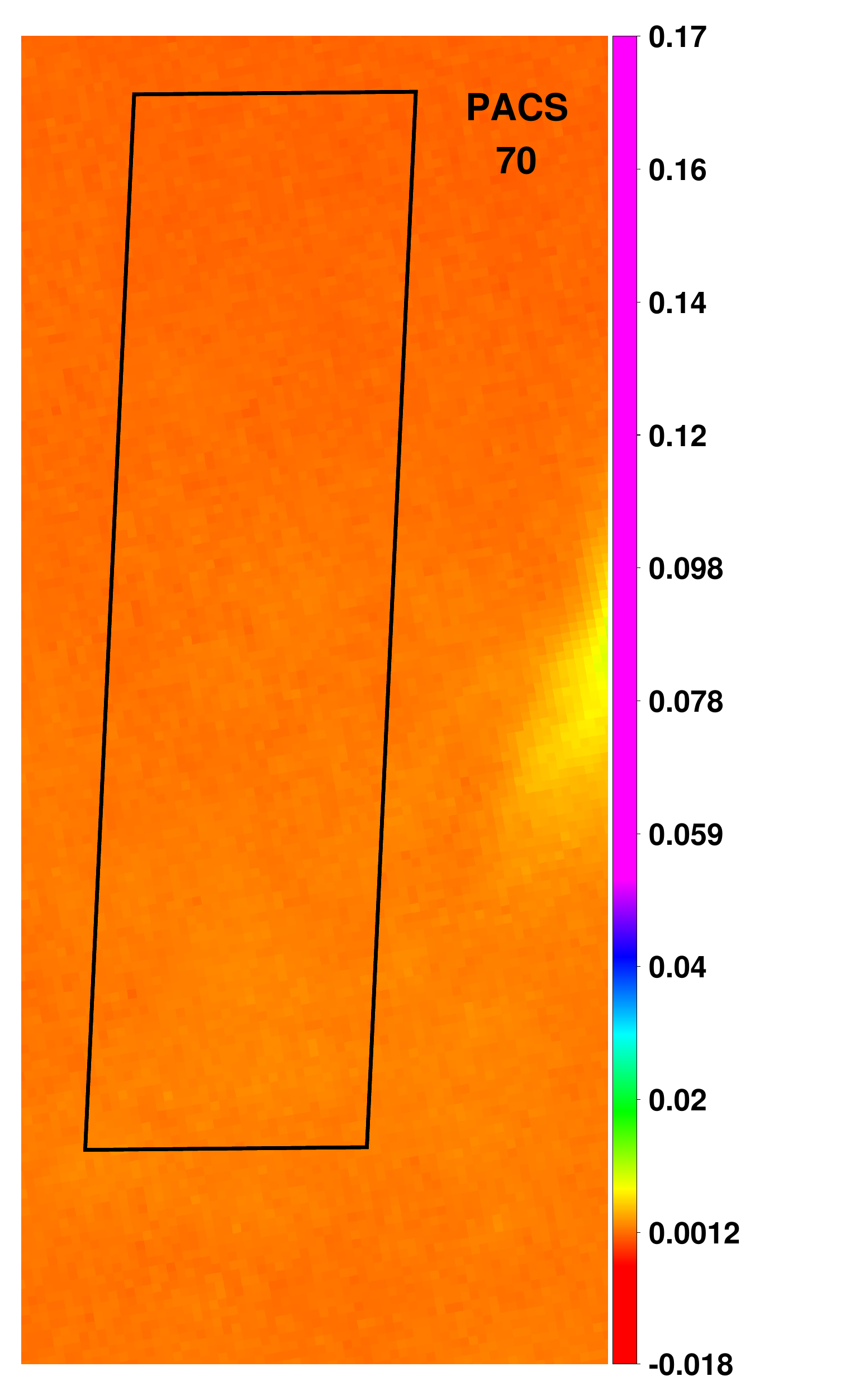}
  \includegraphics[trim=0.8cm 0.05cm 0.8cm 0.45cm,width=2.55cm]{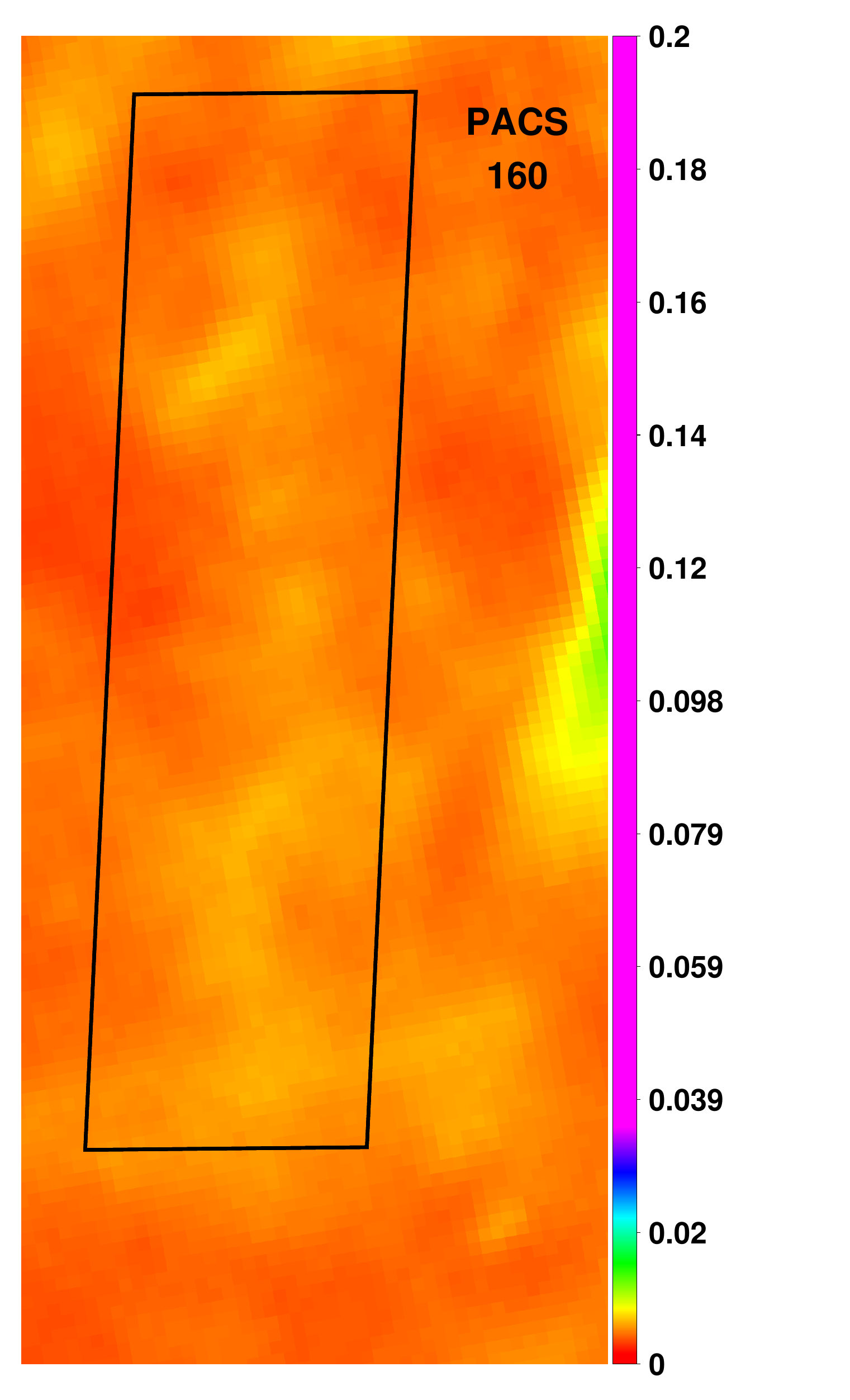}
  \includegraphics[trim=0.8cm 0.05cm 0.8cm 0.45cm,width=2.55cm]{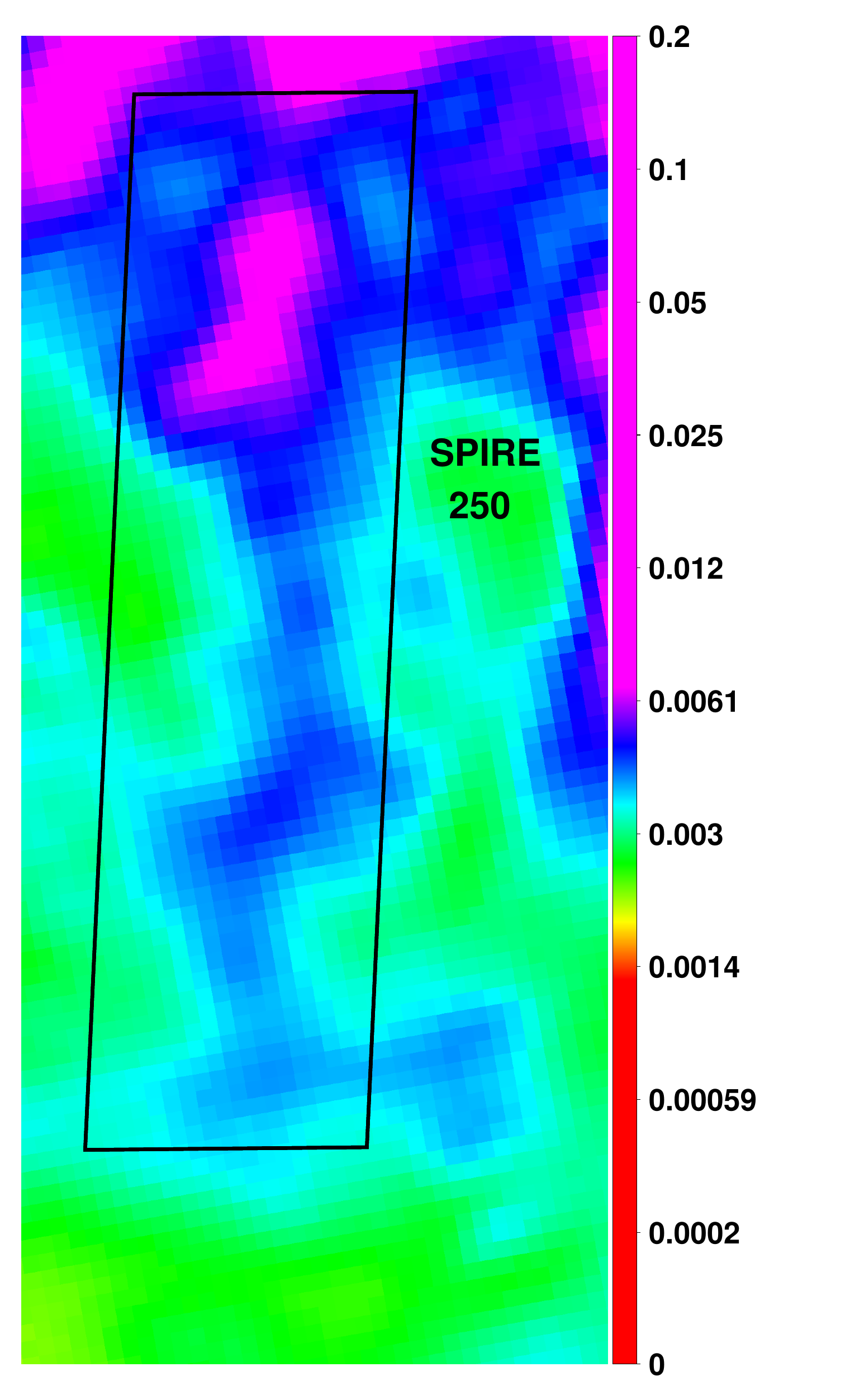}
  \includegraphics[trim=0.8cm 0.05cm 0.8cm 0.45cm,width=2.55cm]{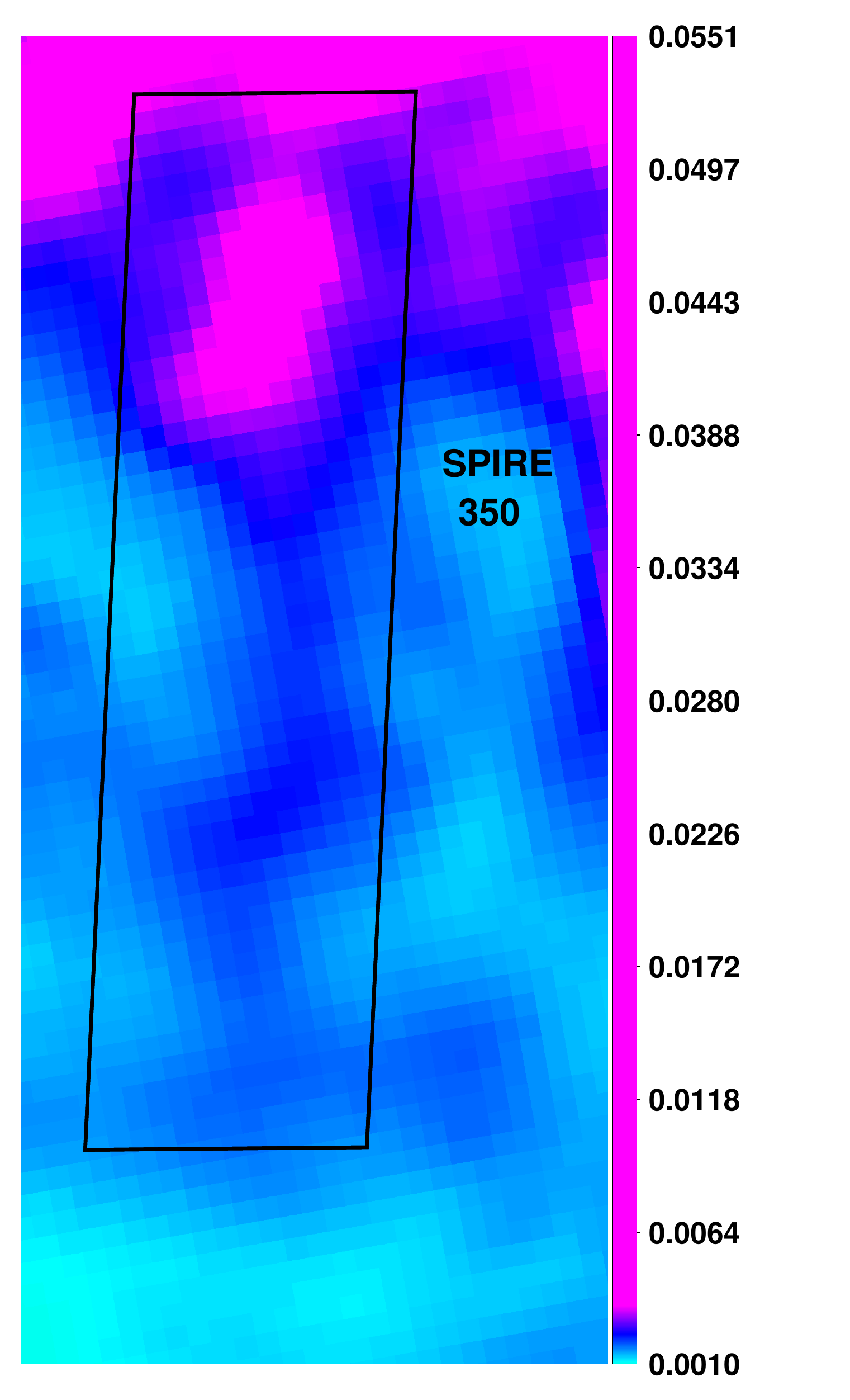}
  \includegraphics[trim=0.8cm 0.05cm 0.8cm 0.45cm,width=2.55cm]{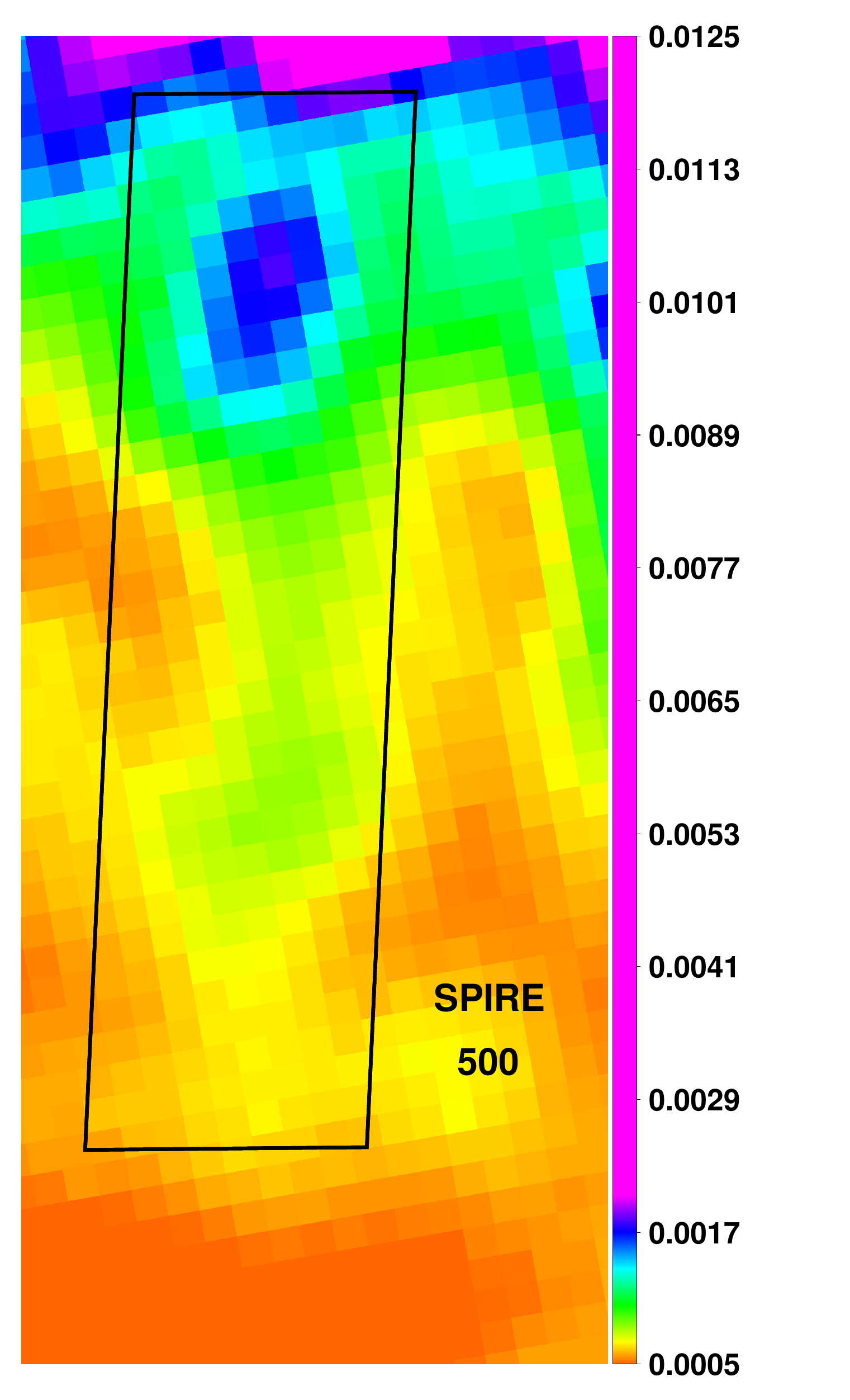}
   \includegraphics[trim=0.8cm 0.05cm 0.8cm 0.45cm,width=2.55cm]{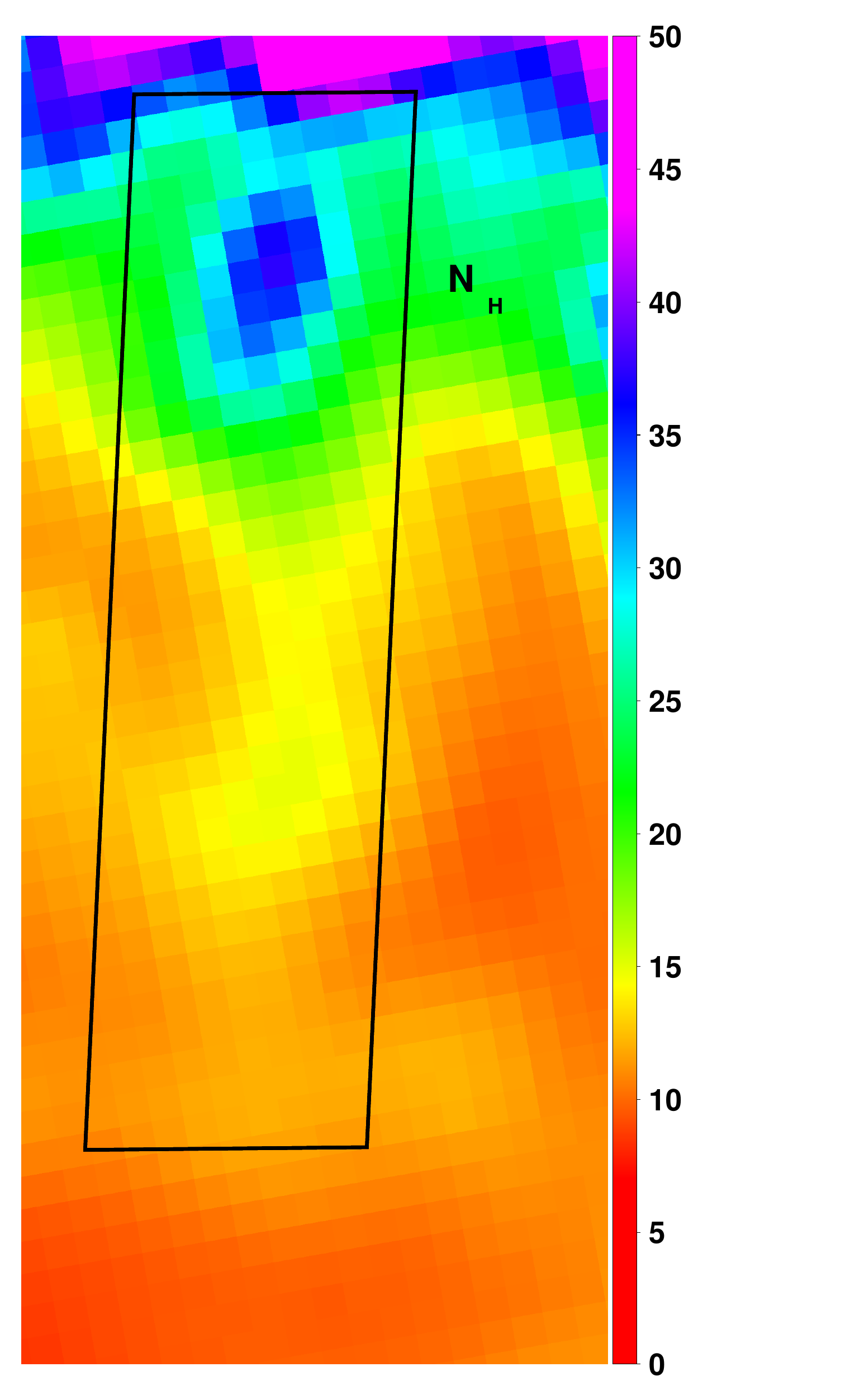}
  \includegraphics[trim=0.8cm 0.05cm 0.8cm 0.45cm,width=2.55cm]{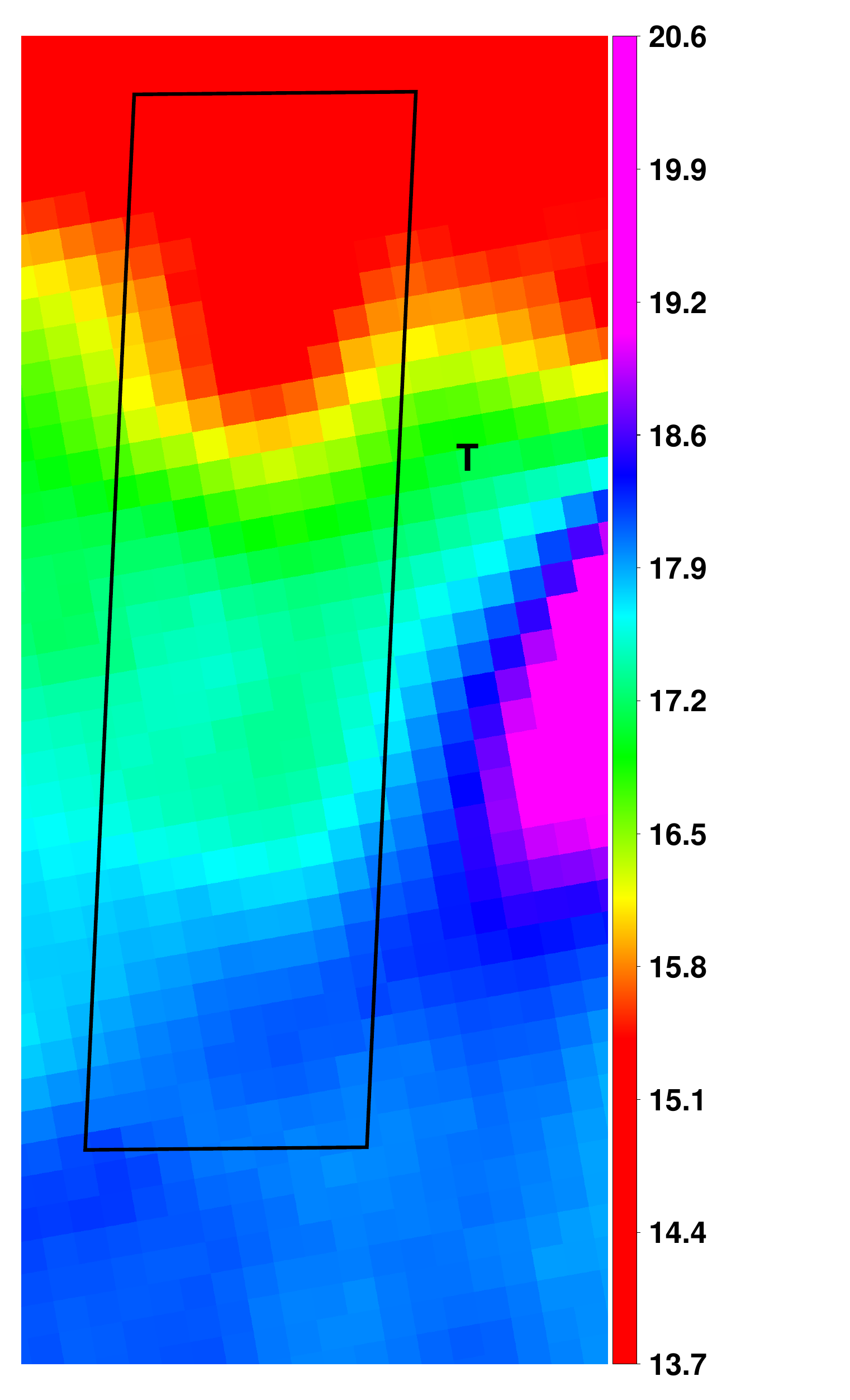}
    \caption{Filament number 3.  The panels and the units are as in Figure~\ref{fil1_on}.}
     \label{fil3}
\end{figure*}
\begin{figure*}
\centering
  \includegraphics[trim=0.8cm 0.05cm 0.8cm 0.2cm,width=2.55cm]{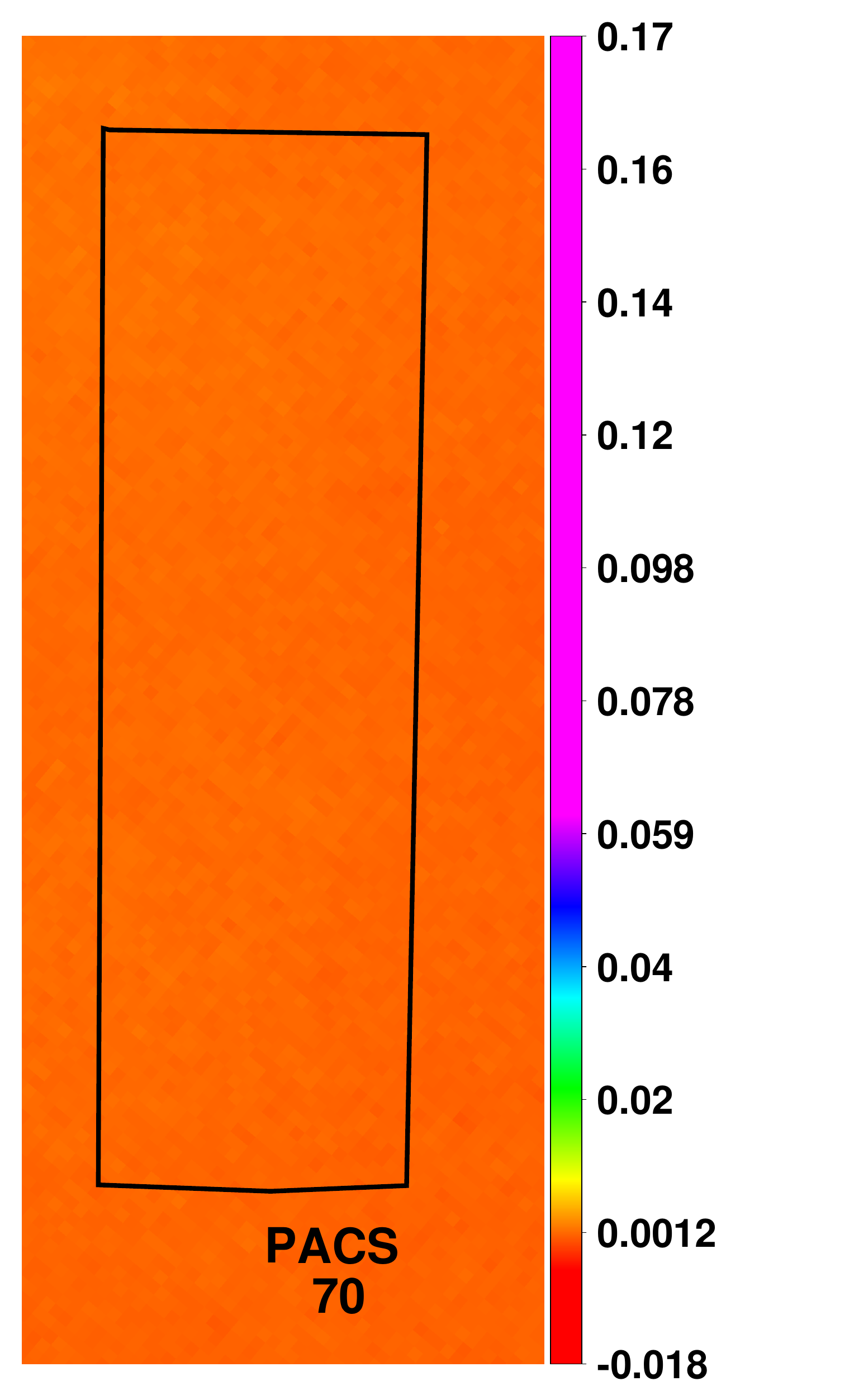}
  \includegraphics[trim=0.8cm 0.05cm 0.8cm 0.45cm,width=2.55cm]{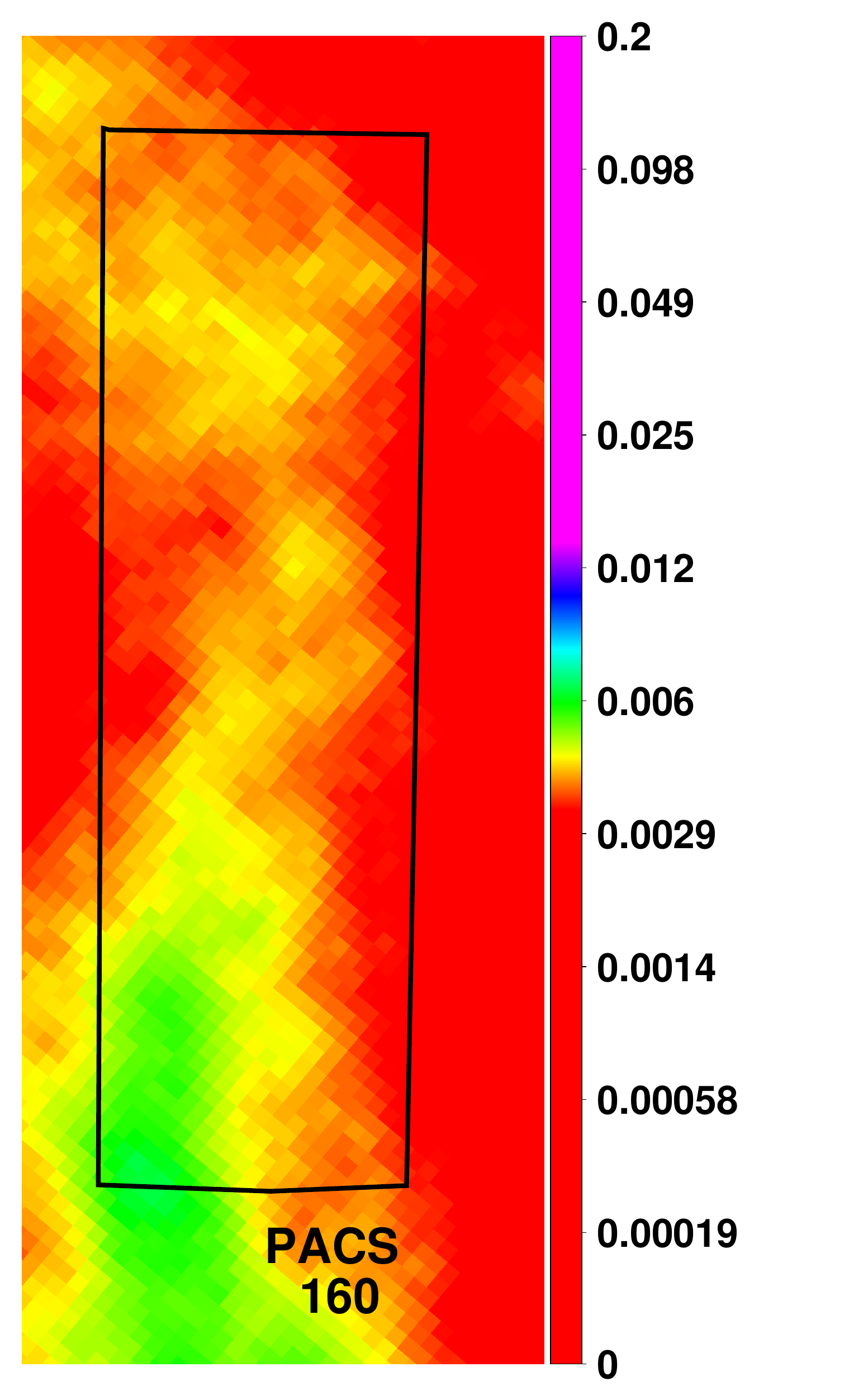}
  \includegraphics[trim=0.8cm 0.05cm 0.8cm 0.45cm,width=2.55cm]{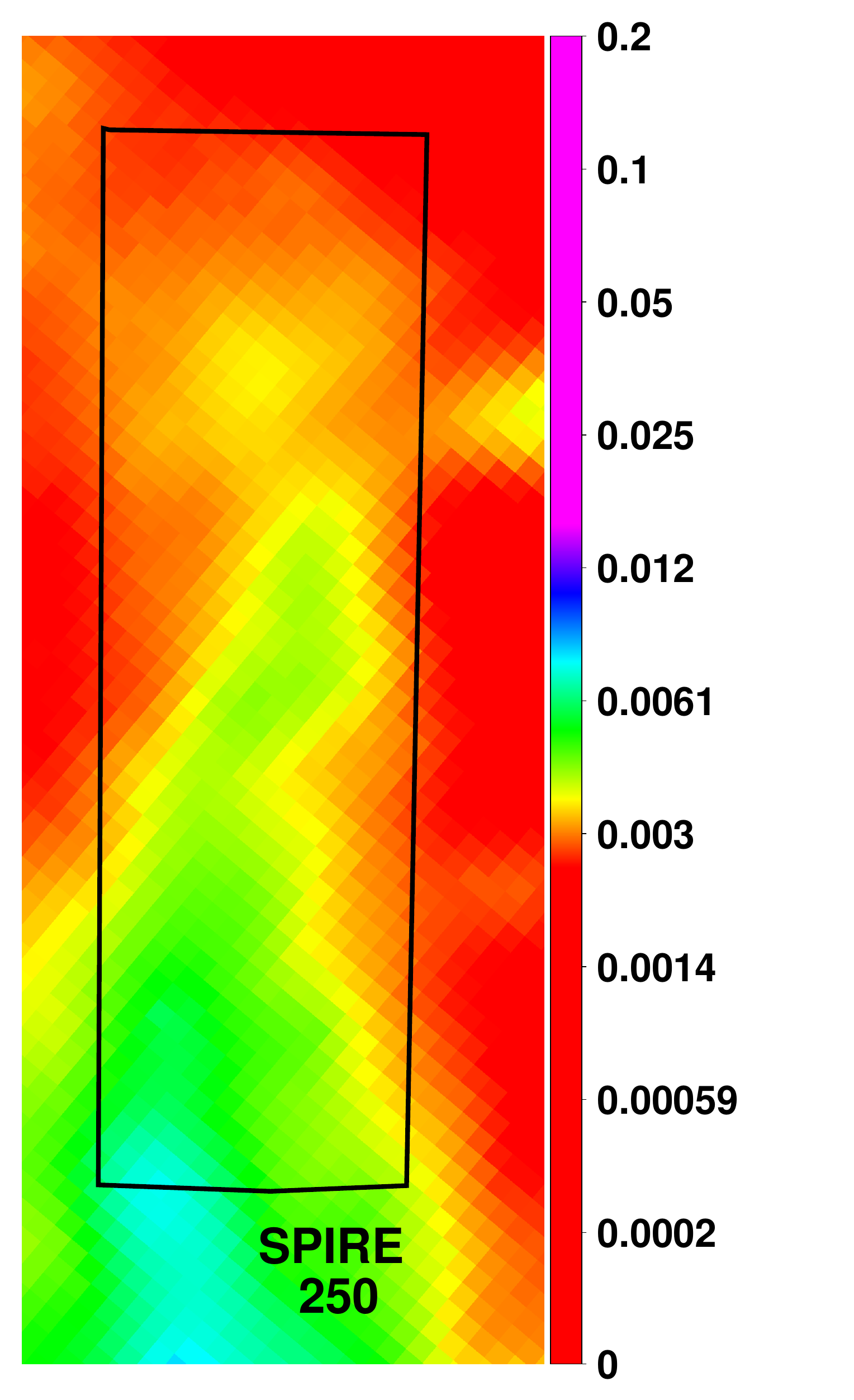}
  \includegraphics[trim=0.8cm 0.05cm 0.8cm 0.45cm,width=2.55cm]{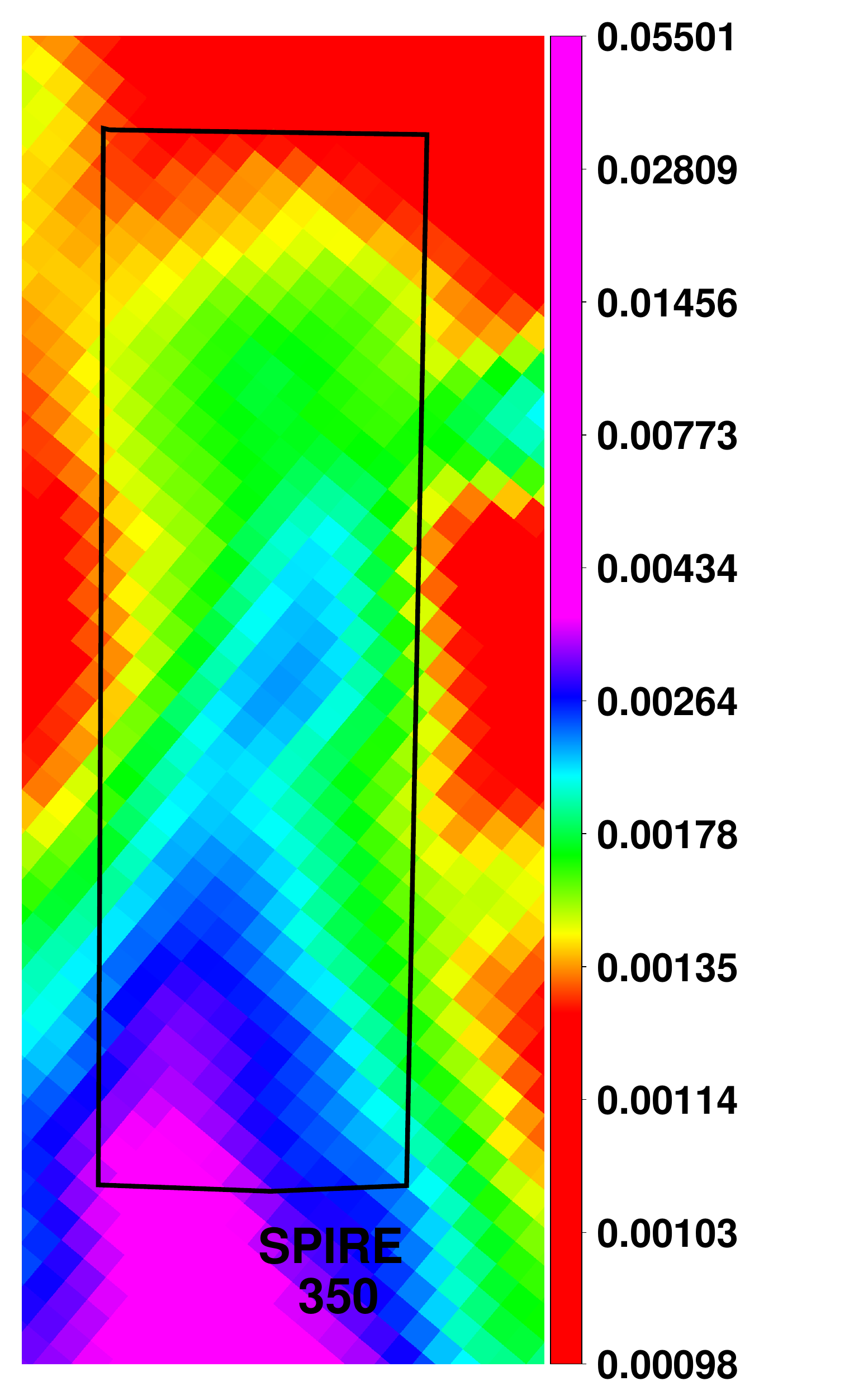}
  \includegraphics[trim=0.8cm 0.05cm 0.8cm 0.45cm,width=2.55cm]{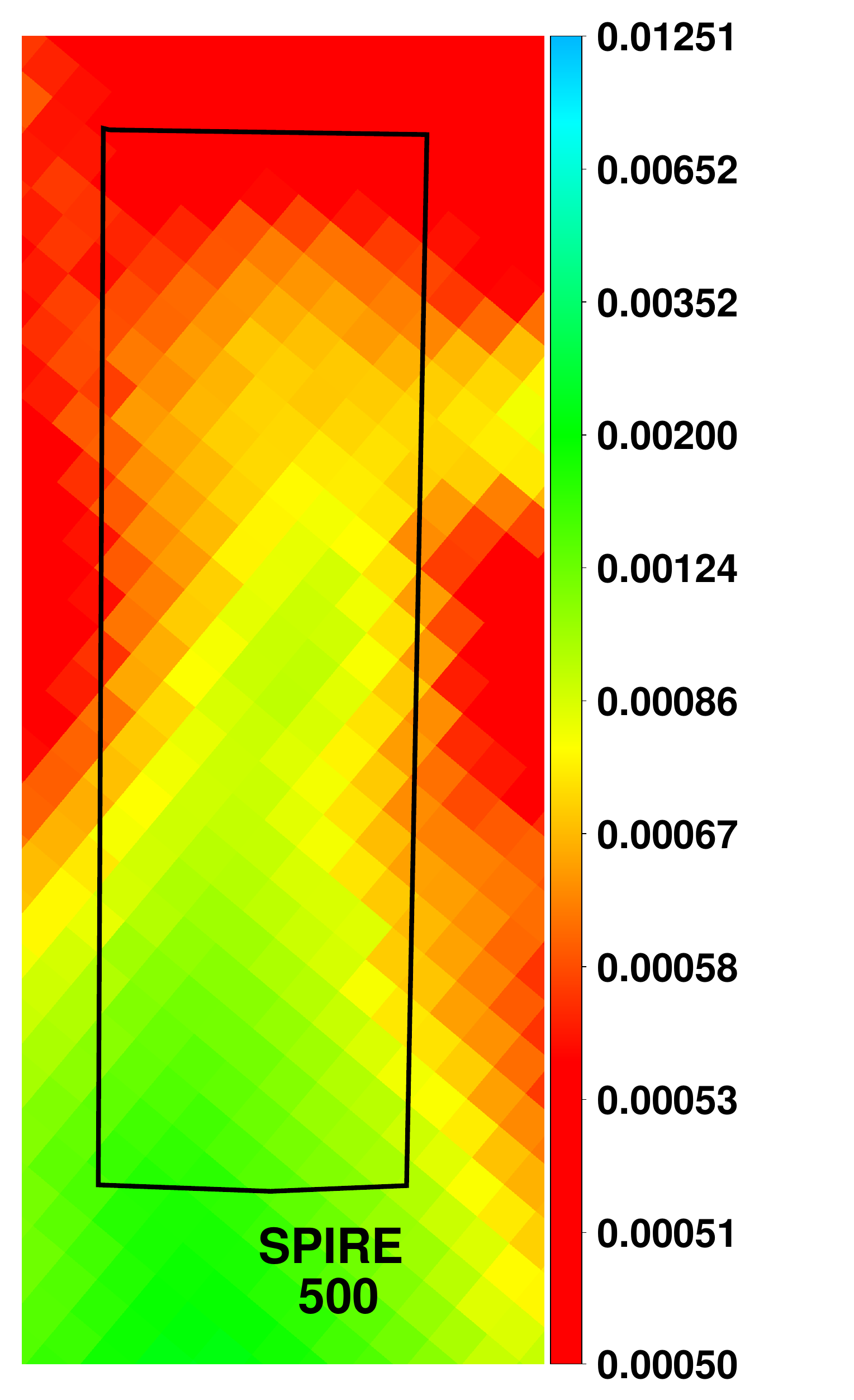}
   \includegraphics[trim=0.8cm 0.05cm 0.8cm 0.45cm,width=2.55cm]{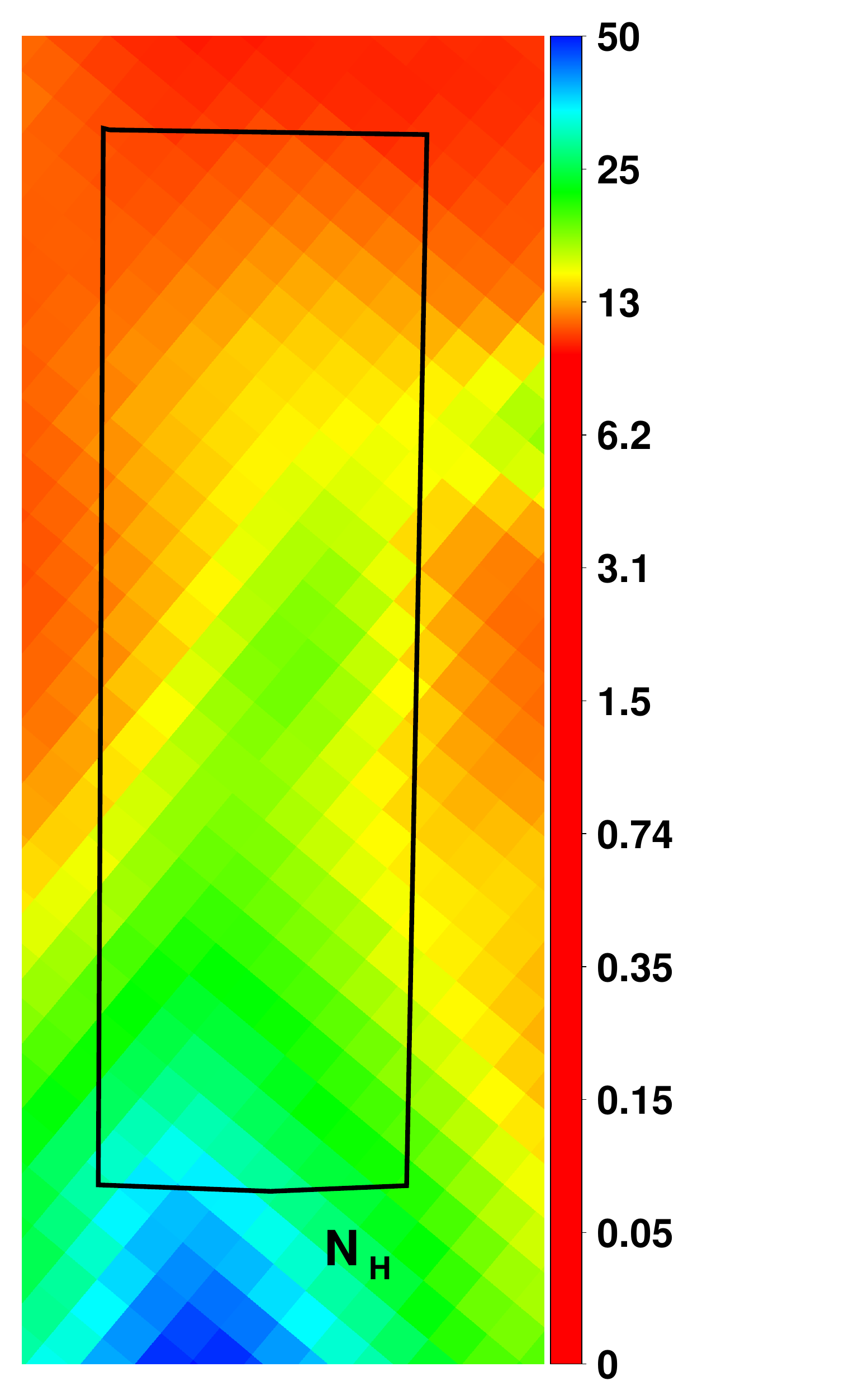}
  \includegraphics[trim=0.8cm 0.05cm 0.8cm 0.45cm,width=2.55cm]{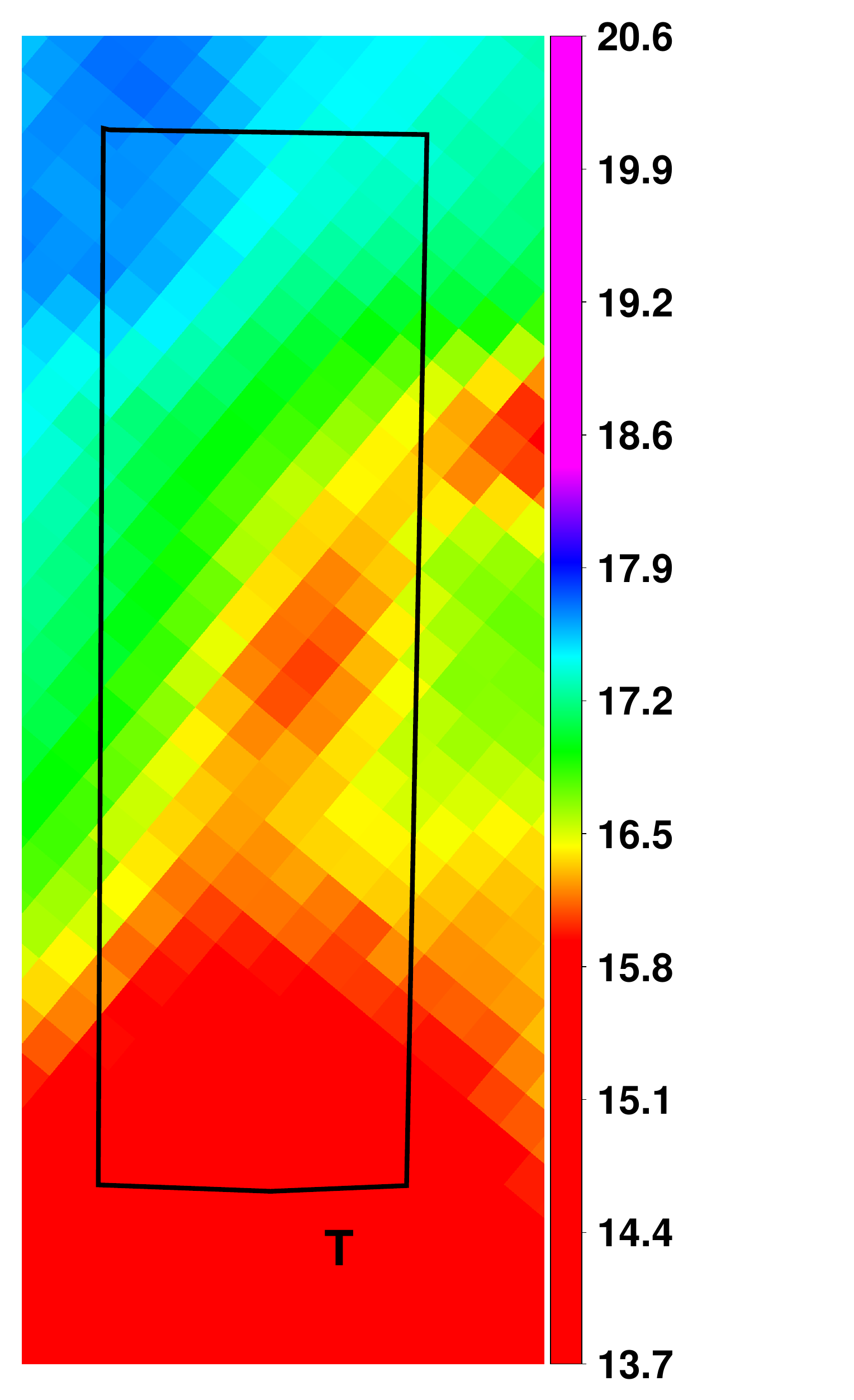}
    \caption{Filament number 4. The panels and the units are as in Figure~\ref{fil1_on}.}
     \label{fil4}
\end{figure*}

\begin{figure*}
\centering
  \includegraphics[trim=0.8cm 0.05cm 0.8cm 0.2cm,width=2.55cm]{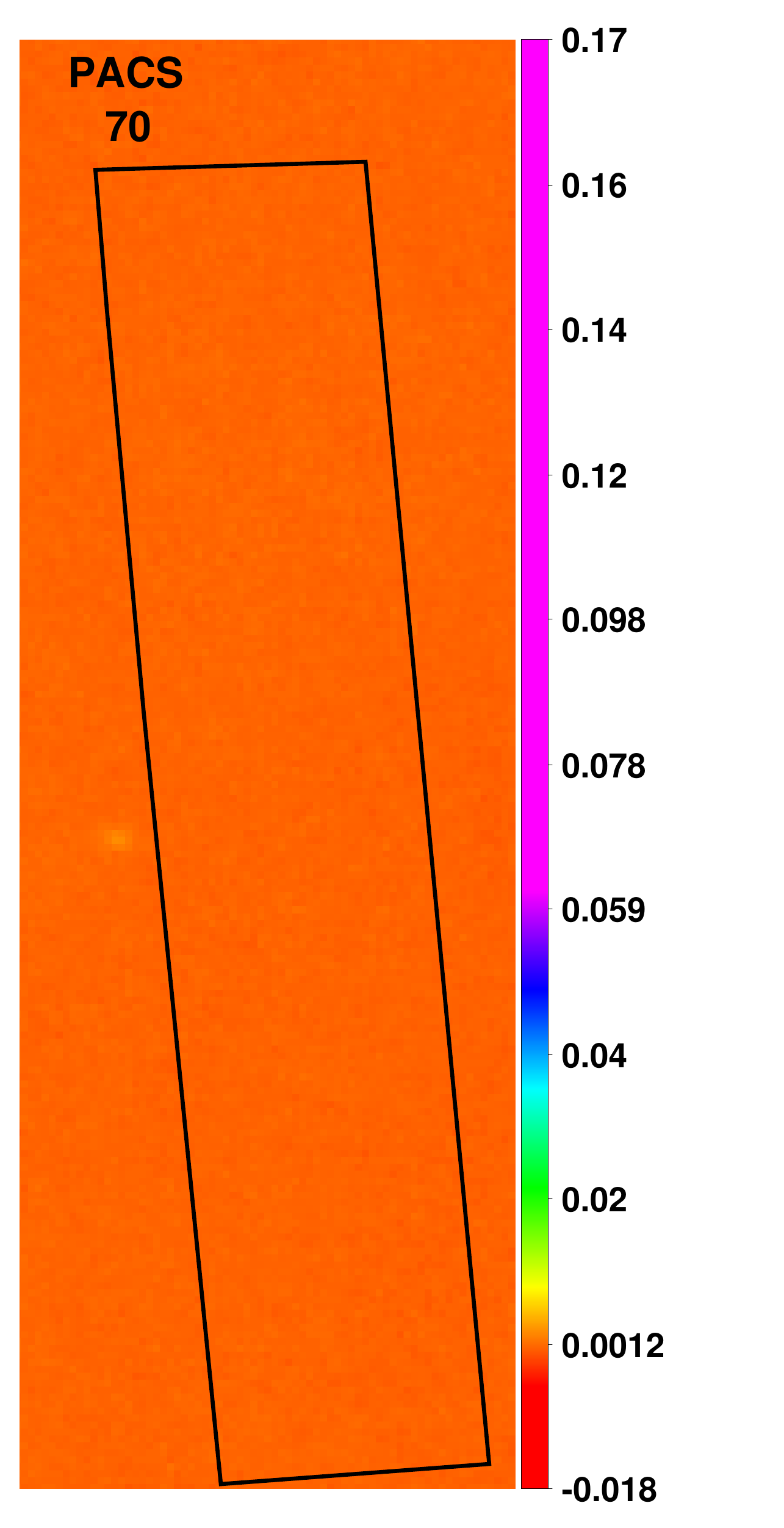}
  \includegraphics[trim=0.8cm 0.05cm 0.8cm 0.45cm,width=2.55cm]{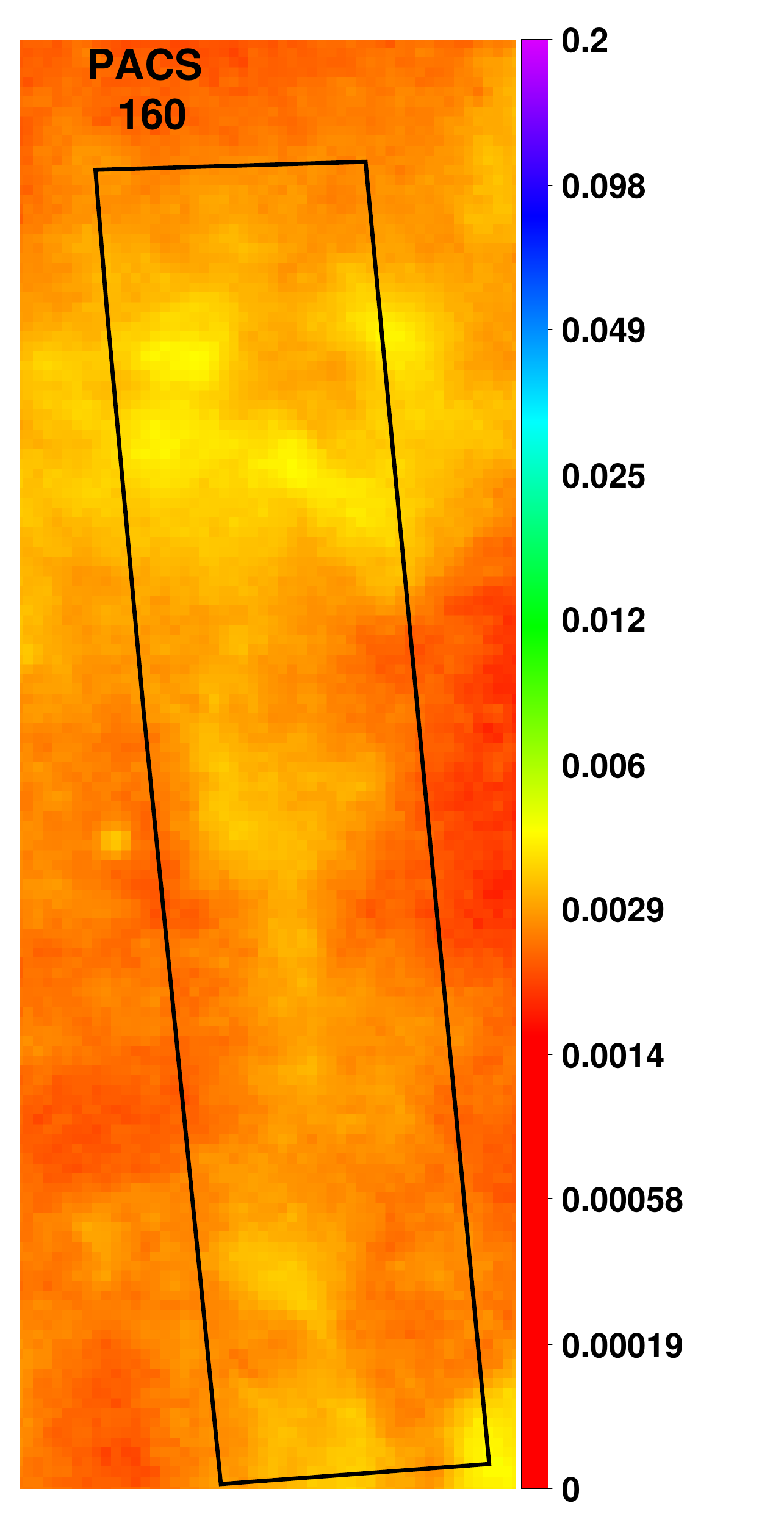}
  \includegraphics[trim=0.8cm 0.05cm 0.8cm 0.45cm,width=2.55cm]{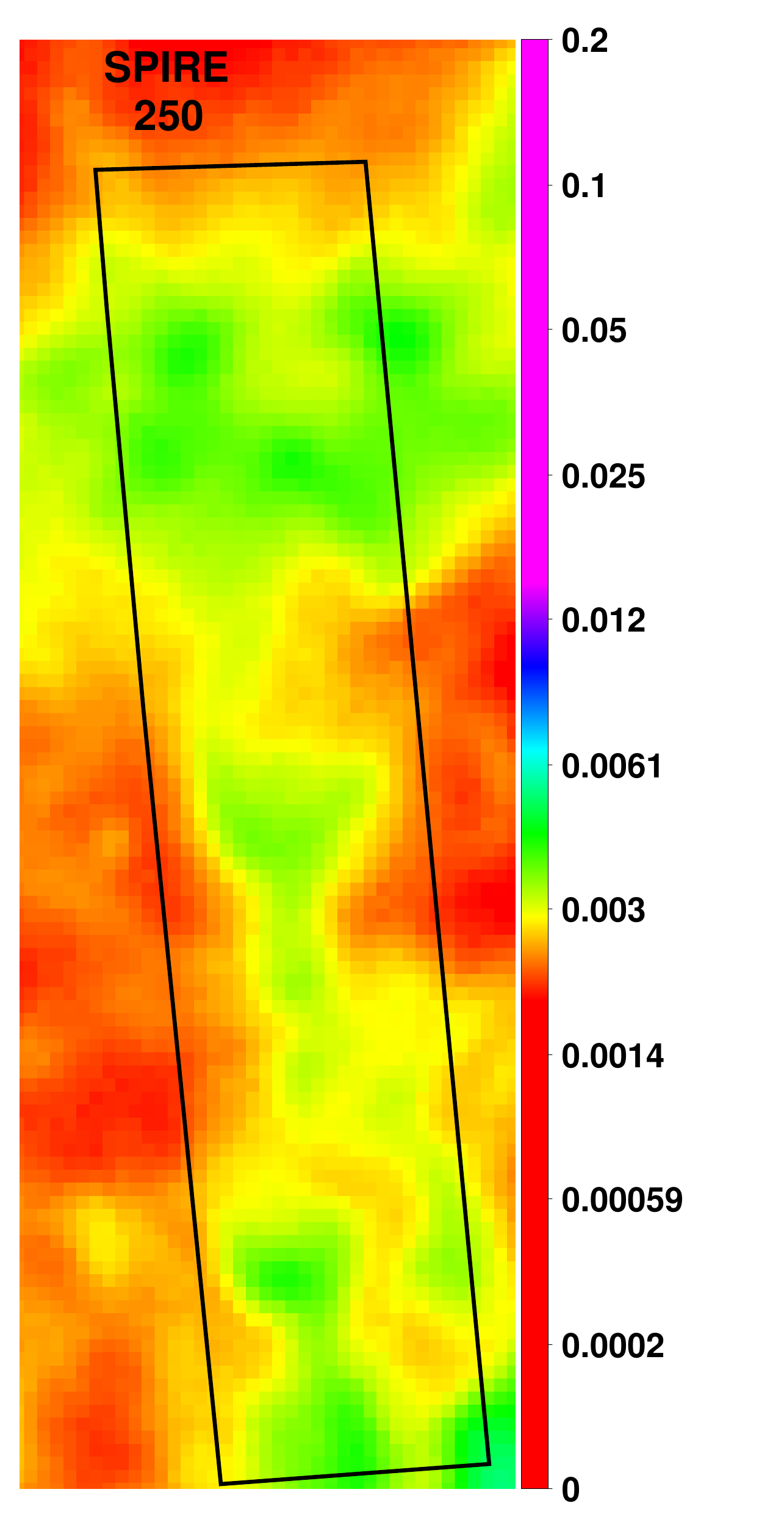}
  \includegraphics[trim=0.8cm 0.05cm 0.8cm 0.45cm,width=2.55cm]{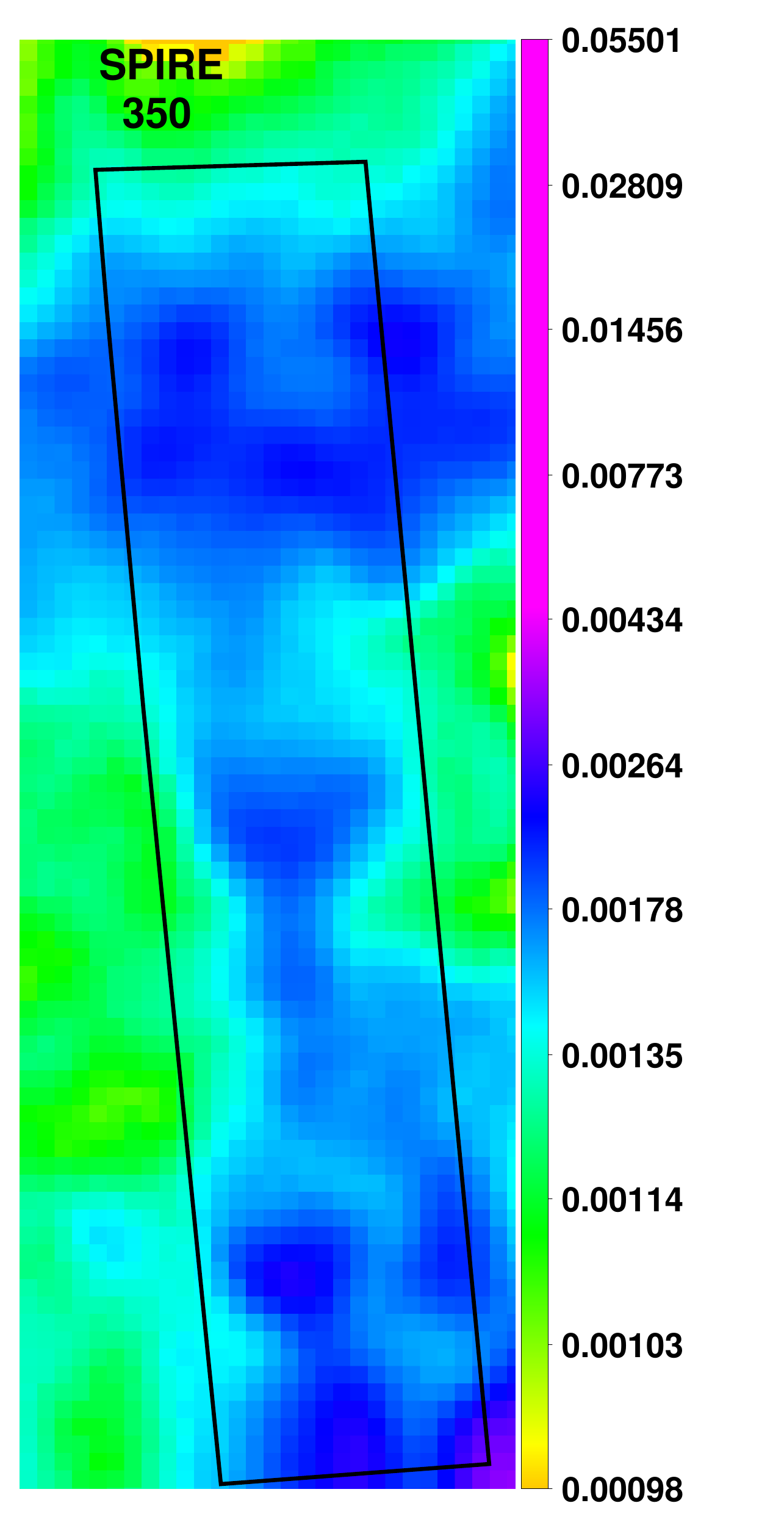}
  \includegraphics[trim=0.8cm 0.05cm 0.8cm 0.45cm,width=2.55cm]{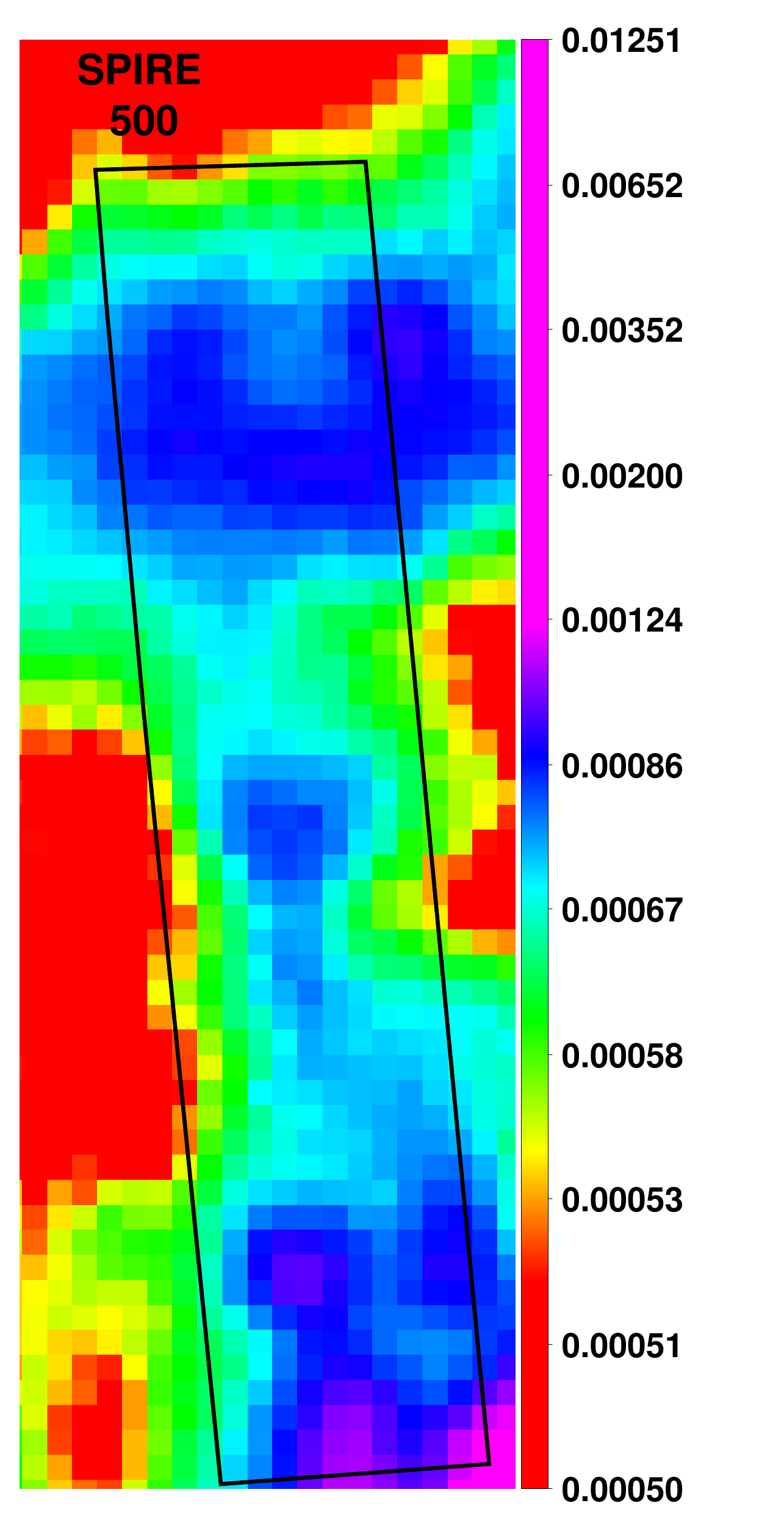}
   \includegraphics[trim=0.8cm 0.05cm 0.8cm 0.45cm,width=2.55cm]{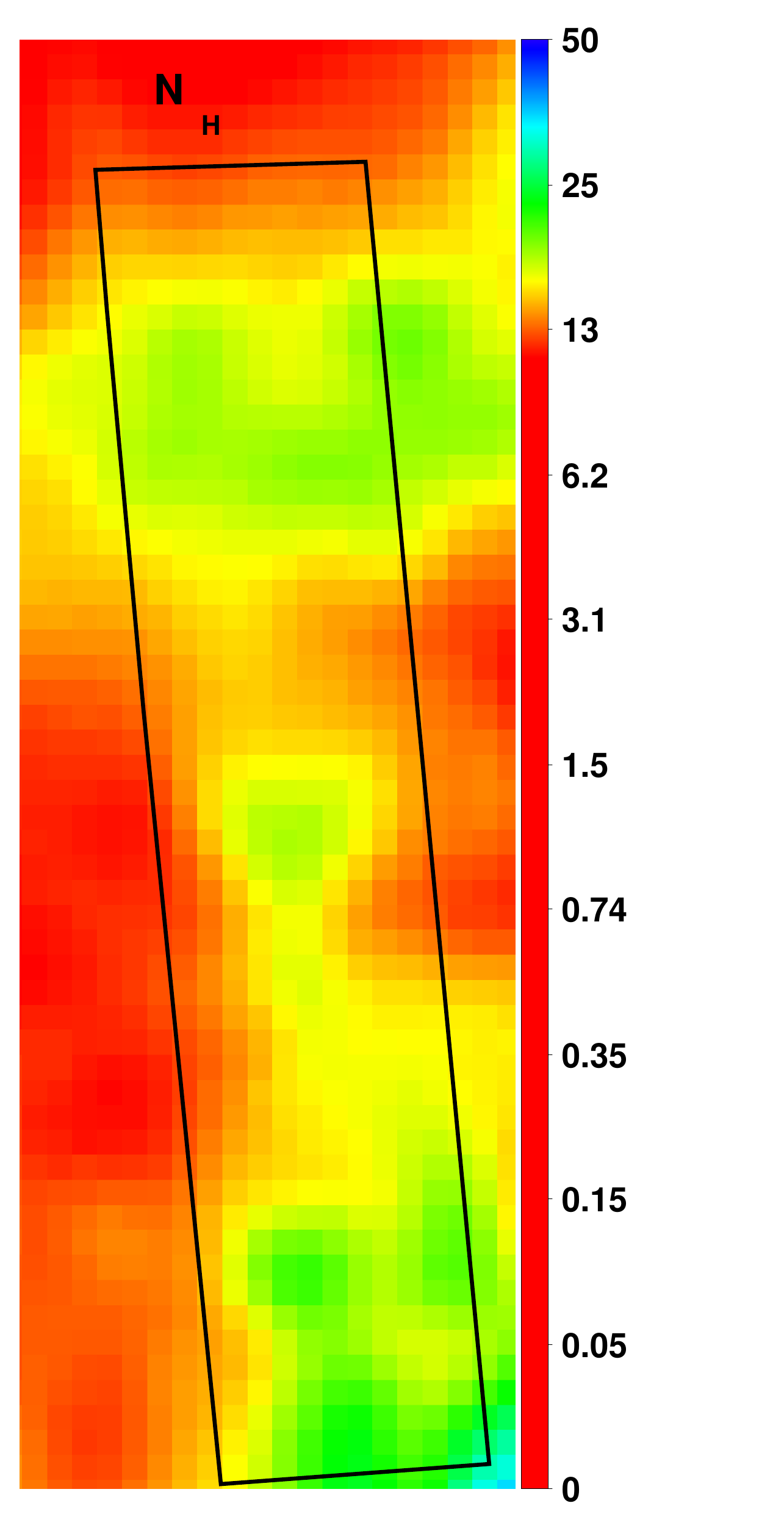}
  \includegraphics[trim=0.8cm 0.05cm 0.8cm 0.45cm,width=2.55cm]{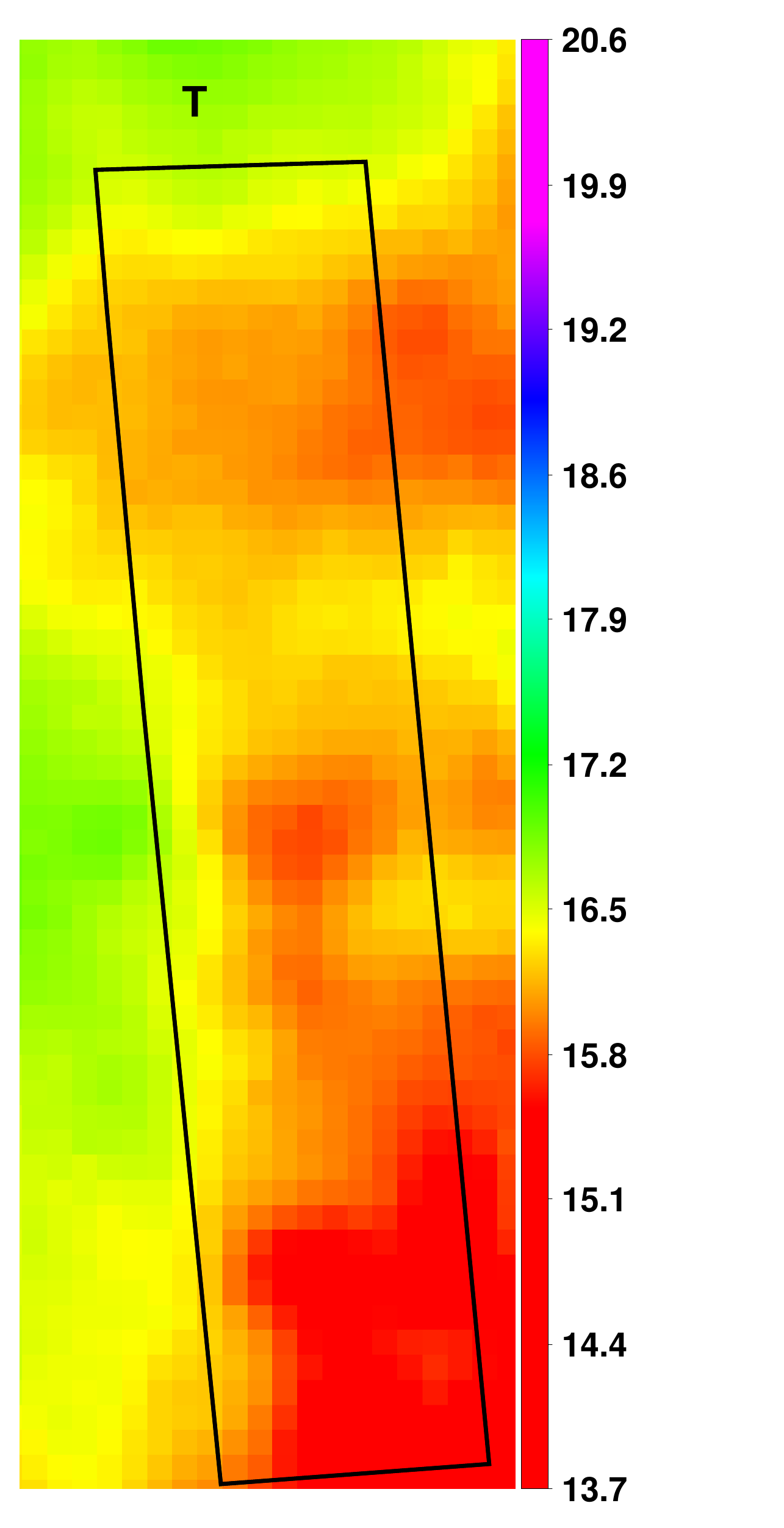}
    \caption{Filament number 5. The panels and the units are as in Figure~\ref{fil1_on}.}
     \label{fil5}
\end{figure*}

\begin{figure*}
\centering
  \includegraphics[trim=0.8cm 0.05cm 0.8cm 0.2cm,width=2.55cm]{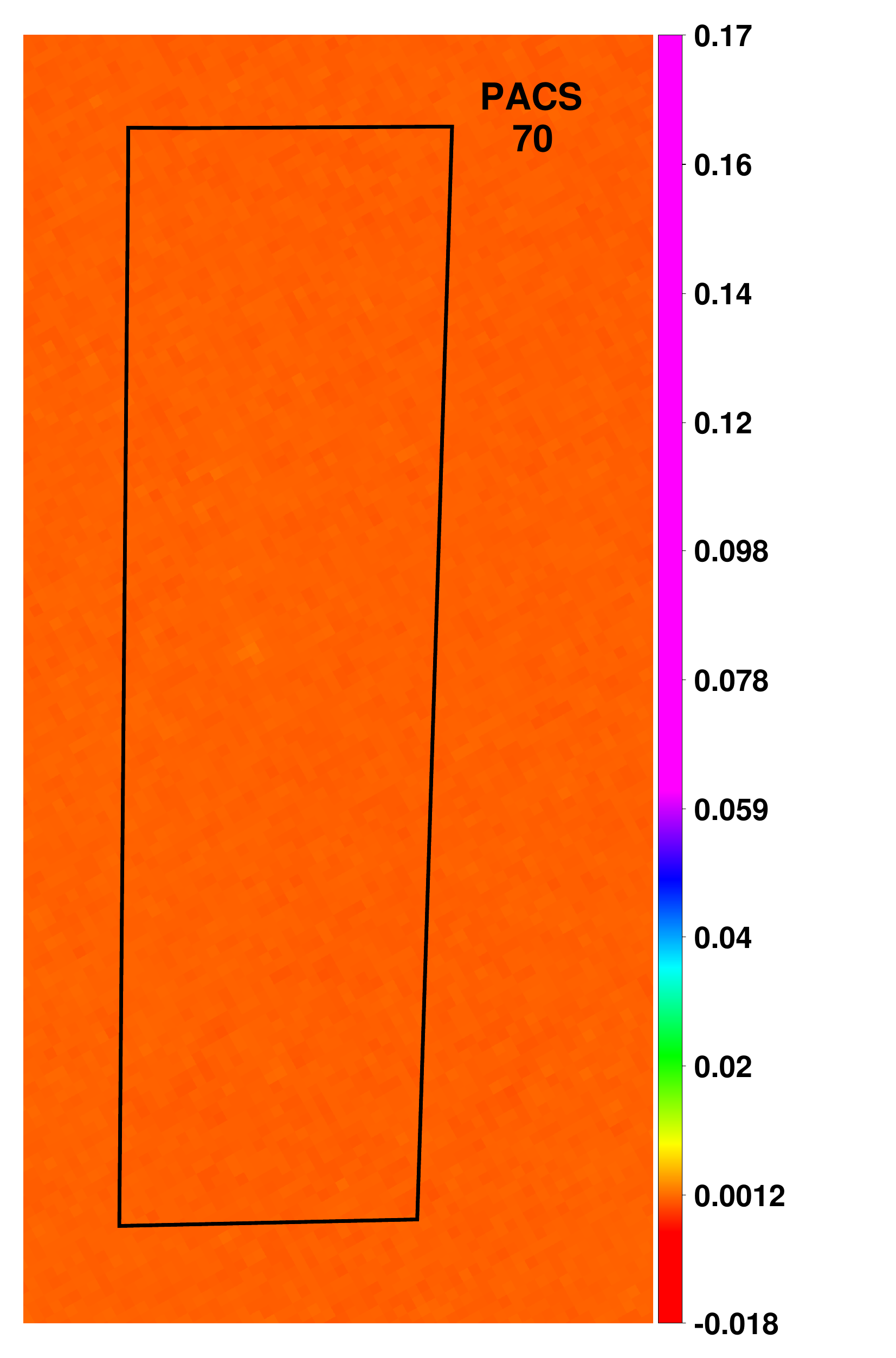}
  \includegraphics[trim=0.8cm 0.05cm 0.8cm 0.45cm,width=2.55cm]{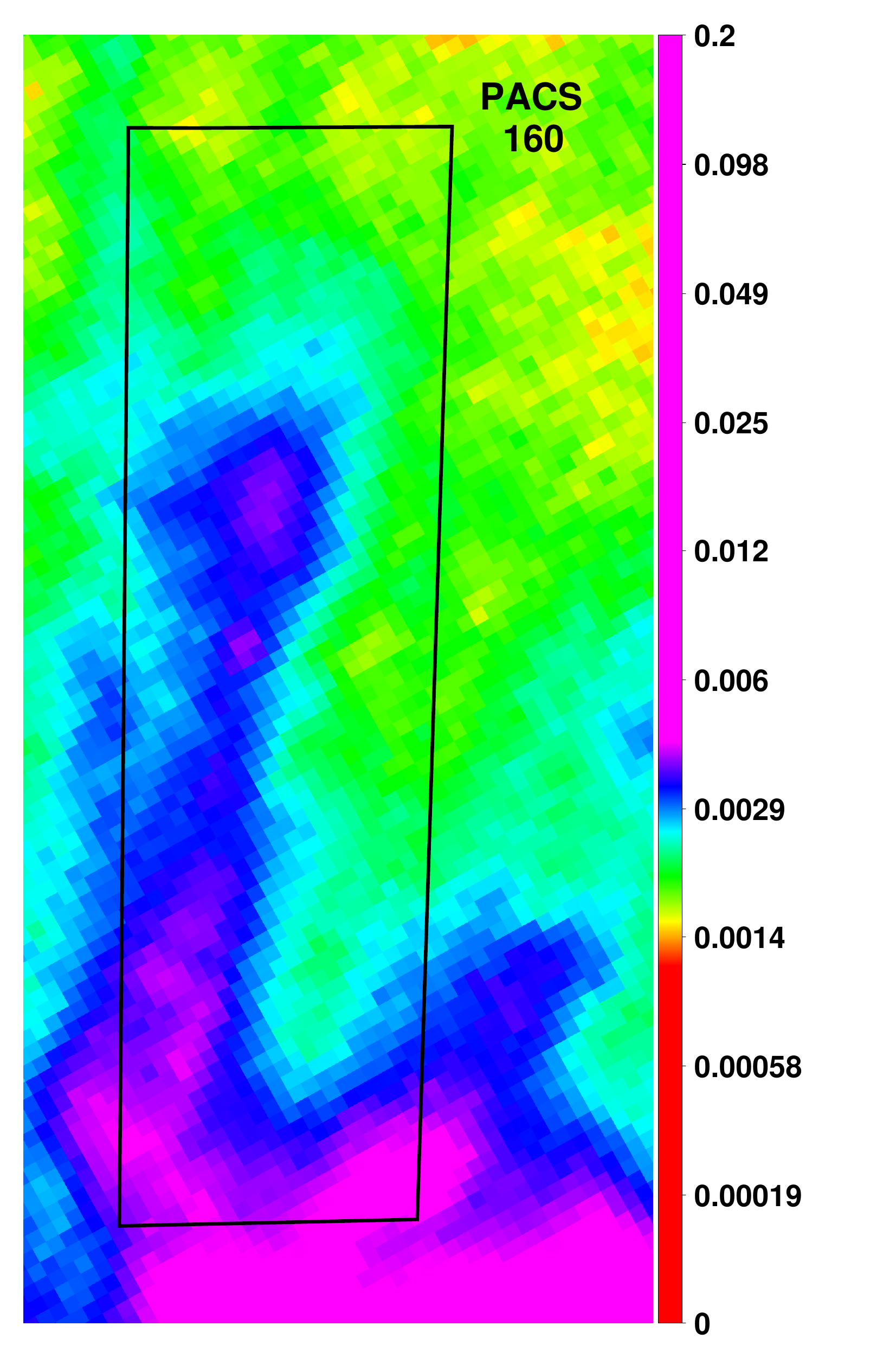}
  \includegraphics[trim=0.8cm 0.05cm 0.8cm 0.45cm,width=2.55cm]{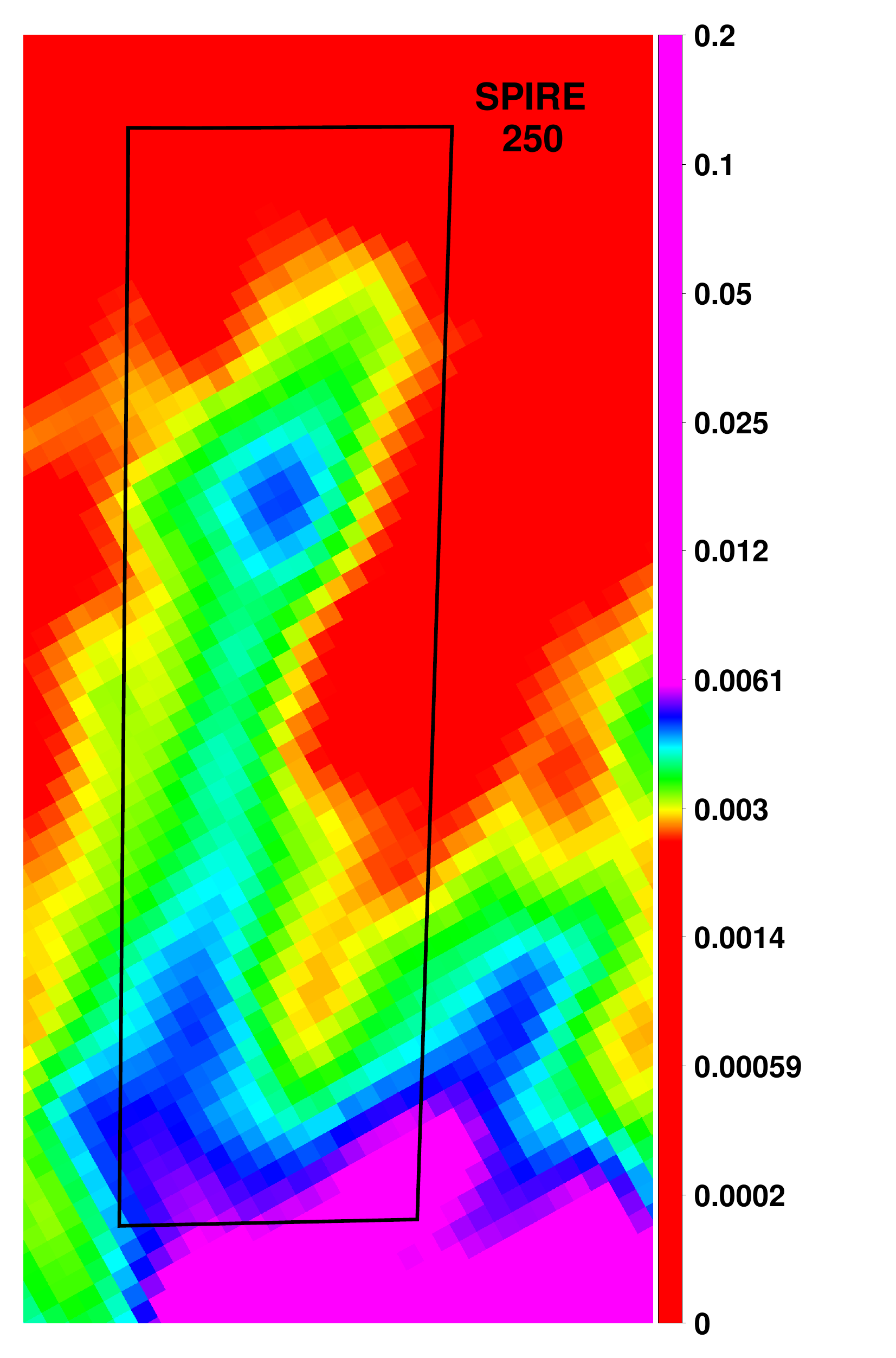}
  \includegraphics[trim=0.8cm 0.05cm 0.8cm 0.45cm,width=2.55cm]{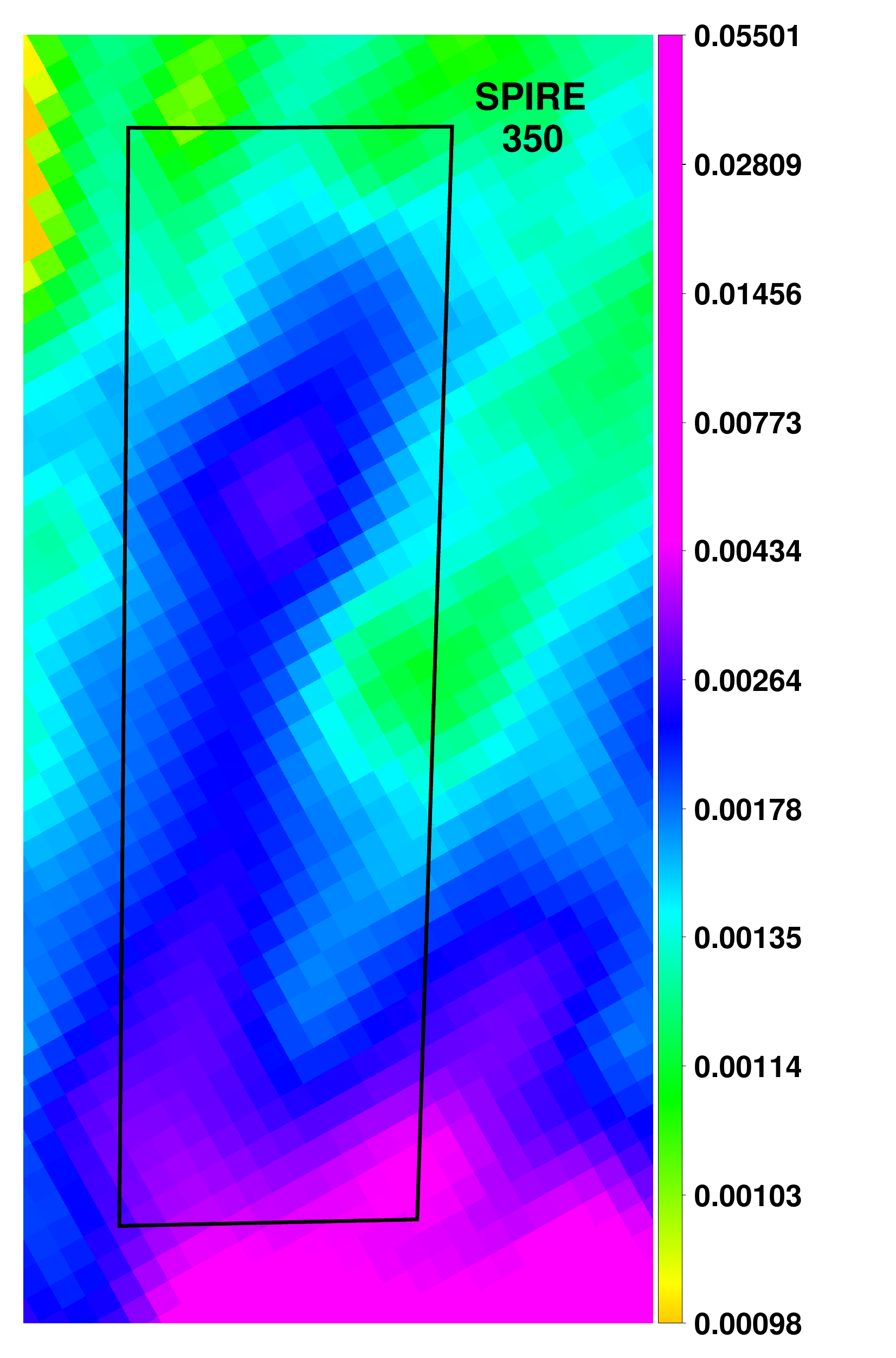}
  \includegraphics[trim=0.8cm 0.05cm 0.8cm 0.45cm,width=2.55cm]{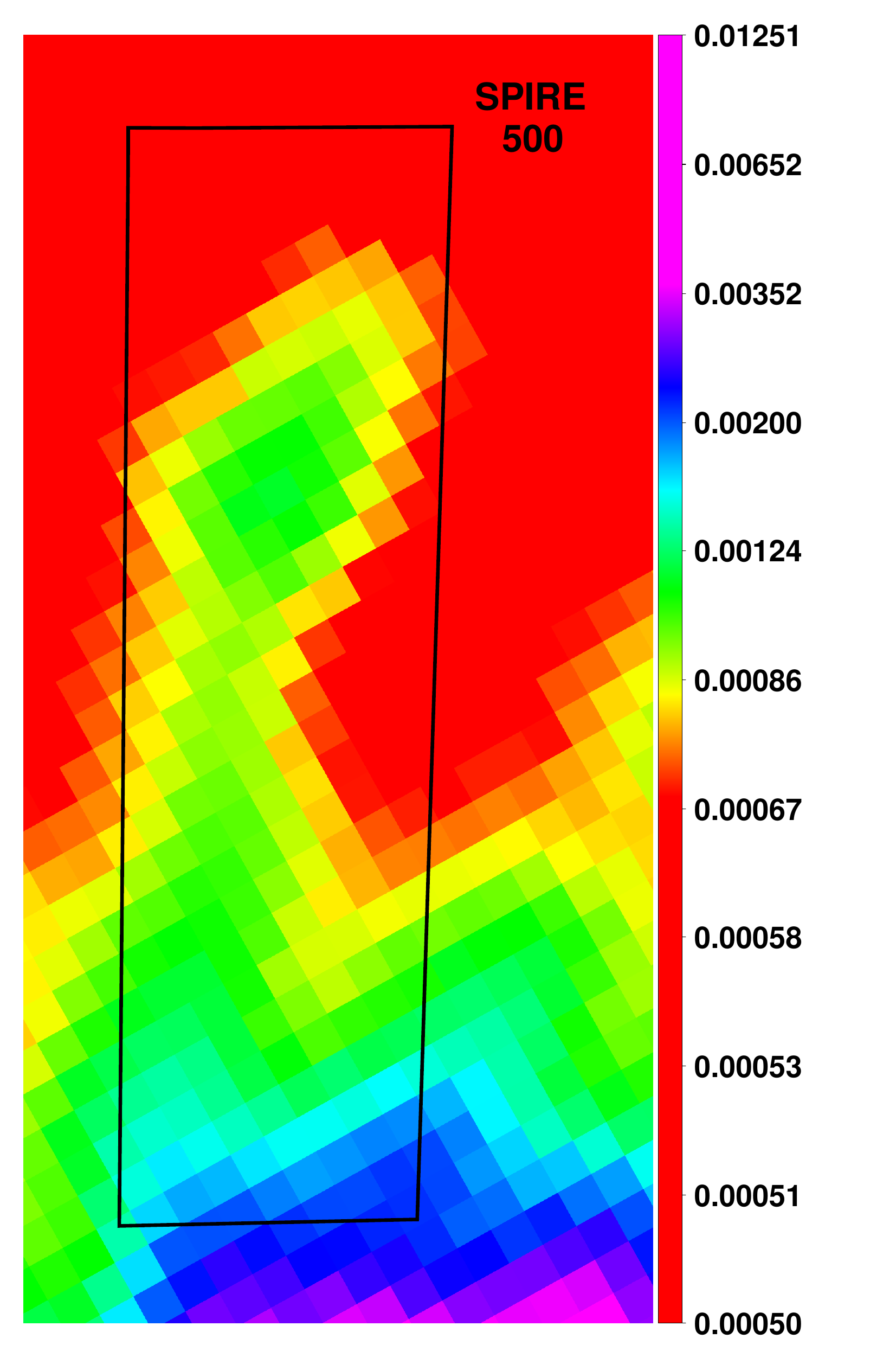}
   \includegraphics[trim=0.8cm 0.05cm 0.8cm 0.45cm,width=2.55cm]{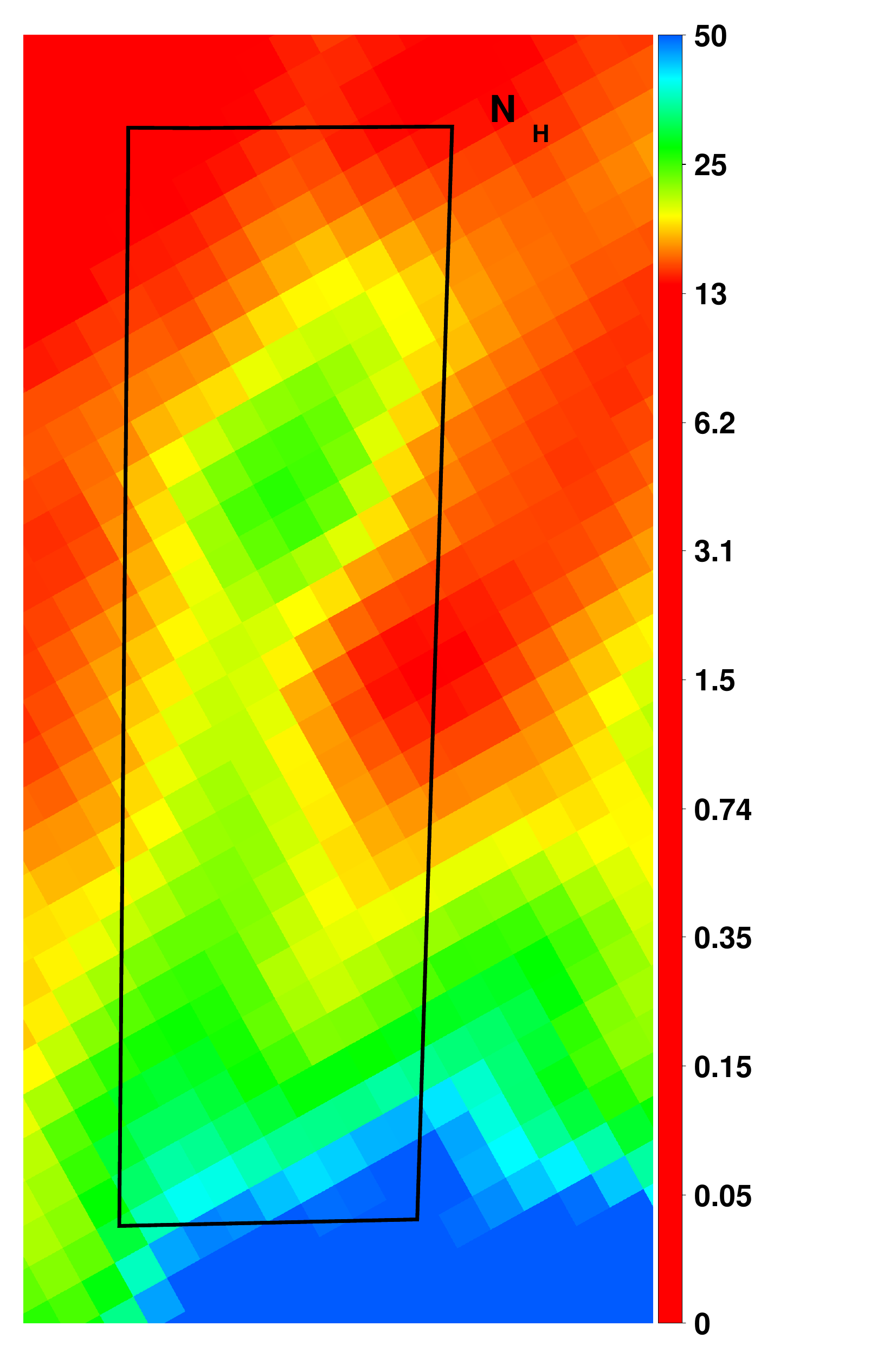}
  \includegraphics[trim=0.8cm 0.05cm 0.8cm 0.45cm,width=2.55cm]{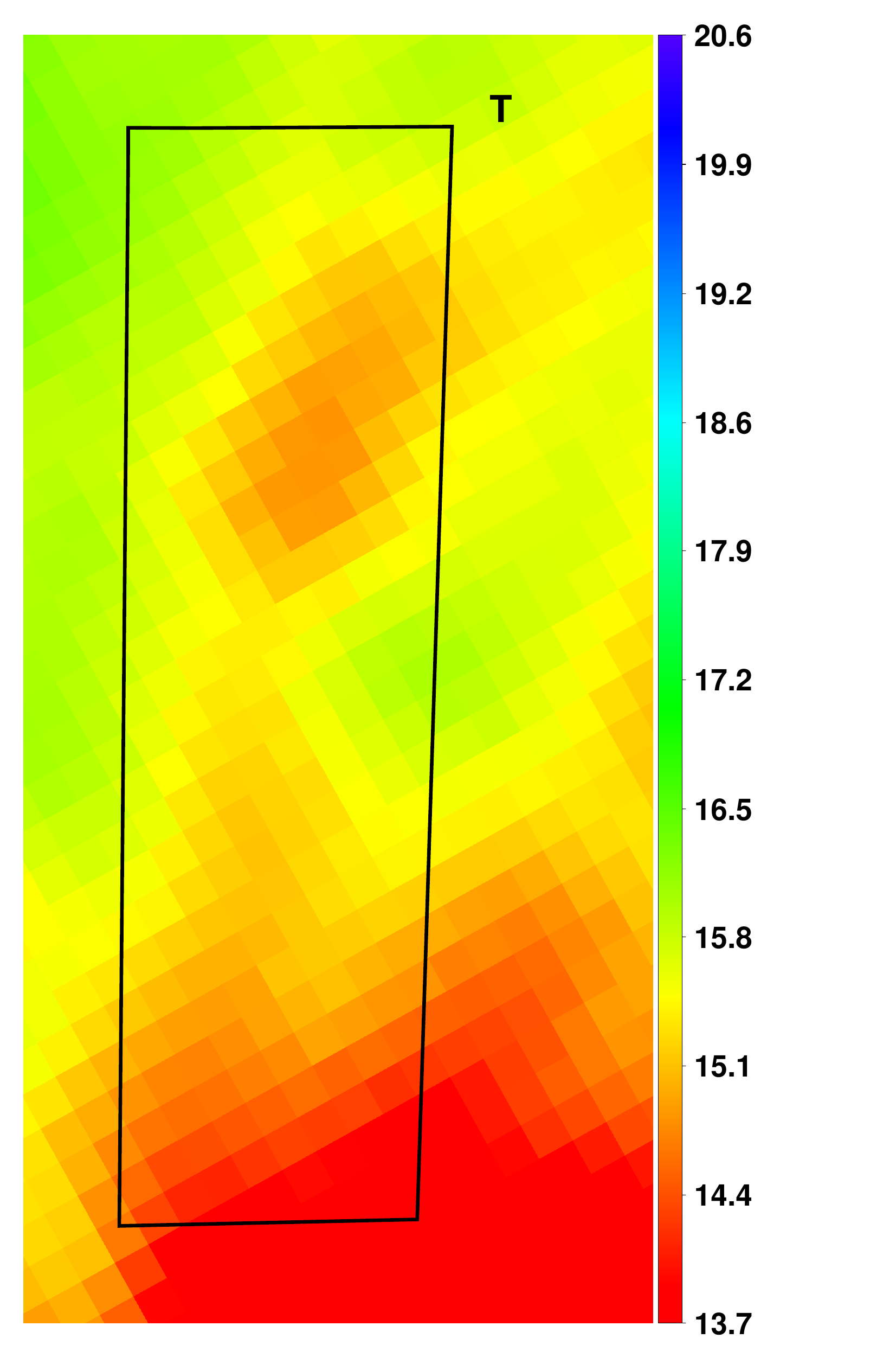}
    \caption{Filament number 6. The panels and the units are as in Figure~\ref{fil1_on}.}
     \label{fil6}
\end{figure*}
\begin{figure*}
\centering
  \includegraphics[trim=0.8cm 0.05cm 0.8cm 0.2cm,width=2.55cm]{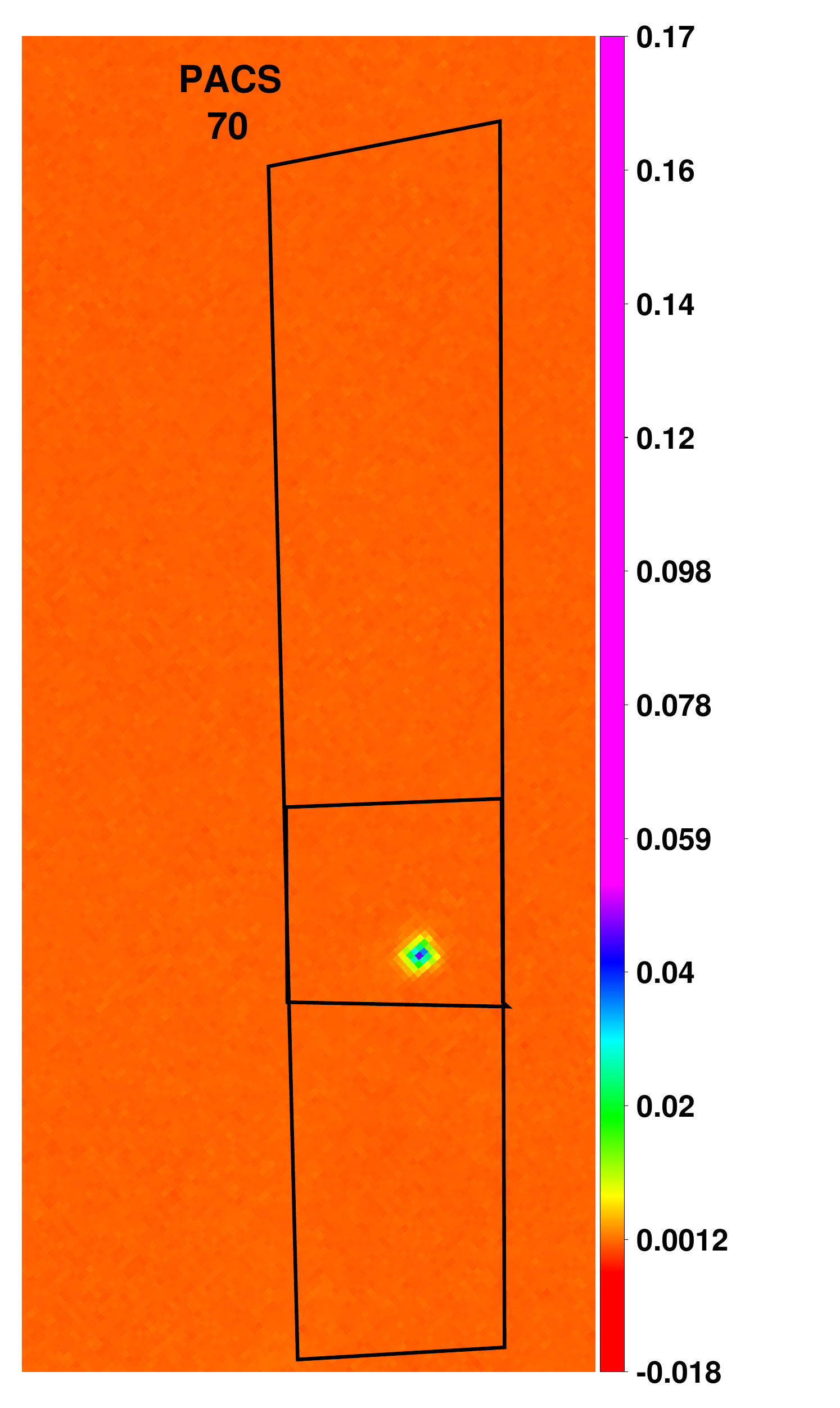}
  \includegraphics[trim=0.8cm 0.05cm 0.8cm 0.45cm,width=2.55cm]{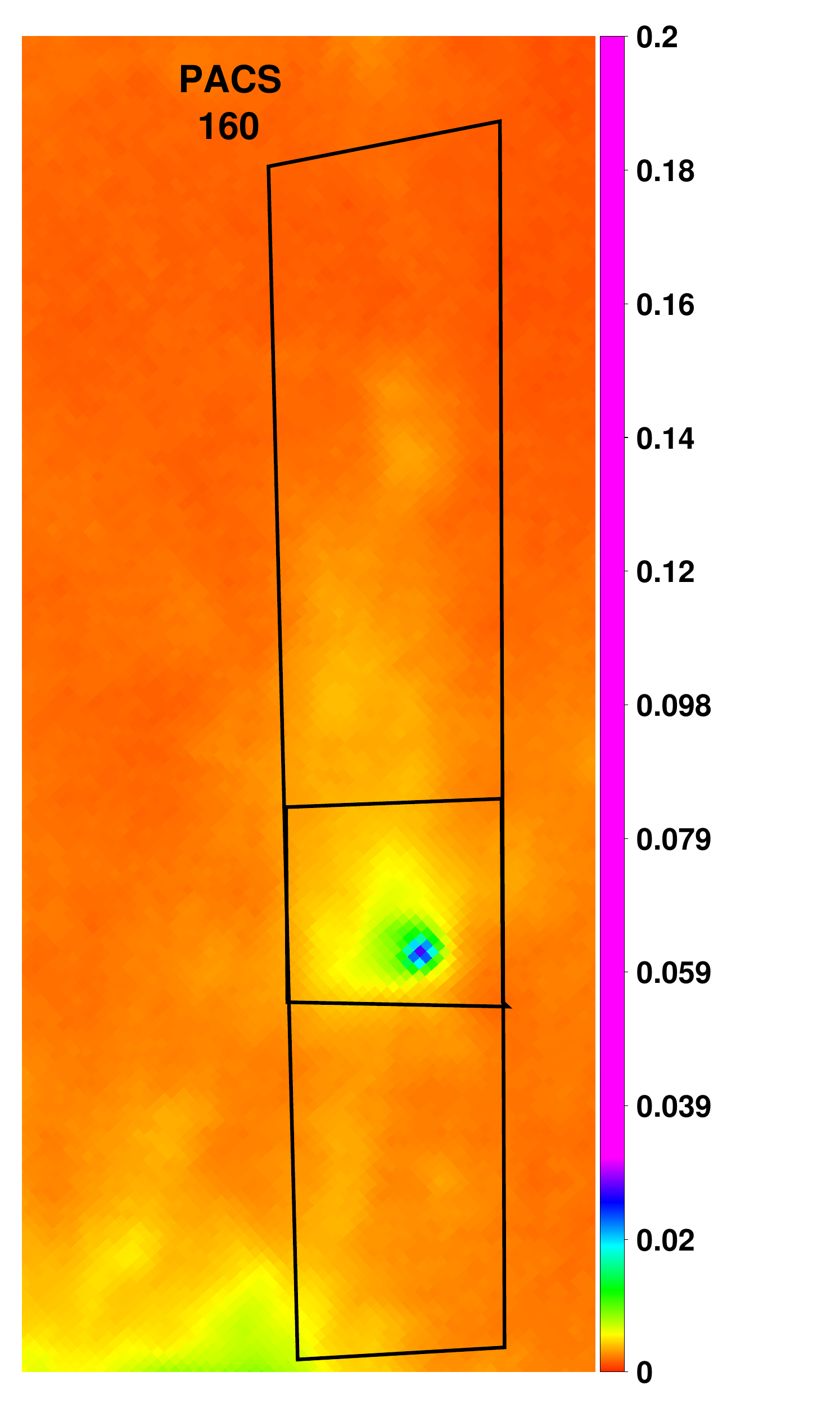}
  \includegraphics[trim=0.8cm 0.05cm 0.8cm 0.45cm,width=2.55cm]{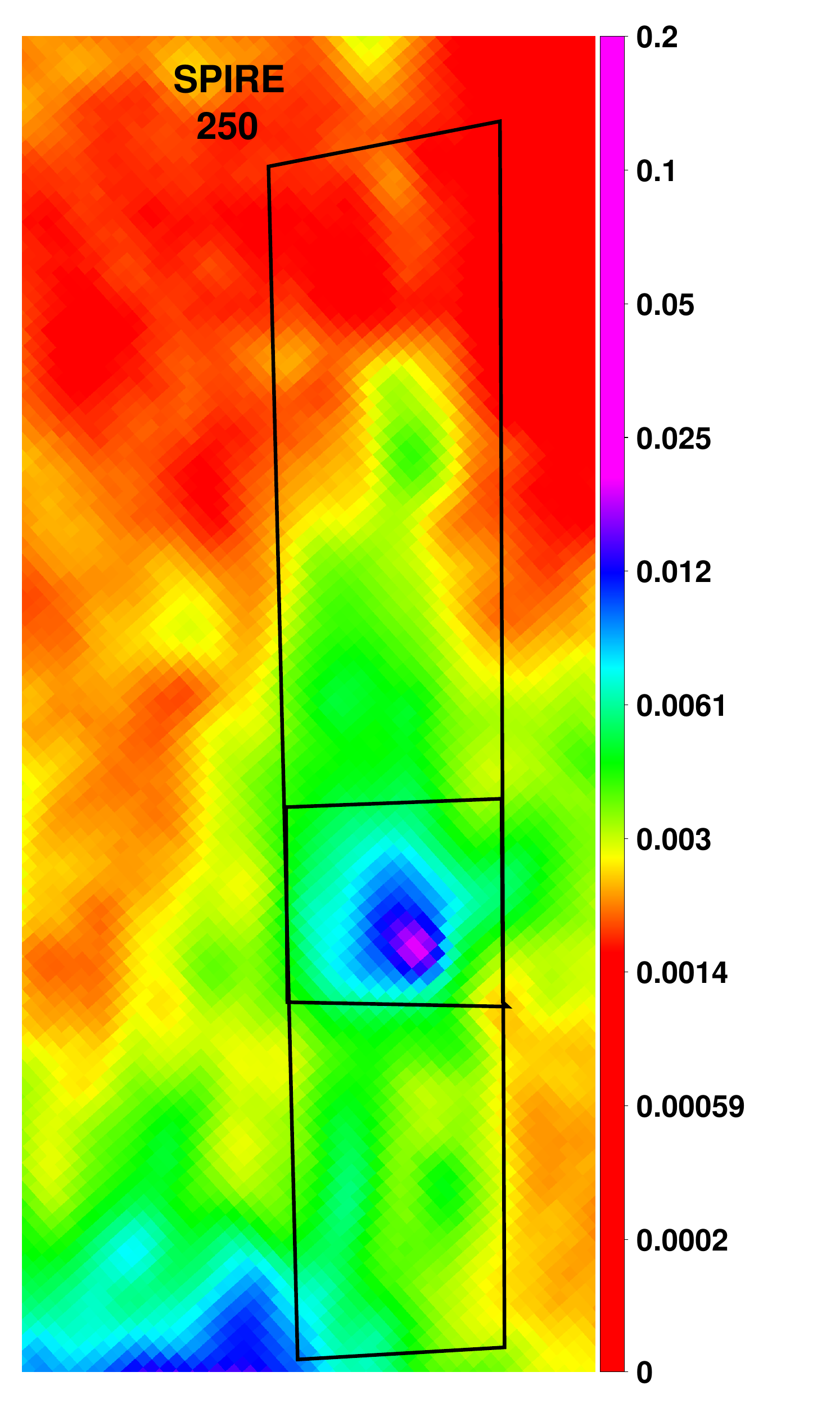}
  \includegraphics[trim=0.8cm 0.05cm 0.8cm 0.45cm,width=2.55cm]{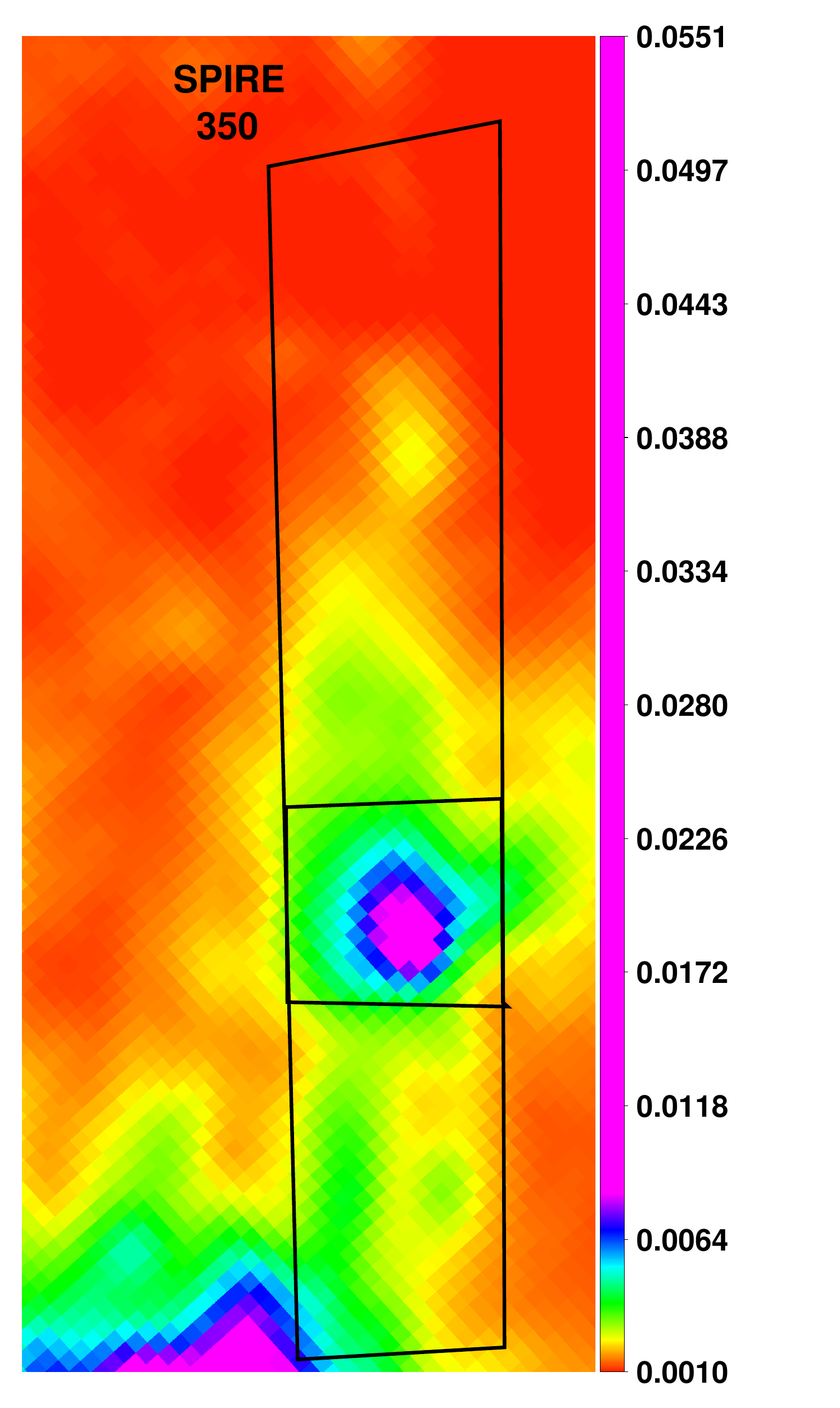}
  \includegraphics[trim=0.8cm 0.05cm 0.8cm 0.45cm,width=2.55cm]{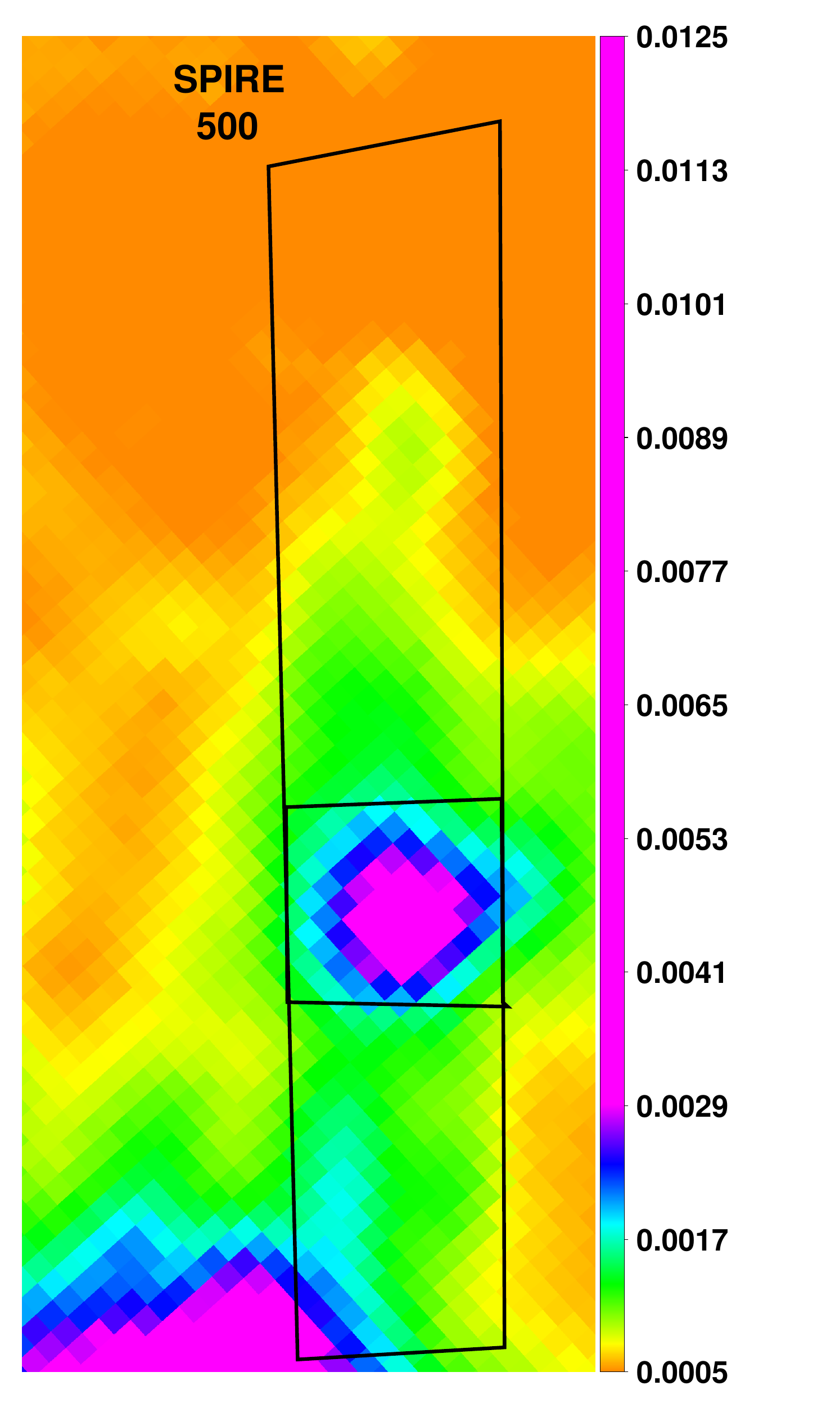}
   \includegraphics[trim=0.8cm 0.05cm 0.8cm 0.45cm,width=2.55cm]{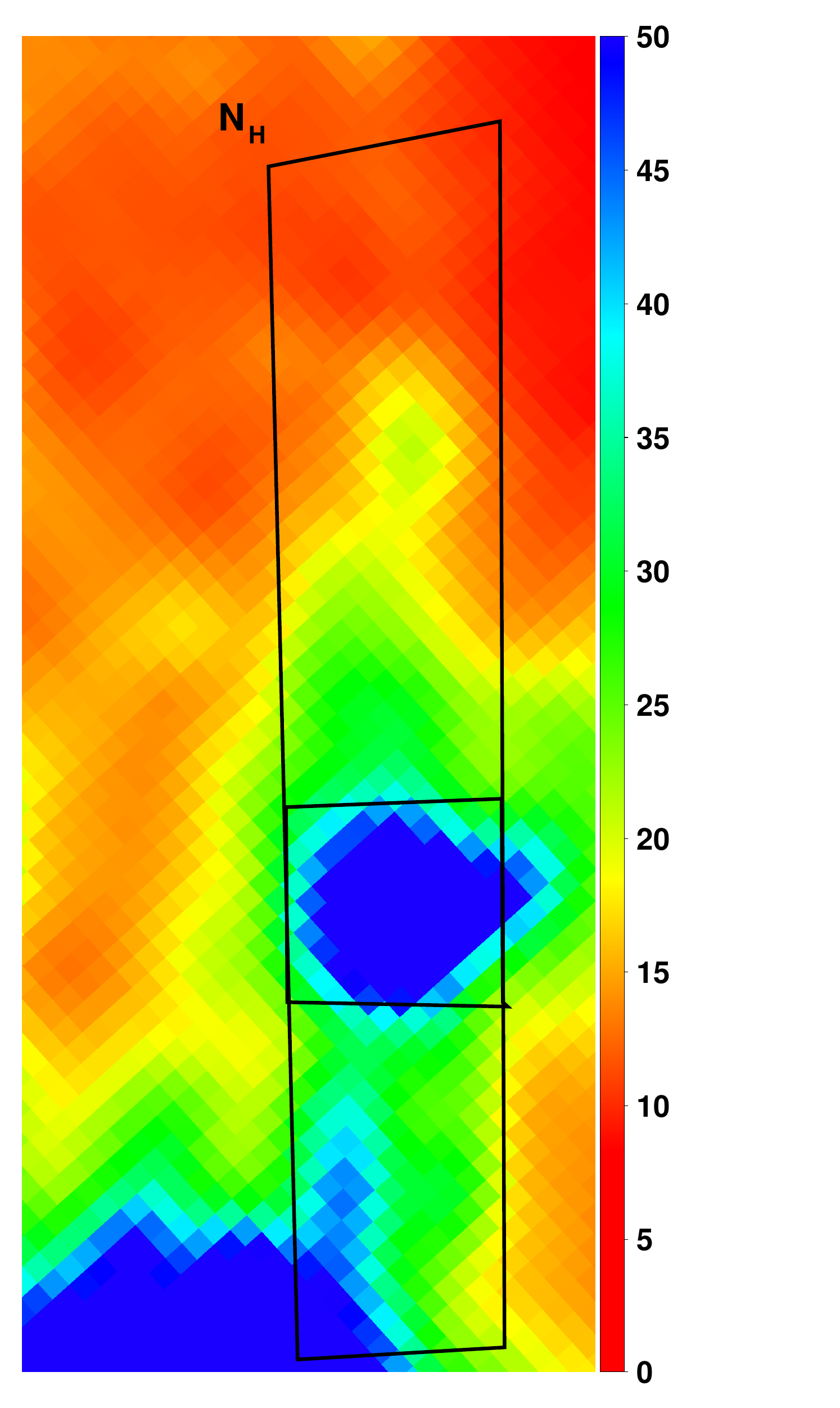}
  \includegraphics[trim=0.8cm 0.05cm 0.8cm 0.45cm,width=2.55cm]{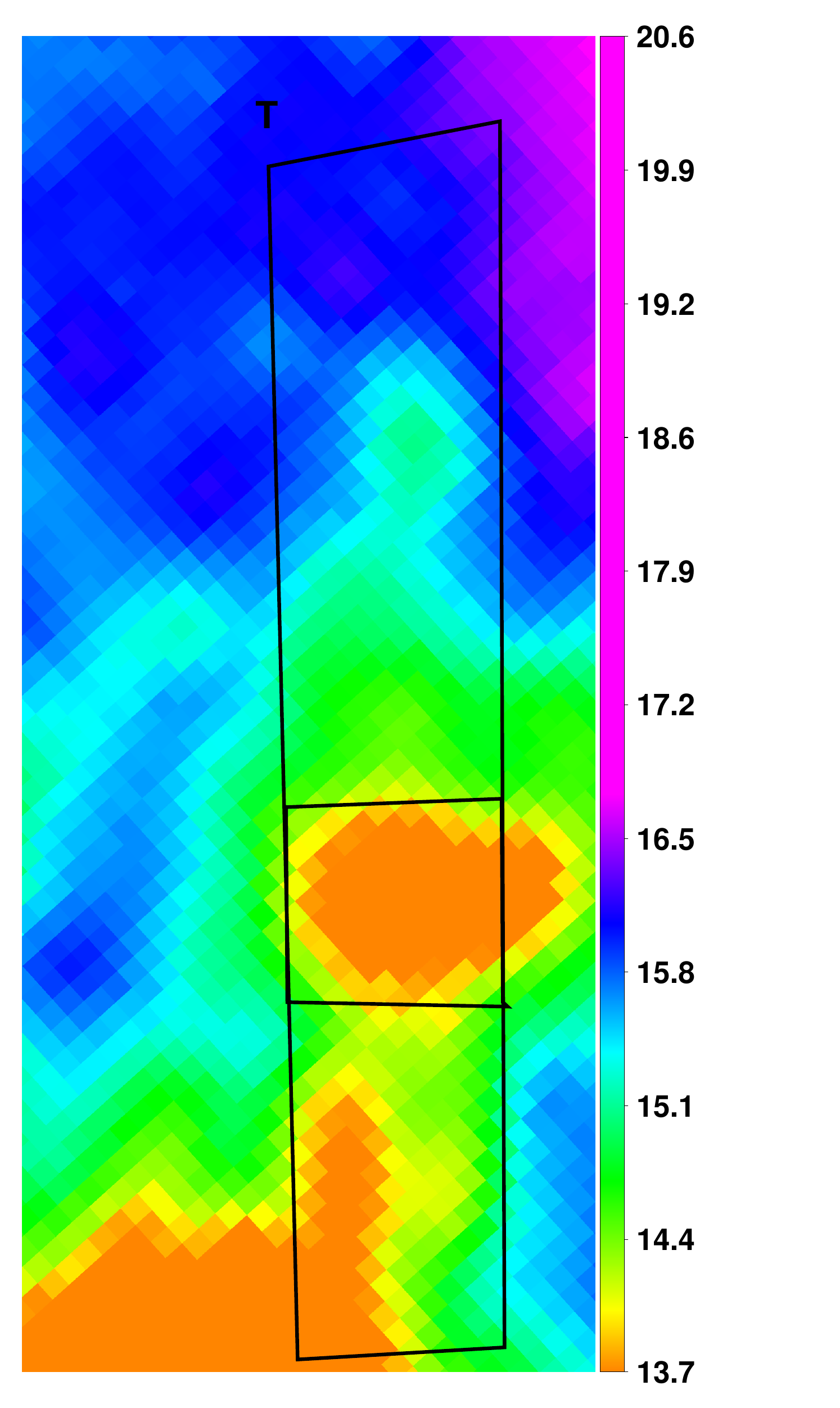}
    \caption{Filament number 7. The panels and the units are as in Figure~\ref{fil1_on}.   The smaller box identifies the 
    position of the core among the filament.}
     \label{fil7}
\end{figure*}

\begin{figure*}
\centering
  \includegraphics[trim=0.8cm 0.05cm 0.8cm 0.2cm,width=2.55cm]{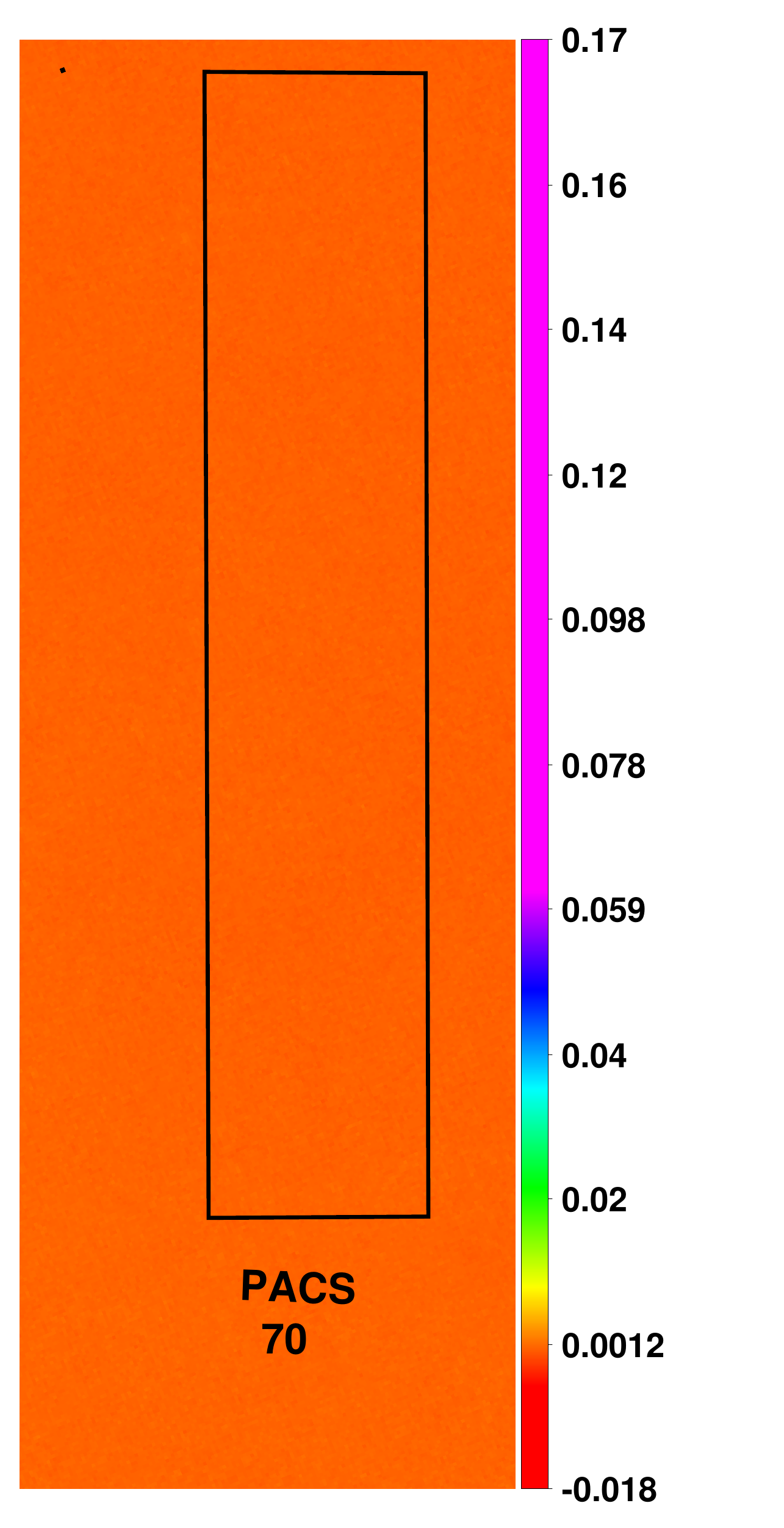}
  \includegraphics[trim=0.8cm 0.05cm 0.8cm 0.45cm,width=2.55cm]{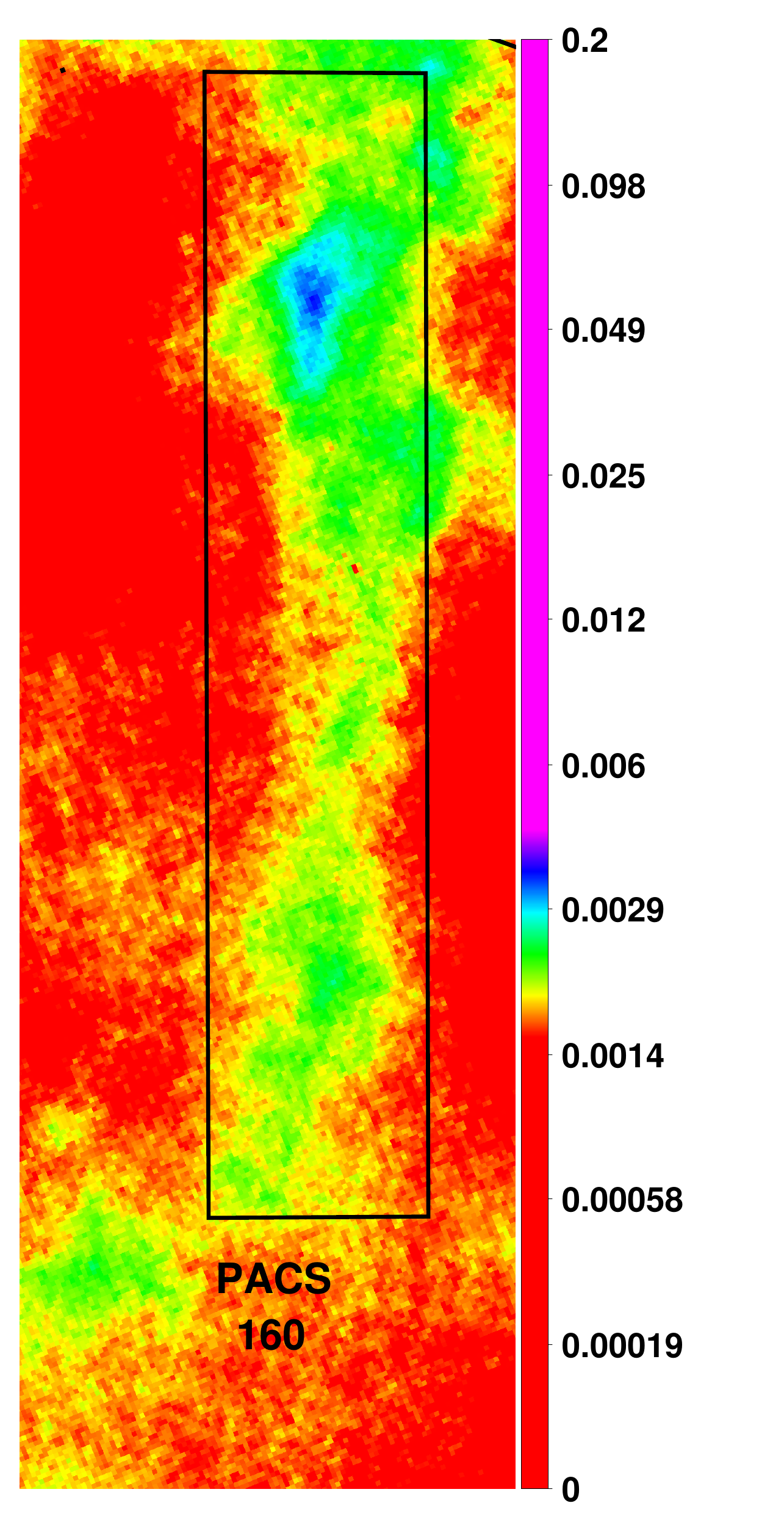}
  \includegraphics[trim=0.8cm 0.05cm 0.8cm 0.45cm,width=2.55cm]{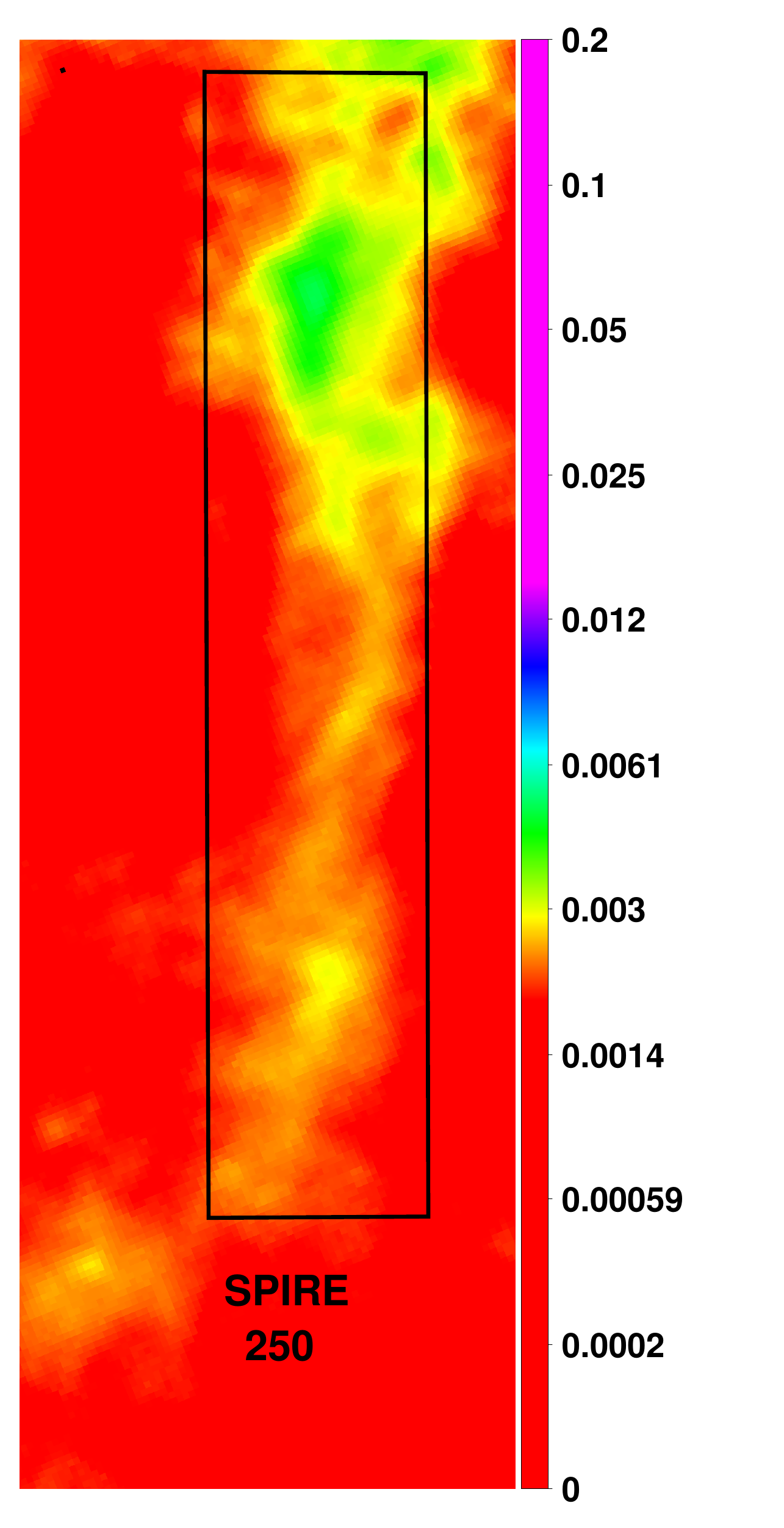}
  \includegraphics[trim=0.8cm 0.05cm 0.8cm 0.45cm,width=2.55cm]{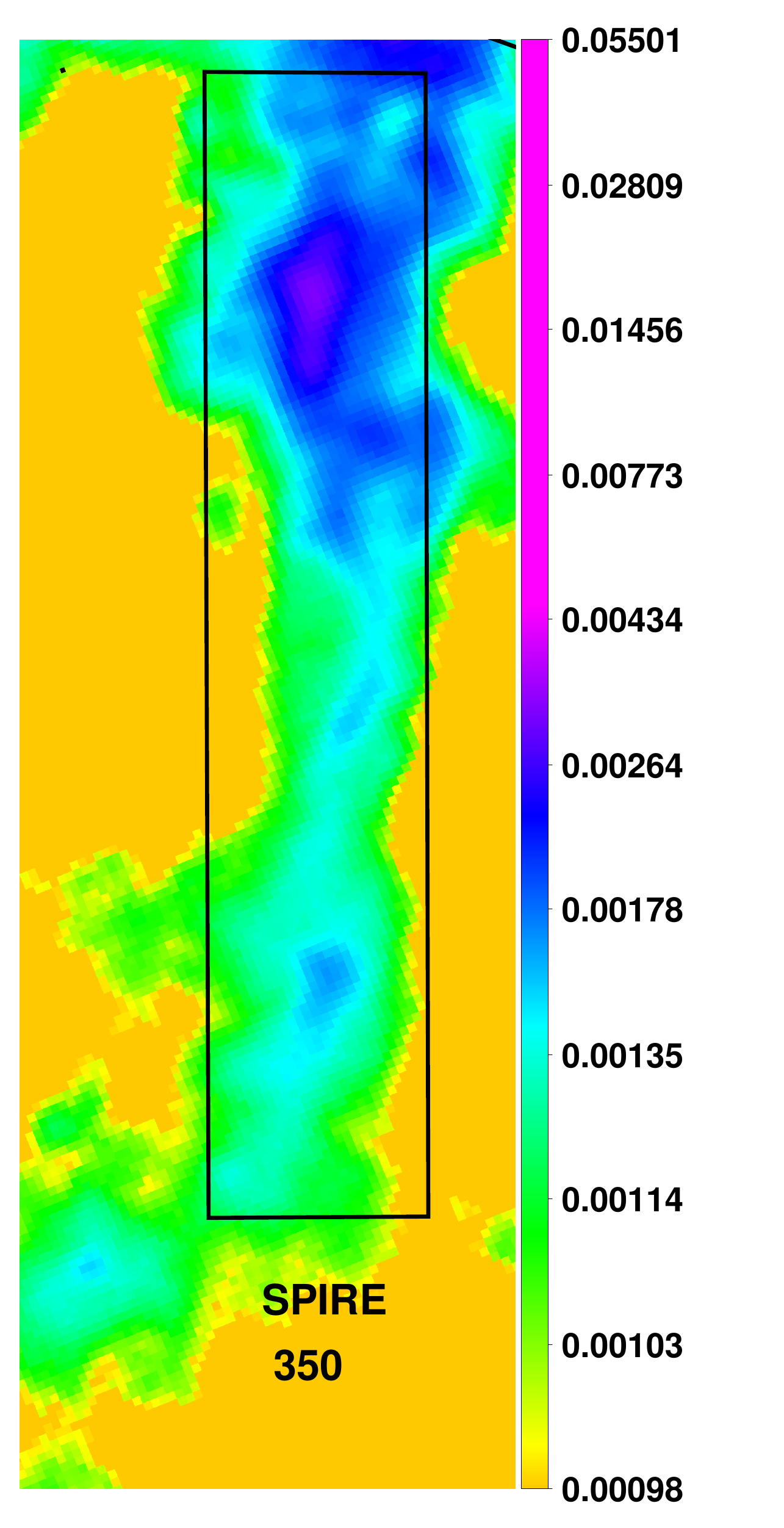}
  \includegraphics[trim=0.8cm 0.05cm 0.8cm 0.45cm,width=2.55cm]{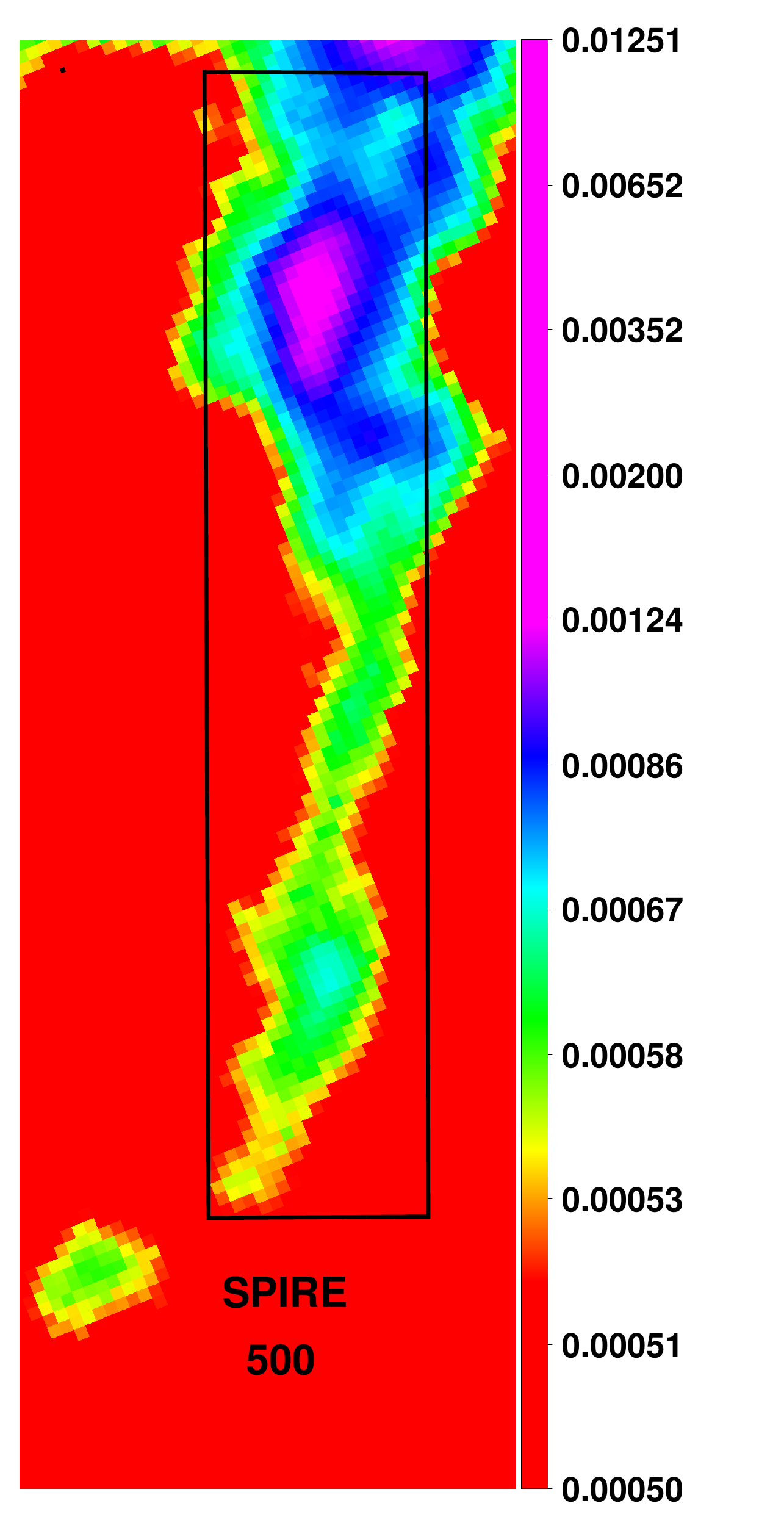}
   \includegraphics[trim=0.8cm 0.05cm 0.8cm 0.45cm,width=2.55cm]{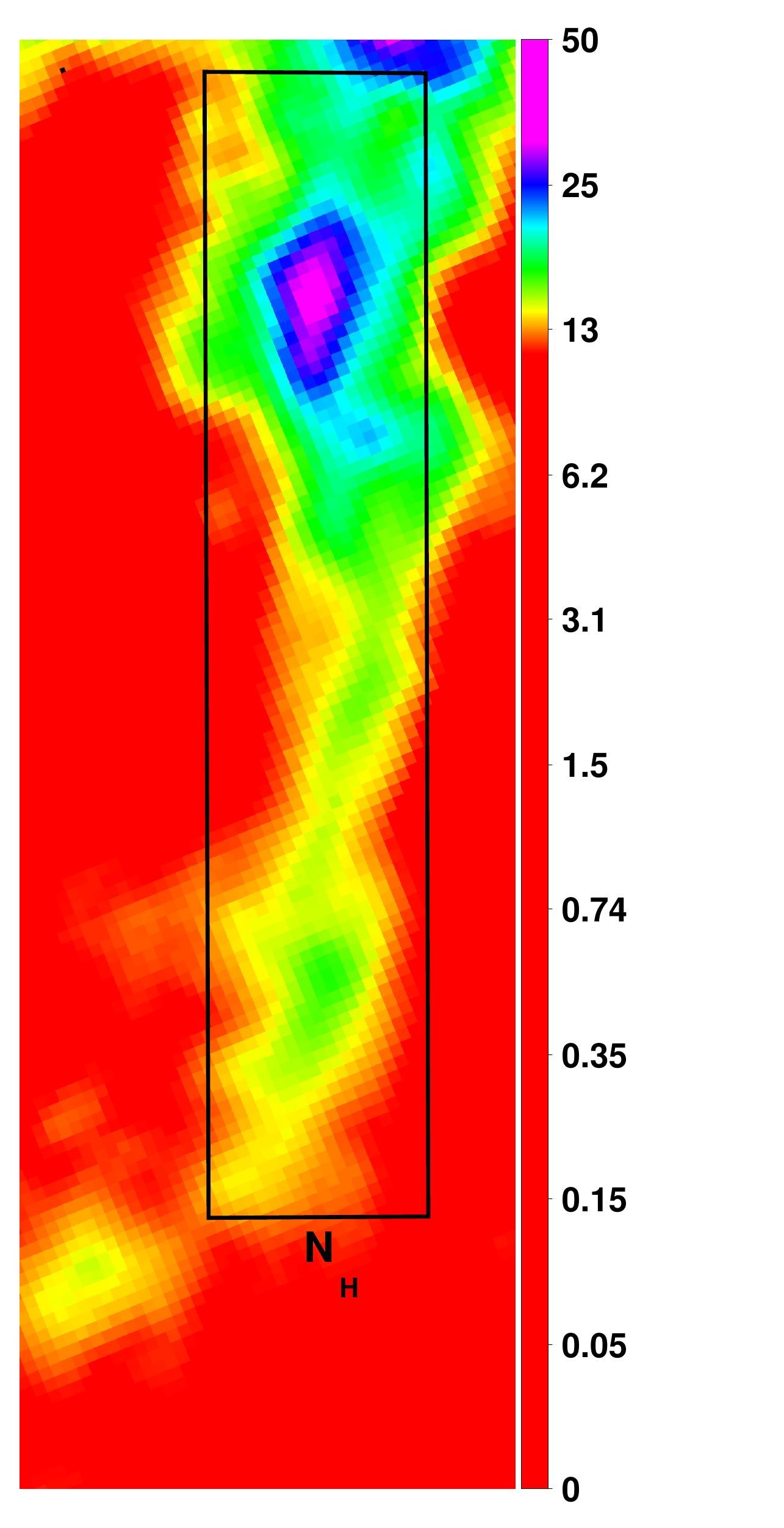}
  \includegraphics[trim=0.8cm 0.05cm 0.8cm 0.45cm,width=2.55cm]{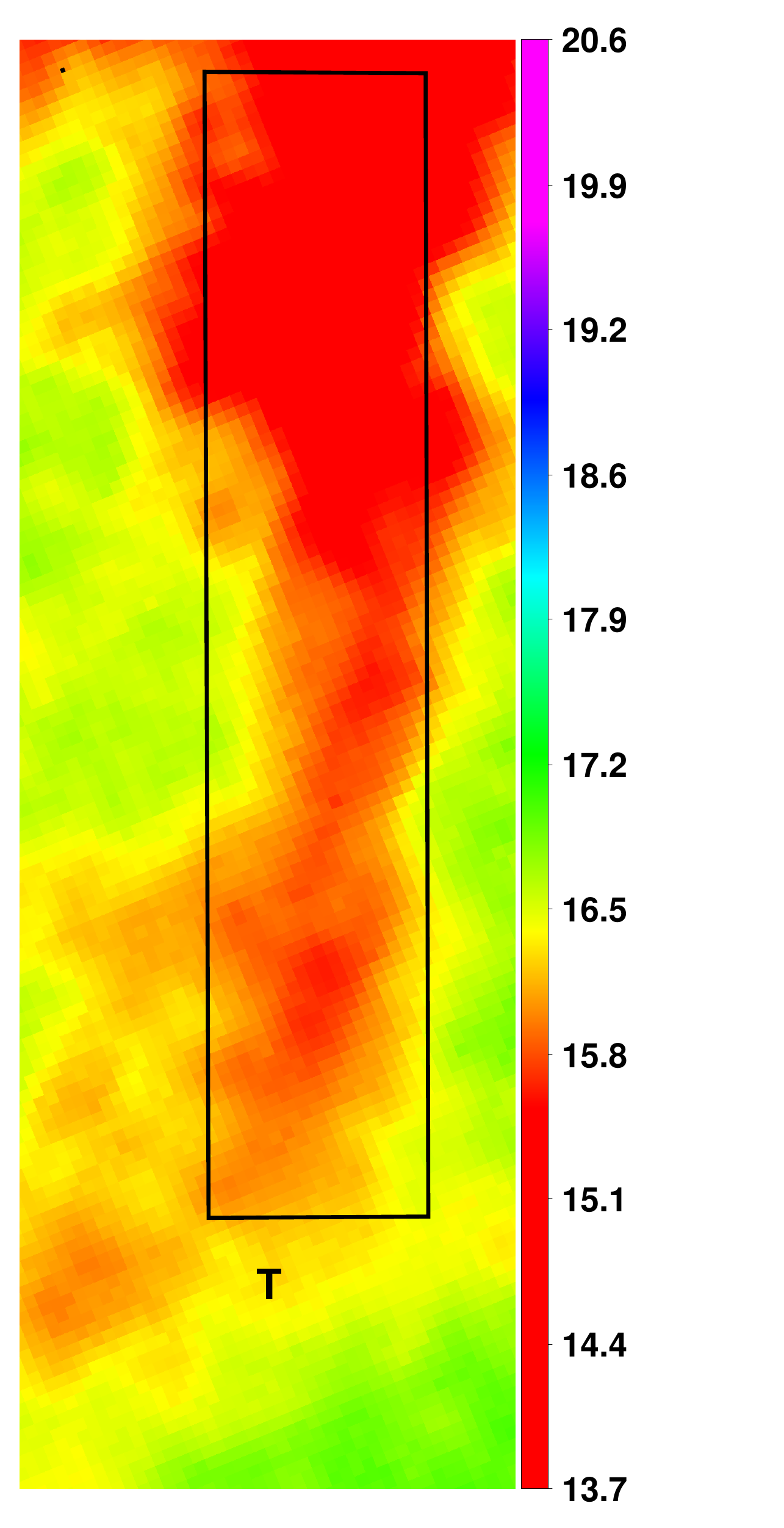}
    \caption{Filament number 8. The panels and the units are as in Figure~\ref{fil1_on}.}
     \label{fil8}
\end{figure*}

\begin{figure*}
\centering
  \includegraphics[trim=0.8cm 0.05cm 0.8cm 0.2cm,width=2.55cm]{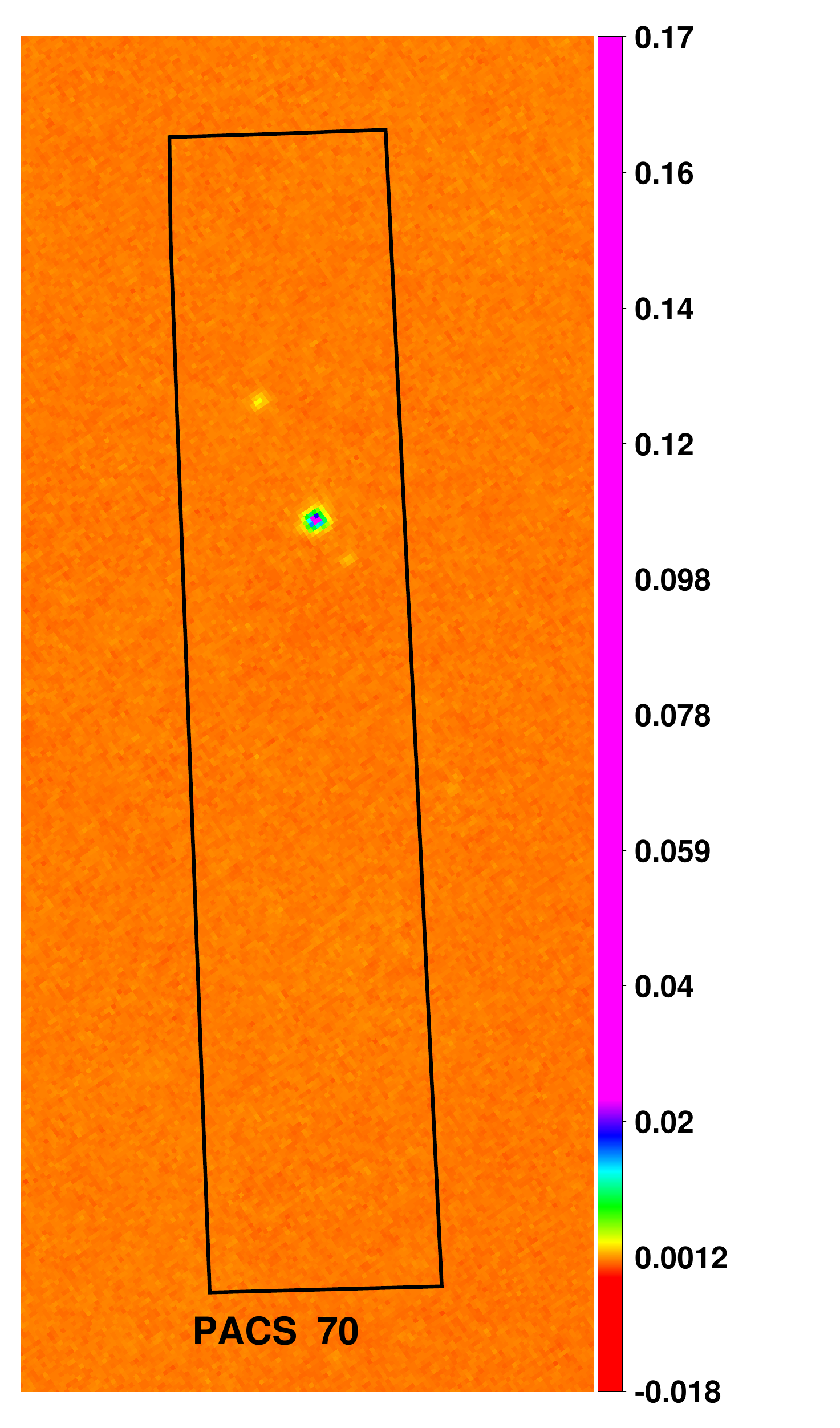}
  \includegraphics[trim=0.8cm 0.05cm 0.8cm 0.45cm,width=2.55cm]{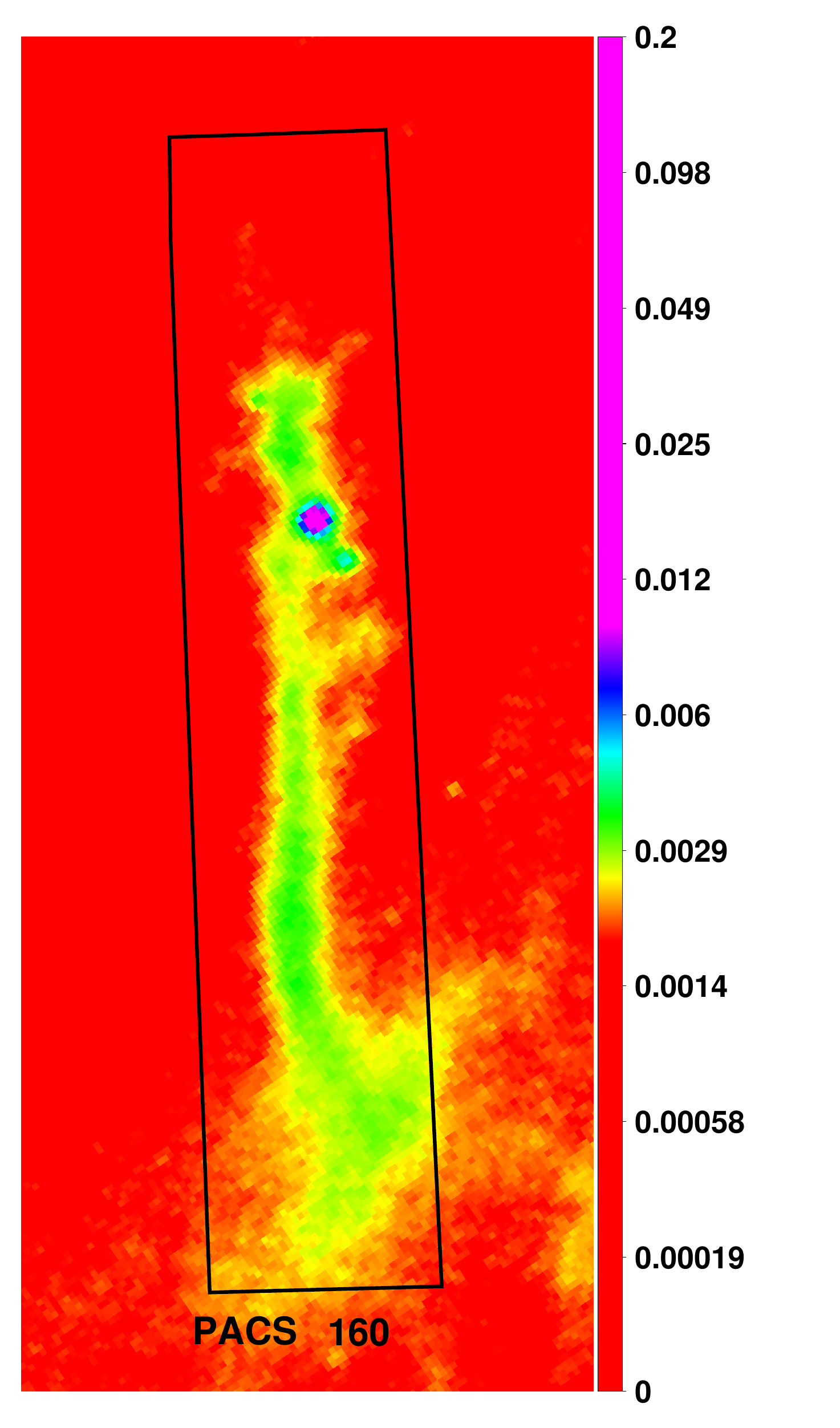}
  \includegraphics[trim=0.8cm 0.05cm 0.8cm 0.45cm,width=2.55cm]{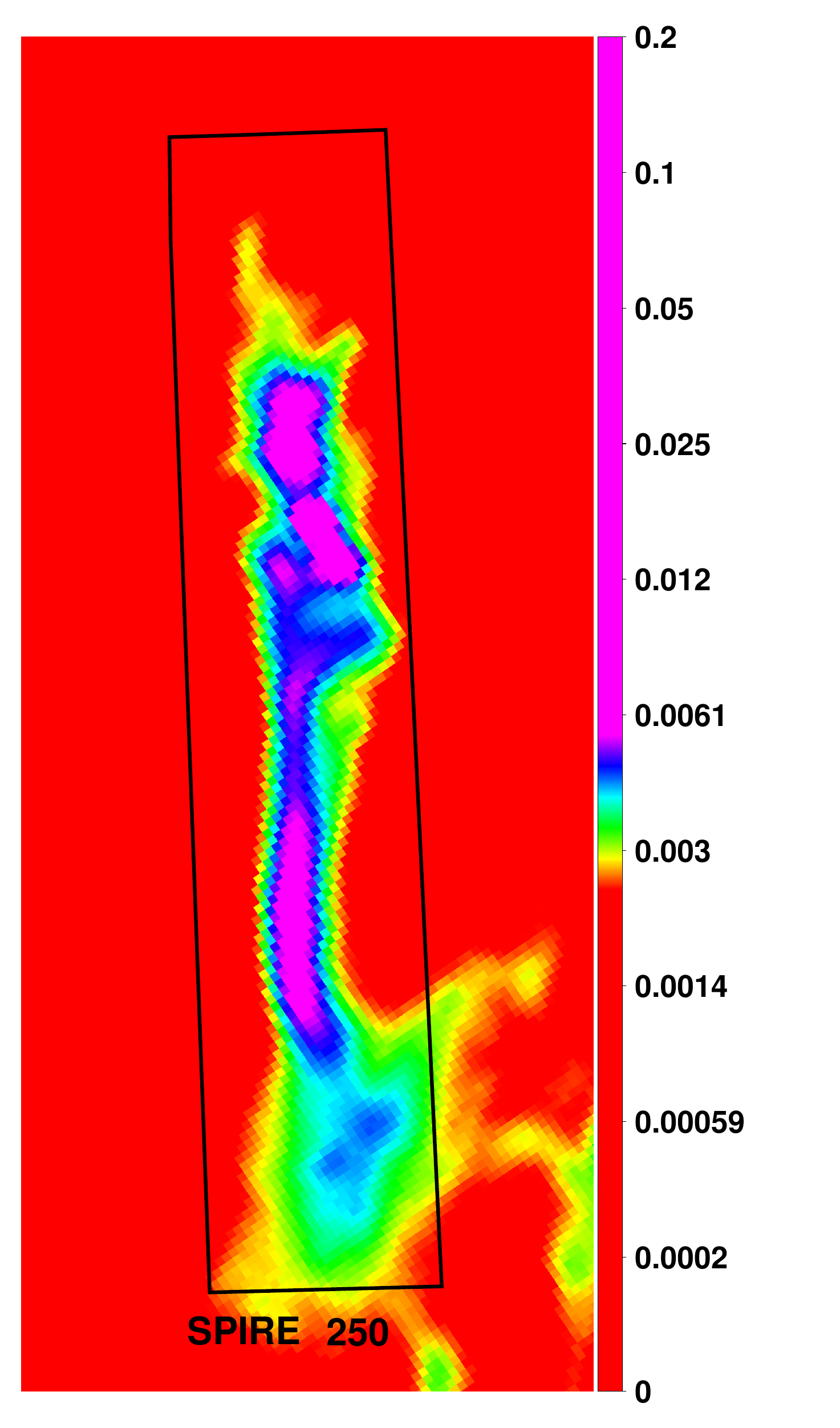}
  \includegraphics[trim=0.8cm 0.05cm 0.8cm 0.45cm,width=2.55cm]{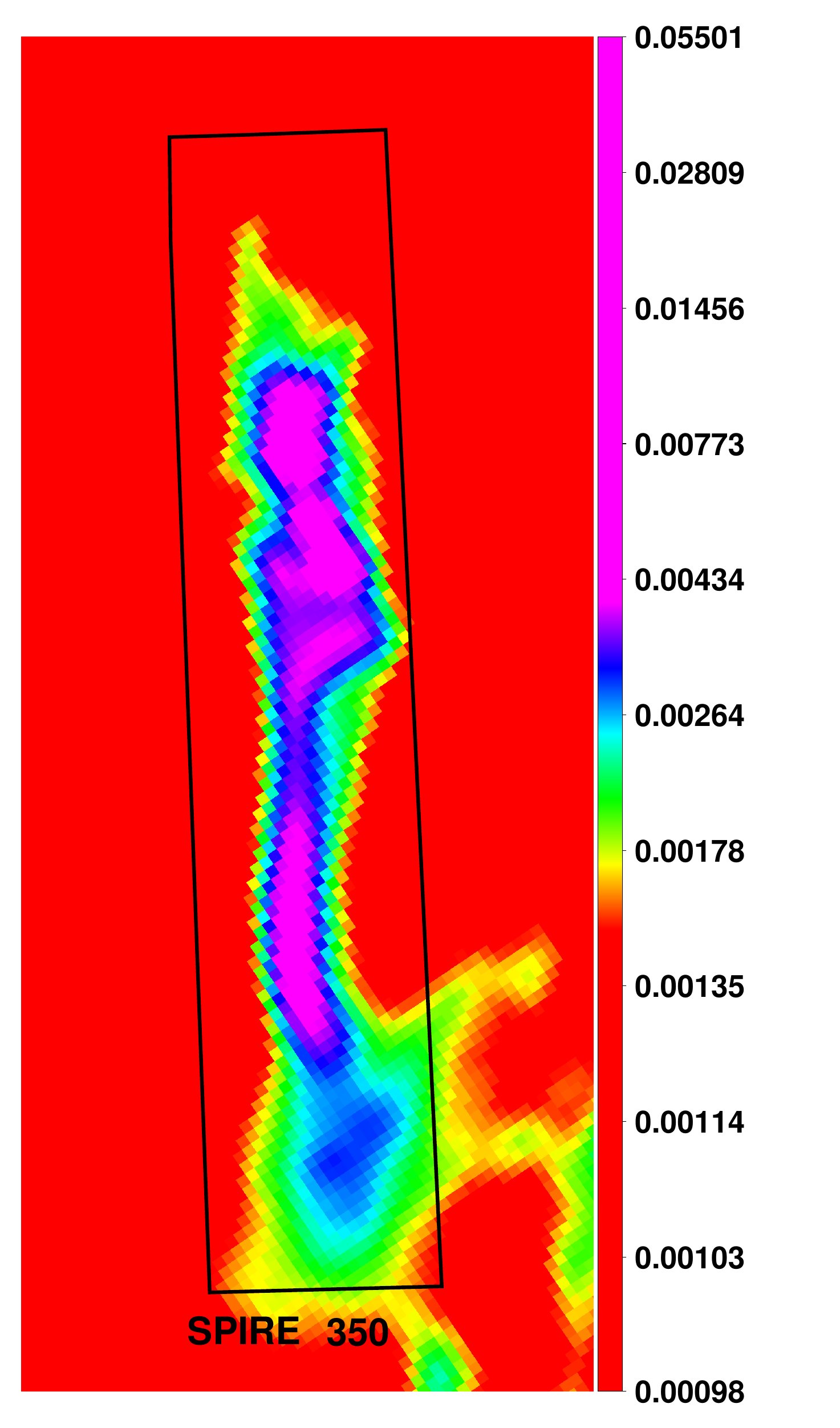}
  \includegraphics[trim=0.8cm 0.05cm 0.8cm 0.45cm,width=2.55cm]{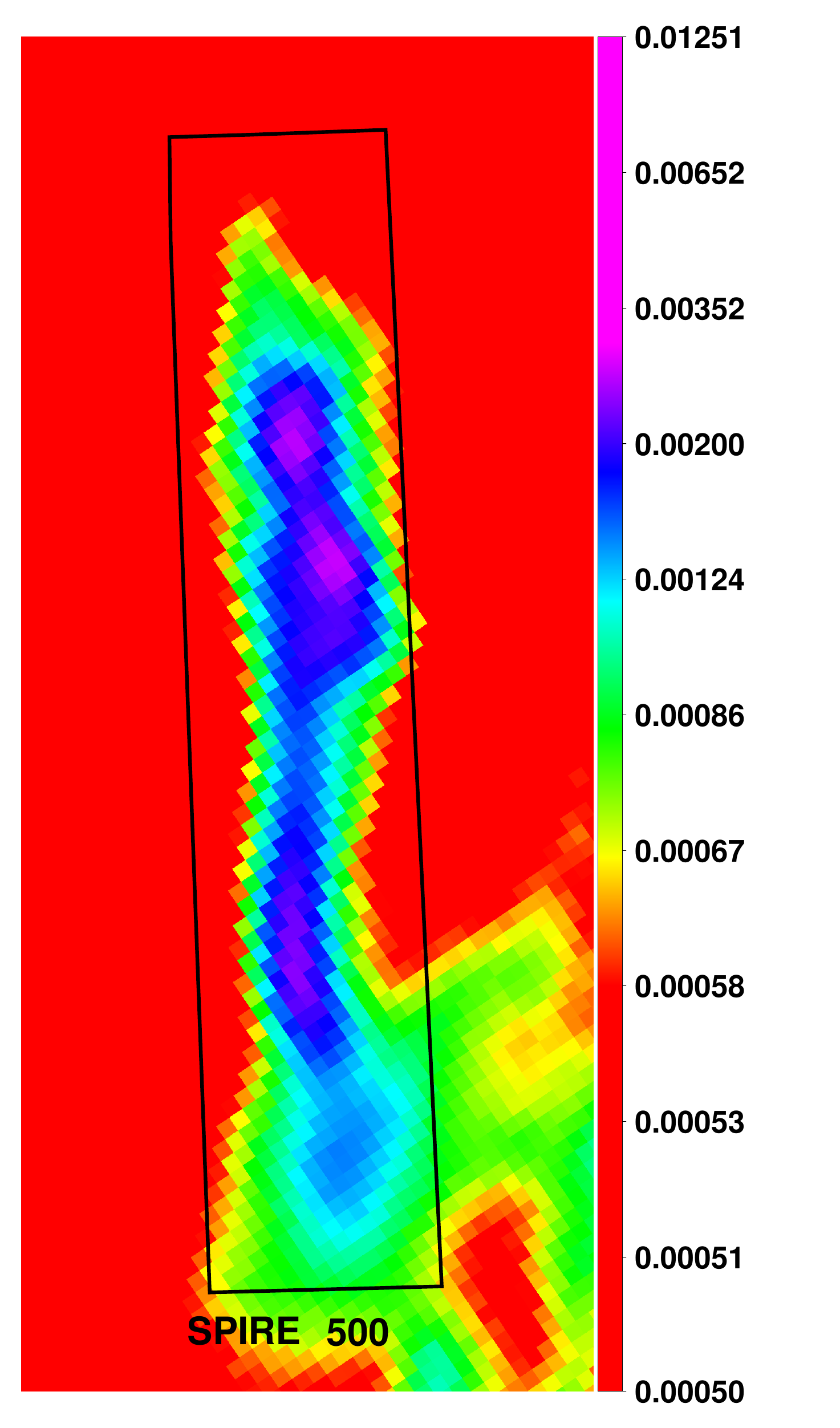}
   \includegraphics[trim=0.8cm 0.05cm 0.8cm 0.45cm,width=2.55cm]{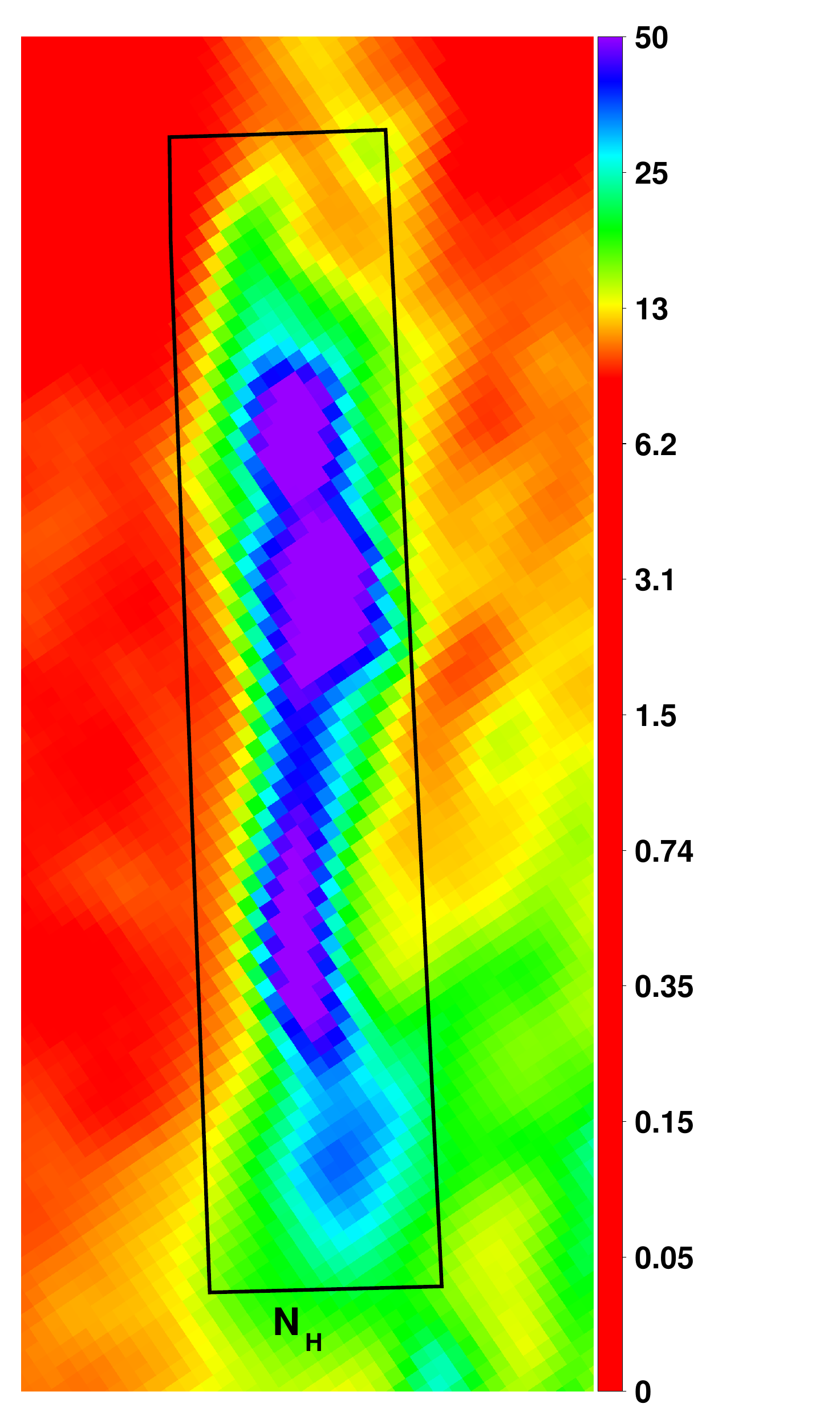}
  \includegraphics[trim=0.8cm 0.05cm 0.8cm 0.45cm,width=2.55cm]{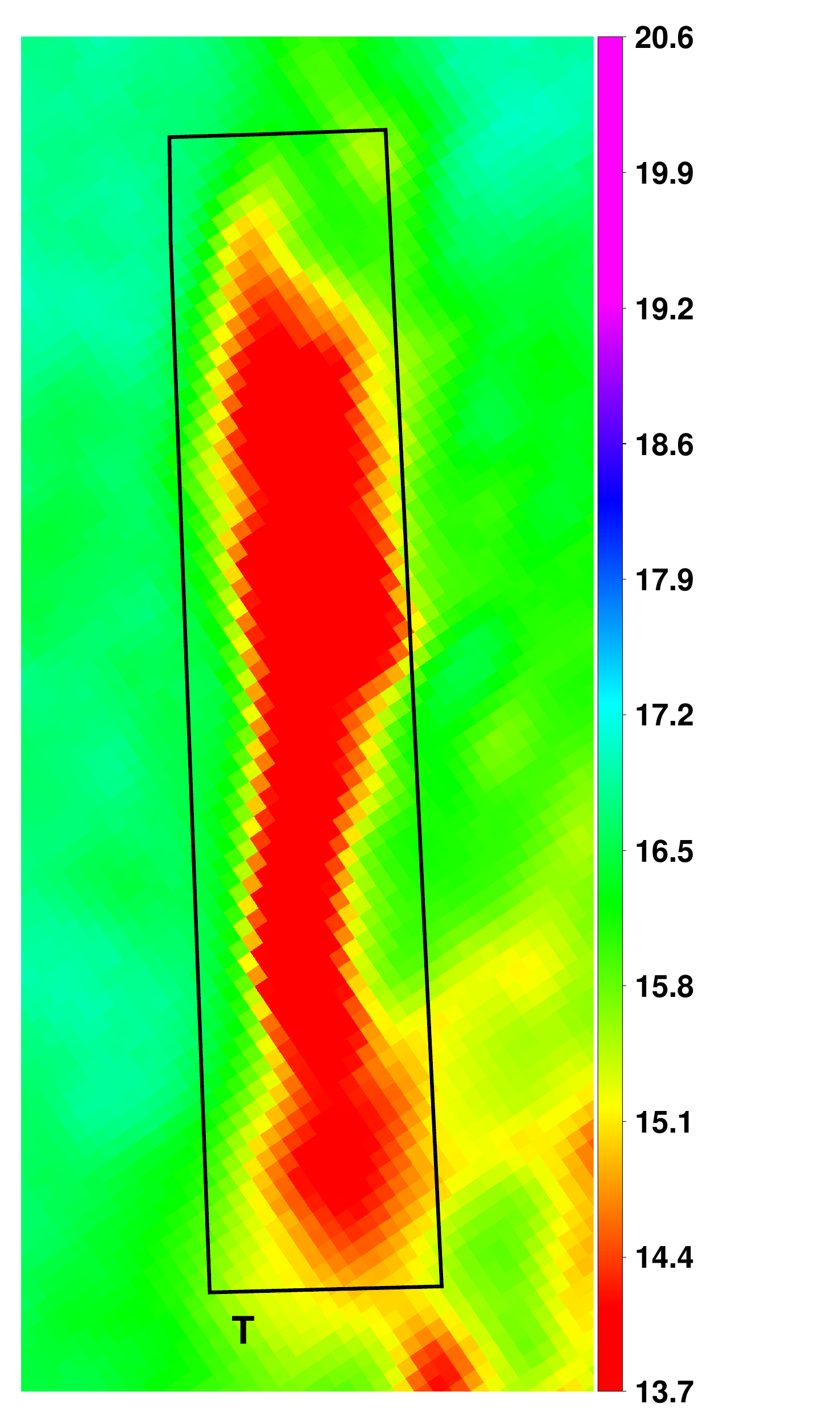}
    \caption{Filament number 9. The panels and the units are as in Figure~\ref{fil1_on}.}
     \label{fil9}
\end{figure*}

}






\end{appendix}

\end{document}